\documentclass[12pt]{article}
\usepackage{natbib}
\usepackage{rotating, amsmath, booktabs, graphicx}
\usepackage{amssymb}
\usepackage{graphicx}
\usepackage{lscape}
\usepackage{tabularx}
\usepackage{ccaption}
\usepackage{dcolumn}
\usepackage{multirow}
\usepackage{nicefrac}
\usepackage{float}
\usepackage{pdflscape}
\usepackage[singlespacing,nodisplayskipstretch]{setspace}
\usepackage{comment}
\usepackage{color}
\usepackage{url}
\usepackage[margin=1in]{geometry}
\usepackage{varioref}
\usepackage{eurosym}
\usepackage[pdfpagemode=UseThumbs,
colorlinks=true,
urlcolor=navy,
citecolor=navy,
linkcolor=navy]{hyperref}
\usepackage{rotating}
\usepackage{xcolor}
\usepackage{wrapfig}
\usepackage[utf8]{inputenc}
\usepackage[T1]{fontenc}
\usepackage{authblk}
\usepackage{soul}

\let\orgautoref\autoref

\renewcommand{\autoref}
{\def\equationautorefname{Equation}%
	\def\figureautorefname{Figure}%
	\def\subfigureautorefname{Figure}%
	\def\sectionautorefname{section}%
	\def\subsectionautorefname{section}%
	\def\subsubsectionautorefname{section}%
	\def\Itemautorefname{item}%
	\def\tableautorefname{Table}%
	\orgautoref}

\newcolumntype{d}[1]{D{.}{.}{#1}}
\newcolumntype{e}{D{.}{.}{-1}}
\newcolumntype{.}{D{.}{.}{3}}
\definecolor{navy}{rgb}{0,0,.5}

\graphicspath{{figs/}}

\begin{document}



\title{Crises and Political Polarization: Towards a Better Understanding of the Timing and Impact of Shocks and Media\thanks{\footnotesize We are grateful to the Rustandy Center for Social Sector Innovation for the generous support, to Salma Nassar for excellent research collaboration, and Marianne Bertrand for priceless guidance throughout this project. We thank Vincent Pons, Giulio Zanella, Nicola Lacetera, and Pol Campos-Mercade for their useful comments. We would also like to thank Greg Saldutte, Kenny Hofmeister, Angela Hsu, and Maggie Li for their support with the analysis and data management; Celia Zhu and Adwaith Kasipur for their excellent research assistance; Patricia Van Hissenhoven, Kelly Ann Hallberg, Carmelo Barbaro, Kirsten Jacobson, and all the staff at the Inclusive Economy Lab and Rustandy Center for Social Sector Innovation for their comments and support. We thank seminar and conference participants of the Economic Science Association Conference, the Behavioral Insights Global, and the European University Institute. This study obtained ethics approval from the University of Chicago IRB No. 20-0471.}

\vspace{0.25cm}}

\author{Guglielmo Briscese\thanks{Corresponding author. University of Chicago \href{mailto:gubri@uchicago.edu }{\texttt{gubri@uchicago.edu}}.}
\and Maddalena Grignani\thanks{Pompeu Fabra University, \href{mailto: Maddalena.Grignani@upf.edu}{\texttt{maddalena.grignani@upf.edu}}.}
\vspace{0.1cm}
\and Stephen Stapleton\thanks{University of Chicago, \href{mailto: sstapleton@uchicago.edu}{\texttt{sstapleton@uchicago.edu}}.}
\vspace{0.1cm}}

\date{\vspace{-5ex}}

\maketitle


\begin{abstract}
\noindent 
We investigate how crises alter societies by analyzing the timing and channels of change using a longitudinal multi-wave survey of a representative sample of Americans throughout 2020. This methodology allows us to overcome some of the limitations of previous studies and uncover novel insights: (1) individuals with a negative personal experience during a crisis become more pro-welfare spending, in particular for policies they perceive will benefit them personally, and they become less trusting of institutions; (2) indirect shocks or the mere exposure to the crisis doesn't have a similar effect; (3) policy preferences and institutional trust can change quickly after a negative experience; and (4) consuming partisan media can mitigate or exacerbate these effects by distorting perceptions of reality. In an experiment, we find that exposing individuals to the same information can recalibrate distorted perceptions with lasting effects. Using a machine learning model to test for heterogeneous treatment effects, we find a negative personal experience did not make individuals more responsive to the information treatment, suggesting that lived and perceived experiences play an equally important role in changing preferences during a crisis.

\end{abstract}

\newpage


\newpage

\begin{spacing}{1.5}
	
\section{Introduction}

Large-scale crises can change societies by shifting citizens' preferences and beliefs, but there is limited evidence on how and when this occurs. Some studies have shown that crises advance support for more generous welfare spending \citep{gualtieri2019repeated, margalit2019, garand2010, cogley2008, piketty1995}, but others claimed they only reinforce pre-existing beliefs, further increasing political divisions \citep{alesina2020, kranton2016, cook2016, lewandowsky2012}. Further, most studies have relied on surveys implemented with large time gaps and on cross-sectional samples, making it difficult to understand the drivers and timing of such changes, if any.


\par We implement a seven-wave survey on a longitudinal representative sample of Americans that we track from April to October 2020, an eventful period characterized by the unfolding of the global COVID-19 pandemic. Across survey waves, we track respondents' preferences for welfare and temporary relief policies, their trust in institutions, and how they processed information about the ongoing crisis. In addition to a rich set of socio-economic and demographic controls, we record respondents' direct and indirect experiences with the crisis, as well as their media diet.

\par We find that support for welfare spending increases after a direct shock, such as losing a significant portion of income or having a family member or close friend hospitalized with the virus. In particular, individuals who were directly affected by the crisis became more supportive of temporary relief policies such as giving money to families and businesses, and protect essential workers. Personal negative experiences also increased support for  more politicized welfare policies, such as universal basic income, greater spending to help the elderly, financial assistance to low-income students, and price controls. At the same time, support for other economic policies, such as subsidies to companies, decreased, suggesting that citizens favored welfare spending over private-sector-led economic policies. Across these policies, however the pattern is less consistent compared to the temporary relief policies, suggesting that political identity played a mitigating role. 

Consuming mostly partisan-leaning news contained the effects of direct shocks and increased polarization of views between Democrats and Republicans on institutions and policies. We show that this media-induced polarization is due to a misperception of the gravity of the crisis. By May 2020, Republicans who consumed predominantly Republican-leaning news were more likely to underestimate the COVID-19 death rate, while Democrats who consumed Democratic-leaning news were more likely to overestimate it. To study whether these beliefs were caused by motivated reasoning or lack of objective information, an experiment was conducted where half of the respondents saw a link to the webpage of the U.S. Centers for Disease Control and Prevention (CDC) to check the official COVID-19 death rate. The results show that this intervention significantly re-aligned respondents' beliefs to the official death rate and changed their judgment of how public authorities handled the crisis. The effect persisted four months later. Using a causal forest methodology to estimate heterogeneous treatment effects \citep{athey2019estimating}, we find a negative personal experience with the crisis did not translate to individuals being more responsive to the treatment, suggesting that correcting media-induced misperceptions can still be beneficial regardless of a person’s direct experience with a crisis. These results are robust to several specifications, alternative measures of shocks, alternative survey weights, and bundling of outcomes.

\par Our study makes several contributions. First, we show that crises can alter citizens' policy preferences mostly when they are directly affected. Second, negative experiences don't increase support for greater welfare spending per se, but rather on policies that can benefit affected citizens directly and that aren't necessarily associated with a political ideology. Third, in contrast with previous studies, these changes can occur very rapidly \citep{fuchs2015}. Fourth, we show that divisions and polarization during a crisis are not the result of negative personal experiences but rather exposure to political-leaning media. In a review of related studies, \citet{margalit2019political} notes that the effect of economic crises on political polarization is puzzling: on the one hand, a crisis can increase support for welfare spending, thus possibly nudging voters to left-leaning parties, while on the other hand, it can reduce trust in institutions, potentially pushing voters towards more populist, right-wing, parties. We argue that previous studies seem to show mixed findings because they didn't control for media consumption. We find that while shocks decrease trust in institutions, the polarization between Democrats and Republicans is better explained by their media diet and that this shaped their understanding of the gravity of the crisis, in line with \citet{bursztyn2020misinformation}. Lastly, we offer methodological contributions to the growing literature on survey experiments \citep{mernyk2022correcting, bursztyn2021misperceptions} showing that these low-cost interventions can have long-term effects regardless of a person's experience with a crisis.

\par The paper is structured as follows. Section 1 describes the methodology and the outcomes we tracked over time - namely, support for welfare and temporary relief policies, trust in institutions, and understanding of the gravity of the crisis. In section 2, we explain how we define the different types of shocks and how we calculated respondents' bias in media consumption. In the third section, we disentangle how shocks and media shaped Americans' beliefs during the crisis. We then zoom in on the effects of biased media consumption on the understanding of the gravity of the crisis and present the results of the survey experiment to correct for misinformation. Section 4 reports a set of robustness checks, showing that our results are consistent across several changes in assumptions and model specifications. We conclude with a summary of our findings in the final section.

\section{Methodology}

We partnered with NORC at the University of Chicago, the organization responsible for administering the General Social Survey (GSS), to implement a survey to a sub-sample of their AmeriSpeak® Panel\footnote{Funded and operated by NORC at the University of Chicago, AmeriSpeak® is a probability-based multi-client household panel sample designed to be representative of the US household population. Randomly selected US households are sampled using area probability and address-based sampling, with a known, non-zero probability of selection from the NORC National Sample Frame. These selected households are then contacted by US mail, telephone, and field interviewers (face-to-face). The panel provides sample coverage of approximately 97\% of the U.S. household population. Those excluded from the sample include people with P.O. Box only addresses, some addresses not listed in the USPS Delivery Sequence File, and some addresses in newly constructed dwellings. While most AmeriSpeak households participate in surveys by web, non-internet households can participate in AmeriSpeak surveys by telephone. Households without conventional internet access but with web access via smartphones are allowed to participate in AmeriSpeak surveys by web.}. We recruited 1,440 U.S. citizens (see Table \ref{tab:summary_stats} in the Appendix for a summary of demographic and socio-economic characteristics), which we interviewed seven times between April and October 2020\footnote{The use of a longitudinal multi-wave panel survey has several advantages. First, we are able to choose the timing and frequency of our survey waves in a way that best allows us to answer our research questions, without the need to wait for two years or more in between data collection periods. Second, we can ask the same set of questions more than twice, thus reducing any possible volatility or inconsistency in respondents' answers. Third, we minimize the risk of recollection bias, as events that occurred in a person's life are more salient, which gives respondents a better opportunity to provide more accurate answers about their economic and health situation during a crisis. At the same time, this methodology doesn't force us to ask questions about preferences and shocks within the same survey wave, which might bias respondents' answers. Fourth, because we follow the same panel of respondents over time, we have baseline data that we can compare against when evaluating changes to their views accounting for their point of departure. This is particularly important when analyzing whether crises lead to convergence (e.g., increasing support for welfare policies among those who were previously not supporting it) or polarization (e.g., decreasing or increasing support for a policy among those who did not have a strong opinion).}. In the first wave of the survey, we collected baseline data on the main outcomes of interest (e.g., policy preferences and trust in institutions) as well as media consumption and beliefs (e.g., political ideology). The subsequent weekly waves allowed us to track respondents' lived experiences during the dramatic first month of the pandemic. The next two waves were administered monthly, on the week commencing May 18 and June 22, 2020, respectively, and recorded respondents' perception of the gravity of the crisis. These waves focused on how Americans were coping in the weeks immediately after a possible health or economic shock, while the event was still vivid in their minds\footnote{In order to minimize possible priming bias, we always left the shock questions at the end of the survey. Further, while we collected information on economic and health shocks in every wave, and in the last wave, we asked respondents to report these shocks on a monthly and more detailed basis}. Lastly, we implemented a seventh and last wave of the survey in the week commencing October 19, 2020. We purposely timed the last wave to track any changes to respondents' beliefs and preferences immediately prior to the Presidential elections. The summary of the questions asked in each wave is presented in \autoref{tab:questions}.

\subsection{Outcomes}

Across survey waves, we collected participants' responses to the following set of outcomes: (i) preferences for welfare policies, (ii) preferences for temporary relief policies, (iii) trust in institutions, and (iv) how respondents perceived the gravity of the crisis. \\

\textbf{Preferences for welfare policies}. To study how the crisis affected preferences for welfare policies, we administered a module of questions based on the GSS questionnaire, which asks respondents whether they think it should be the government's responsibility to intervene in a series of policy areas. Respondents can provide an answer for each of these policies on a 4-point scale from ``\textit{Definitely should not be}'' to ``\textit{Definitely should be}.'' The policy areas are the following: (1) provide mental health care for persons with mental illnesses, (2) help individuals affected by natural disasters, (3) keep prices under control, (4) provide a decent standard of living for the old, (5) provide a decent standard of living for the unemployed, (6) provide everyone with a guaranteed basic income, (7) provide the industry with the help it needs to grow, (8) reduce income differences between the rich and the poor, (9) give financial help to university students from low-income families\footnote{In our survey we replicate the exact wording of the GSS survey. Later we compare our baseline findings to previous GSS waves.}. We asked these questions in waves 1, 4, and 7. In addition, we also asked respondents a question about universal healthcare. The question read as follows: ``\textit{Do you favor or oppose a universal health care system covered by the government so that every American can have equal access to health care, even if this means that you will have to pay higher taxes?}''\footnote{Response options were on a 5-point scale from ``\textit{Strongly oppose}'' to ``\textit{Strongly favor}''}\footnote{We purposely asked this question in a way that encouraged respondents to think carefully about costs and benefits of a universal healthcare system, and limited saliency bias that might arise from the ongoing crisis on universal health care.}. We also asked this question in waves 1, 4, and 7 of our survey. \\ 

\indent \textbf{Preferences for temporary relief policies}. In addition to tracking Americans' preferences for the role of government in the economy, we also tracked their support for the less politicized welfare policies that federal and state governments considered adopting to respond to the crisis \citep{druckman2021, druckman2013}\footnote{Indeed, recent surveys suggest that, despite deepening partisan divisions, Americans tend to agree on several policy areas. See, for instance: \url{https://cgoap.net/ and https://vop.org/wp-content/uploads/2020/08/Common_Ground_Brochure.pdf}}. These questions were asked in waves 4 and 7, and they elicited respondents' agreement on the following statements: (1) ''\textit{the government should transfer money directly to families and businesses for as long as lockdown measures are kept in place}'', (2) ''\textit{the government should do more to protect essential workers from contracting the virus}'', (3) ''\textit{the government should spend more on public healthcare to reduce the number of preventable deaths}''. \\

\indent \textbf{Trust in institutions}. Since previous studies have documented that economic crises result in loss of trust in institutions \citep{algan2017, dotti2016}, we measure how trust in institutions might have changed during this crisis by asking our respondents the following set of questions, replicating the wording of the GSS: ''\textit{How much confidence do you have in the people running the following institutions?}''\footnote{Like the GSS questions, response options were on a 5-point scale, from ''\textit{Complete confidence}'' to ''\textit{No confidence at all}''}. The list of institutions was the following: (1) U.S. Congress and Senate, (2) White House, (3) scientific community, (4) banks and financial institutions, (5) private sector, (6) hospitals and healthcare professionals, (7) health insurance companies. We asked all of these trust questions in waves 1, 4, and 7.


\subsection{Shocks}
We focus on four types of shocks: direct and indirect, economic, and health. Direct shocks refer to major life events that affected the respondents personally, while indirect shocks refer to exposure to a crisis in the areas where they lived.

\indent \textbf{Economic shocks}. To measure direct economic shocks, we asked all respondents in the last wave of the survey to report their (and their spouse, if present) monthly gross income between February and October\footnote{In addition to asking in most waves whether respondents incurred any economic or health shock, in the last wave, we asked them to report the exact amount of household income for every month as well as if they knew anyone hospitalized each month. This allows us to have a more granular and quantifiable measure of economic shock beyond the timing of our survey waves}. Further, we also asked respondents' (and their spouse, if present) monthly additional sources of income, the monthly number of hours worked, and whether they received any financial support from the government or non-government organizations at any time during the crisis. This data allows us to estimate both the timing and the magnitude of the economic shocks incurred by respondents' households between waves. We measure direct income shocks in two different ways, and we show that they provide comparable results. In our main specification, we consider whether respondents have lost more than 20\% of their income (combining both incomes from work and other sources) between any two months between February (or the baseline month) and October (or the outcome month) 2020 to capture the effects of a sudden large drop in income. In the Appendix, we show that the results remain unchanged when adopting a less stringent measure of 10\% income loss between any two months. 

$$shock_1 = \begin{cases} 1, & \mbox{if } \frac{income_{t}-income_{t-1}}{income_{t-1}}\leq -0.20 \\ 0, & \mbox{otherwise } \end{cases} $$

In our sample of respondents who participated in the first and the last survey waves (i.e., \textit{n}=1,076), we find that about 38\% of respondents lost at least 20\% of their household income, between any two months, between February and October 2020\footnote{As reported in Table \ref{tab:balancetable_shock} in the Online Appendix, respondents who lost at least 20\% of their household income between any two months from March to October 2020 are more likely to be young, with a low baseline income and to belong to a racial minority group. Furthermore, women, Democrats, and those who live in a metropolitan area have incurred such a negative income shock with a marginal significantly higher probability, while co-habitation (or marriage) seems to smooth the financial impact of the pandemic. We control for all these characteristics in our analysis and show how using different specifications does not change our main results}.

In addition to measures of personal economic shocks, we also control for indirect economic shocks since it is possible that many Americans changed their preferences just by mere exposure to the crisis, such as by knowing someone who got affected economically by the crisis or living in an area that suffered a relatively higher economic distress compared to others \citep{dyer2020covid, wright2020poverty}. Measuring economic variations between two months of the same year, however, is a challenge. Many macroeconomic indicators, such as unemployment rate or business closures, are rarely available at the county level, and often they are only released at an aggregate level or on a frequency that is less regular than the timing of our survey waves, making any meaningful comparison difficult. Therefore, we use data collected and updated in real-time by Harvard's Opportunity Insights team on the weekly percentage variations in consumer expenditures with respect to the first week of January 2020 \citep{chetty2020}. This variable is seasonally adjusted and is available at the county level, which we match with the respondents' residential information\footnote{The Opportunity Insights team uses anonymized data from several private companies to construct public indices of consumer spending, employment, and other outcomes. See \citet{chetty2020} for further information on series construction.}.

\indent \textbf{Health shocks}. Our main measure of direct health shock is whether respondents had a family member, a friend, or an acquaintance who was hospitalized with COVID-19\footnote{We consider this combined measure, as 2.4\% of the respondents has a family member who has been hospitalized, 9.8\% has a relative, 14.1\% a friend and 14.9\% an acquaintance. To control for additional direct health shocks, we also asked respondents their type of health insurance (e.g., public or private), whether they have caring responsibilities towards an elderly or someone with disabilities, which are at greater risk of complications from contracting the virus, and if they knew a healthcare professional who had been working closely with COVID-19 patients}. About 30\% of our respondents knew someone (among family, friends, or acquaintances) who was hospitalized with COVID-19, while 69\% knew someone who tested positive. About 33\% tested positive for COVID-19 themselves.

Similarly to measuring indirect economic shocks, collecting information at a zip code or country level on indirect health shocks, such as COVID-19 cases and deaths, is complicated. In the early stages of the pandemic, U.S. states followed different guidelines for testing and recording COVID-19 cases and deaths. While our survey asked also whether respondents knew a family member, friend, or acquaintance who tested positive, we complemented it with data on the number of COVID-19 cases in their county of residence. While this measure might be subject to different protocols depending on the State, these figures were likely to be the same ones reported by the local media. We consider COVID-19 cases\footnote{We exploited the data collected by the New York Times from state and local governments and health departments, and available here \url{https://github.com/nytimes/covid-19-data}.} at the county level reported by the middle of each week. We then consider the population size at the county level in 2019 and construct the following measure\footnote{We multiply this measure by 100 to ease the interpretation of the coefficients in our regressions} of increase in cases between week \textit{t} and \textit{t-1} in county \textit{c}: $\frac{cases_{ct} - cases_{ct-1}}{population_c}$. When, instead, we consider an outcome that is not in changes, we focus on the logarithm of the cumulative number of cases weighted by the county population: $ln \left( \frac{cases_{ct}}{population_c} * 100,000 \right) $. 

\subsection{Media consumption} To understand how the media might have shaped Americans' views, we collected information on respondents' preferred news sources (including international news and social media) and the number of hours of news they consumed\footnote{The question asked: ``Do you get your news from any of these sources (either on television or on the internet)?'', and the multiple option answers were: ``ABC, CBS, NBC, CNN, Fox News, MSNBC, and 'other, please specify''' (e.g., some respondents added The NY Times, The Washington Post, BBC, NPR, and PBS). While there is no exact methodology to measure the partisan bias of news sources \citep{budak2016fair, groseclose2005}, and since within each source different programs might cover the same news in different tones \citep{bursztyn2020misinformation}, we measured whether respondents were exposed to different points of view during the crisis.}. Based on the news sources indicated by our respondents, we constructed a ``bias score'' using the ``\textit{AllSides.com}'' platform, one of the most commonly used sources of partisan media bias analysis\footnote{\url{https://www.allsides.com/unbiased-balanced-news}}. The website assigns a score from 1 (Extremely Left) to 5 (Extremely Right) to major sources of news by analyzing their written content on a regular basis.

\par Matching the scores by Allsides\footnote{We use the scores of the first week of April 2020, our baseline wave} to the respondents' choices, we create an index summing the scores of each source consulted by an individual and divided by the maximum number of possible points. 

\begin{center}
\textit{Media slant index, for an individual consuming N sources of news} = $ \frac{\sum_{n =1}^{N} score_{n}}{N} $
\end{center}

This variable measures how politically homogeneous the news sources that respondents consumed are, by taking any value between 1 and 5: the closer a respondent is to 1, the more they consume homogeneous (i.e., less politically diversified) left-leaning media, while the closer they are to 5 the more homogeneous and right-leaning is their media consumption. A score towards the middle indicates either that respondents consume unbiased news, or that they consume news that is biased in both directions, and so that they are exposed to both sides. Based on this specification, we see that 51\% of the Republicans consume Republican-leaning news, and 46\% of the Democrats consume Democratic-leaning news. Among independents and non-voters, around 25\% (24\%) consume Republican-leaning news (Democratic-leaning news). 

\subsection{Estimation}

To estimate changes in main outcomes (i.e., preferences for welfare and economic policies, preferences for temporary relief policies, and trust in institutions), we rely on the same estimation approach. For brevity, we present the approach referring to trust in institutions as an example. 
Since most of our outcomes are measures in a Likert-scale, we construct a variable equal to one if the respondent decreased (increased) their confidence in a given institution (their support in a policy), between the first and the last wave\footnote{In the Online Appendix, in Tables \ref{tab:decr_policy}, \ref{tab:decr_covid_policy}, and \ref{tab:incr_trust} we replicate the same analyses with the inverted binary variables, i.e., decreased support for policies and increased trust for institutions, and show that the results are unchanged.}. This approach allows us to overcome some of the limitations of survey-based measures previously highlighted by \citet{bond2019sad} and \citet{zanella2019}. We flag respondents who could not have further decreased (increased) their trust (policy preference), since they had already given the minimum (maximum) value on the given scale in the baseline (i.e., wave 1). We then estimate the following linear probability regression, considering only the respondents who participated in both waves:

$$Y_{ic}= \alpha + X_i \beta + S_i\theta_1 + Z_c \theta_2 + Yb_{i}\gamma + \epsilon_{ic} $$

\noindent with $Y_{ic}$ being a dummy variable equal to 1 if respondent \textit{i}, living in county \textit{c}, decreased (increased) their level of trust in a certain institution (or support for a policy) between the first and the seventh wave (and between the fourth and seventh wave for temporary relief policies). $X_i$ is a vector of time-invariant demographic characteristics; $S_i$ is a vector including the direct health and economic shocks that affected respondents between survey waves when we collected outcome measures; $Z_c$ is a vector of indirect health or wealth shocks at the county (or zip code) level, reported in variation between the first and the last wave (and the fourth and last wave for the temporary policies). $Yb_{i}$ is a dummy variable equal to 1 if the respondent was at the lower bound in wave 1, i.e., if they already gave the highest or lowest score, i.e., could not possibly further decrease (or increase) their score. Further, we flag respondents who completed the surveys in a time equal to or shorter than the first percentile of sample duration, which we consider a proxy of limited attention during the survey. Given that we consider multiple outcomes, in our robustness checks\footnote{See tables \ref{tab:AES_policy}, \ref{tab:AES_covid_policy}, and \ref{tab:AES_trust}, in the Online Appendix}, we replicate our analyses using Average Effect Sizes (AES), as in \citet{kling2004moving, clingingsmith2009estimating, heller2017thinking}. We also replicate the regressions using a fixed effect model with data in panel format\footnote{See Tables \ref{tab:fe_policy_1}-\ref{tab:fe_trust_2}, in the Online Appendix}, and we vary how we measure shocks\footnote{See tables \ref{tab:covid_policy_inc23}-\ref{tab:trust_inc3} in the Online Appendix}, showing that the main results remain unchanged. All regressions presented throughout the paper use survey weights\footnote{In all our regressions, we apply survey weights, making our sample representative of the U.S. population, and we adjust the standard errors considering the primary sampling units (PSUs) and strata that the population was divided into. Survey weights are recalculated in every wave to keep the sample representative of the population. In section C1 (Tables \ref{tab:balancetable_attr} and \ref{tab:predict_attrition}) in the Online Appendix, we present the analyses on survey attrition and show that these are not correlated with the outcomes.}.

\section{Results}
We begin by looking at the overall support for policies and institutional trust across survey waves. 
We implemented the first survey wave shortly after COVID-19 cases started soaring in the U.S. In Tables  \ref{tab:GSS_policies}, \ref{tab:GSS_trust1} and \ref{tab:GSS_trust2}, in the Appendix, we report the share of respondents, by political party, that supported each policy and trusted each institution, which will serve as the baseline for our analysis. From the tables, one can also notice that our baseline values already differ from the last available GSS survey from 2016. This can be due to many reasons, including four years of Trump presidency that might have increased political polarization on several topics, the start of the 2020 Presidential campaign that dominated the media discourse, or the first signs of a crisis-triggered polarization. As we will show in this section, most of these values fluctuated throughout the course of the year, suggesting that the implementation of our first wave was still timely. This is also supported by the fact that most of our respondents had not incurred a direct shock yet at the time of our first wave. Further, these baseline levels are another testament to the importance of using longitudinal repeated surveys with shorter time gaps to study how preferences change during a crisis.
As we discuss in greater detail in sections \ref{sec:policies} and \ref{sec:trust_inst}), we check for pre-trend effects by running a series of comparison tests on all welfare policies and institutional trust in wave 1 between respondents who later incurred a shock and those who didn't (see Tables \ref{tab:policy_pretrends} and \ref{tab:institution_pretrends}) and find no significant differences. These preliminary insights give us confidence in the external validity of our data and confirm the timeliness of the implementation of our study.

\par In the next subsections, we report the results of the regressions estimating changes in preferences for policies and trust in institutions, followed by a subsection on how an information experiment can help explain the polarizing effects of partisan media. 

\subsection{How shocks and media change support for policies}\label{sec:policies}

\begin{figure}[H]
	\caption{The effect of shocks and media on welfare policy preferences}
	\label{fig:welfare_preferences}
	\begin{center}
		\includegraphics[height=19cm]{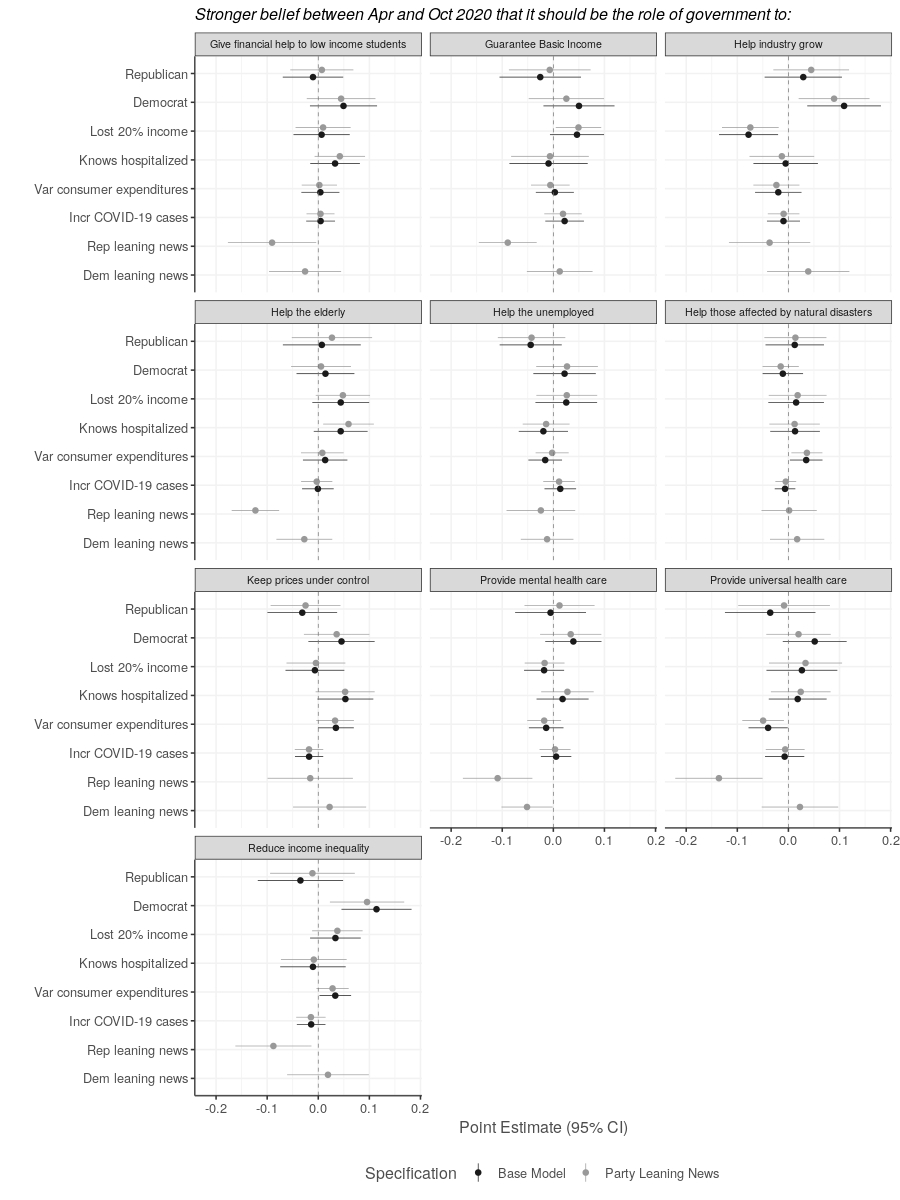}
	\end{center}
\begin{minipage}{1\linewidth \setstretch{0.75}}
	{\scriptsize{\textit{Notes}:} 
		\scriptsize All regressions are OLS regressions that account for population survey weights and the sampling procedure. The dependent variable is a dummy=1 if the respondent increased their belief that it should be the government's responsibility to provide the following policies. The control variables include: gender, race, age, education, parental status, caring responsibilities for an elderly or a person with a disability, baseline income in February 2020, cohabitation with a partner, labor force participation and employment status in February 2020, health insurance provider, if the respondent had financial difficulties before the pandemic, macro-region, metro vs. rural, the population density at the zip code, and two dummy variables indicating if they consume at least 30min a week of international news and if they have at least one social media account. We also control for whether respondents completed the survey in a shorter time than the 99$^{th}$ percentile as well as ceiling effects.}
\end{minipage}	
\end{figure}

In Figure \ref{fig:welfare_preferences}, we show the relative effect of each variable in changing preferences for policies between April and October 2020 (for full a specification, see Tables \ref{tab:media_policies_A}, \ref{tab:media_policies_B}, and \ref{tab:media_policies_C} in the Appendix). Losing at least 20\% of income is associated with a marginal increase in support for the introduction of a guaranteed basic income, and assistance for the elderly. At the same time, an income loss decreases the belief that the government should help industry grow. The income shock coefficient is even larger on the increase in support for greater intervention by the government in all the temporary relief policies, as shown in Figure \ref{fig:covid_preferences} (see Table \ref{tab:covid_policies} in the Appendix for a full specification). Similarly, knowing someone who was hospitalized with COVID-19 led to an increase in support for a greater government intervention to assist the elderly, presumably because most hospitalizations occurred among older Americans who were more vulnerable to the virus, as well as a marginal increase in support for helping low-income students, and keeping prices under control. 

\par An indirect economic shock, namely living in a county that recovered faster its consumer expenditure, is associated with stronger support for a reduction in income inequality, helping citizens affected by natural disasters, and keeping prices under control. This measure of indirect shock is also correlated with stronger support for all temporary relief policies. Our interpretation of these correlations is that whether a shock affected a person directly or indirectly changes the type of policies they support. A person who incurred a direct shock might now be more appreciative of welfare policies that are targeted at the individual level and can improve the livelihood of their own families, while respondents who have not been directly affected but lived in areas that witnessed a faster economic recovery will be more appreciative of economic policies that can boost internal demand and restart the economy. This interpretation is in line with the analysis by \citep{chetty2020}, who noted that economic policies during a pandemic have different effects on households based on their income level. Thus, it is possible that families who lost part of their income during the crisis would now favor more social insurance policies that help mitigate the economic hardship they lived through, while higher-income households might be more likely to assume that more traditional macroeconomic policies aimed at stimulating internal demand would still be effective at reducing the unemployment rate.

\begin{figure}[H]
	\caption{The effect of shocks and media on temporary relief policies}
	\label{fig:covid_preferences}
	\begin{center}
		\includegraphics[height=6cm]{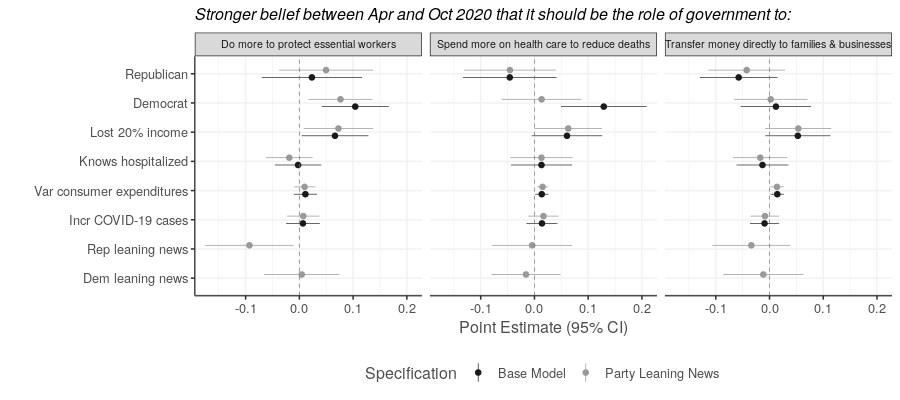}
	\end{center}
\begin{minipage}{1\linewidth \setstretch{0.75}}
	{\scriptsize{\textit{Notes}:} 
		\scriptsize All regressions are OLS regressions that account for population survey weights and the sampling procedure. The dependent variable is a dummy=1 if the respondent increased their belief that it should be the government's responsibility to provide the following policies. The control variables include: gender, race, age, education, parental status, caring responsibilities for an elderly or a person with a disability, baseline income in February 2020, cohabitation with a partner, labor force participation and employment status in February 2020, health insurance provider, if the respondent had financial difficulties before the pandemic, macro-region, metro vs. rural, the population density at the zip code, and two dummy variables indicating if they consume at least 30min a week of international news and if they have at least one social media account. We also control for whether respondents completed the survey in a shorter time than the 99$^{th}$ percentile as well as ceiling effects.}
\end{minipage}	
\end{figure}

\par Across all outcomes, we also note important differences between Democrats and Republicans. As reported in Tables \ref{tab:media_policies_A}, \ref{tab:media_policies_B}, and \ref{tab:media_policies_C}, the sign of the Republican party dummy variable is almost always negative while the opposite is true for the Democratic party variable. In the second column of each outcome, we see that this polarizing effect can be mostly explained by respondents who consumed politically biased media, in line with other studies \citep{gentzkow2011newspapers, dellavigna2007, allcott2020polarization, grossman2020political, simonov2020persuasive}.

\subsection{How shocks and media change trust in institutions} \label{sec:trust_inst}

We now look at the impact of the crisis on people's trust in institutions. In Figure \ref{fig:covid_preferences}, we see that while demand for government spending increased among those who have been affected by this crisis, respondents were also more likely to reduce their confidence in the people running most institutions (see Tables \ref{tab:trust_a} and \ref{tab:trust_b} in the Appendix for full specifications). We also find that losing at least 20\% of the household income in any two months during the crisis significantly decreased trust in financial institutions and in the private sector - two closely related entities - as well as in the Congress and Senate, and hospitals. As shown in the Appendix, some of these effects are even stronger among respondents whose income in October was at least 20\% lower than in April - that is, those who did not recover from the economic shock by the last wave of our survey. Looking at our measures of indirect shocks, we don't see large effects besides that an increase in consumer expenditures between April and October is positively correlated with a decrease in confidence in the White House. We explain this with the fact that this measure is sensitive to its baseline: indeed, the larger the initial drop, the larger the possible subsequent increase in consumer expenditures. Conversely, we see that respondents who lived in counties that recovered more quickly from the initial drop in consumer spending were less likely to have reduced their confidence in health insurance companies and hospitals, presumably as they associated the economic recovery with better crisis response by institutions. 


\par We note again substantial differences across parties. Compared to the Independents and non-voters, Republicans were less likely to have decreased trust in the U.S. Congress and Senate and in the White House, while the exact opposite is true for Democrats (by October, only about 3\% of Democrats had a lot of confidence in the White House, compared to 52\% of Republicans and 18\% of Independent and non-voters). Democrats were also less likely to have decreased their trust in the scientific community and in hospitals, regardless of whether they incurred any shock. In early April, 67\% of the Democrats, 50\% of the Republicans, and 51\% of the other respondents declared to have a ``great deal'' or ``complete'' confidence in the scientific community, whereas by the end of October, the percentage of respondents reporting the same trust had increased to 69\% for the Democrats, but it had dropped to 44\% for the Independents and to 36\% for the Republicans. 

\begin{figure}[H]
	\caption{The effect of shocks and media on trust in institutions}
	\label{fig:institution_preferences}
	\begin{center}
		\includegraphics[height=16cm]{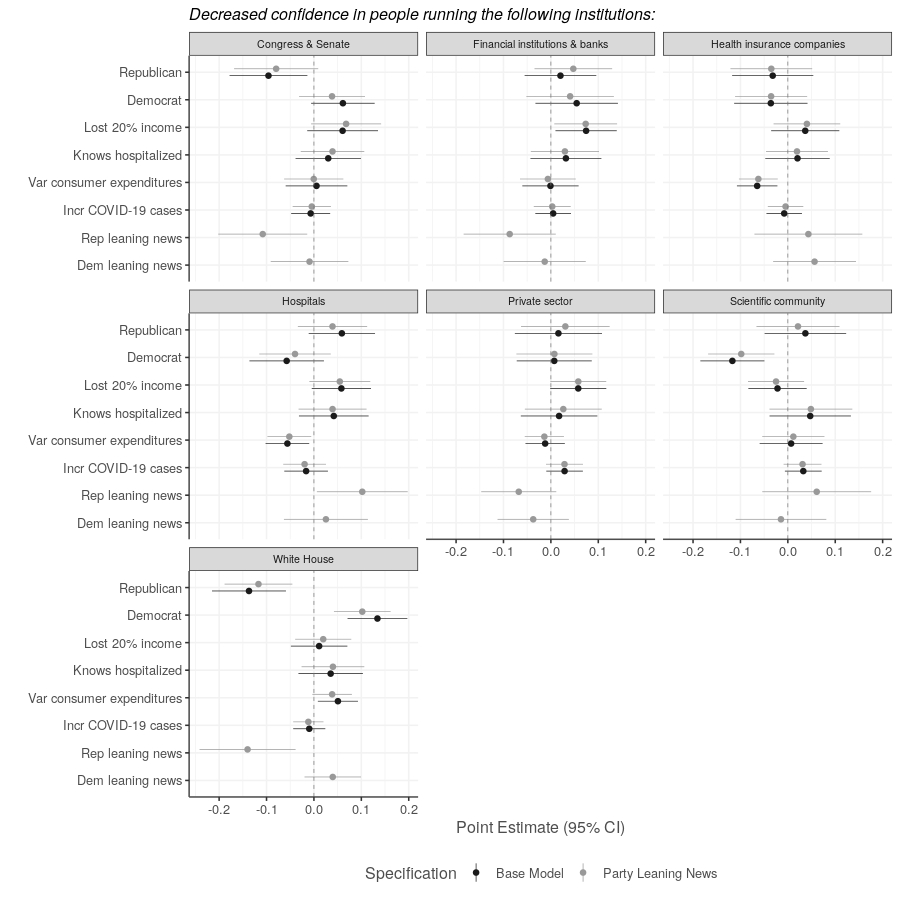}
	\end{center}
\begin{minipage}{1\linewidth \setstretch{0.75}}
	{\scriptsize{\textit{Notes}:} 
		\scriptsize All regressions are OLS regressions that account for population survey weights and the sampling procedure. The dependent variable is a dummy=1 if the respondent increased their belief that it should be the government's responsibility to provide the following policies. The control variables include: gender, race, age, education, parental status, caring responsibilities for an elderly or a person with a disability, baseline income in February 2020, cohabitation with a partner, labor force participation and employment status in February 2020, health insurance provider, if the respondent had financial difficulties before the pandemic, macro-region, metro vs. rural, the population density at the zip code, and two dummy variables indicating if they consume at least 30min a week of international news and if they have at least one social media account. We also control for whether respondents completed the survey in a shorter time than the 99$^{th}$ percentile as well as ceiling effects.}
\end{minipage}	
\end{figure}

These results suggest that direct negative experiences during a crisis play an important role in increasing support for welfare policies and greater government spending, as well as reducing trust in institutions. We have also shown that these effects can occur very rapidly, sometimes over a period of one to six months, and rarely return to pre-crisis levels in an equally short time. These effects are robust to several specifications and a rich set of controls, as shown in greater detail in Section \ref{robust}. We also find that political party affiliation per se doesn't fully explain the polarizing trends, and that Democrats and Republicans who lived through similar negative experiences might tend to react in similar ways when it comes to policy support and confidence in institutions. We find, instead, that consuming mostly politically biased media is associated with stronger polarization. This bears the question of whether citizens might be more likely to converge their views on several issues in the absence of polarizing media outlets. In the next section, we study more closely the mechanisms through which partisan-leaning media consumption might have increased polarization, and we show that most of the Democrats-Republicans gap can be explained by a different understanding of the gravity of the crisis. 


\subsection{How media can influence beliefs during a crisis}


In the previous section we showed how personal negative experiences can have a converging effect on policy preferences and trust in institutions between Democrats and Republicans, and how this trend was mitigated by the consumption of partisan leaning news\footnote{see \autoref{sec: partisan_gap} in the Online Appendix for a visual summary of the partisan gap over time across respondents' clusters}. This bears the questions whether exposure to new information can overcome this media-driven gap, and to what extent incurring a shock during a crisis would make citizens more or less responsive to this new information. To study these questions, we focus on respondents' understanding of the COVID-19 death rate, arguably the key indicator of the gravity of the crisis. 

\par In the fifth wave of the survey (week of May 18th), we asked respondents to forecast the COVID-19 death rate in the U.S. by the end of the year, after presenting them with the latest official death rate from the CDC, and asked for their judgment on how the government handled the crisis\footnote{The questions asked: \textit{By May 17, the U.S. Centers for Disease Control and Prevention (CDC) stated that about 90,000 Americans have so far died from COVID-19 (coronavirus). In addition to this, how many more Americans do you think will die by the end of this year due to coronavirus?} and \textit{Looking again at your estimated number of total coronavirus deaths in the U.S. by the end of the year, and considering how public authorities in the country have been managing the pandemic crisis, do you think the estimate you expect can be defined as a: Great success/ Success / Failure / Great Failure.} We specifically chose the wording 'public authorities' to partly reduce political priming effects.}. Our goal was to understand the impact of partisanship on perceptions and gauge expected death rates, controlling for information gap by providing the latest statistics\footnote{\citet{gaines2007} studies a similar setting showing results of a survey where Americans were asked to state the need and support for the Iraqi war in 2003: while the majority of all respondents thought it was unlikely that the U.S. would ever find weapons of mass destruction, Democrats were more likely to concluded that they simply did not exist while Republicans were more likely to state that they believed the Iraqi government moved or destroyed the weapons.}. Results showed a significant difference in expected death rate between Democrats and Republicans, but a less polarized view among Independents and non-voters\footnote{24\% of Republicans believed the rate would be 10,000 deaths or fewer (the lowest available option) compared to just 9\% of Democrats. The trend is reversed for the high bound estimates, where 10\% of Republicans believed there were going to be additional 100,000 deaths or more, compared to 31\% of Democrats. A Kruskal-Wallis equality-of-populations rank test confirms that these differences are statistically significant ($\chi^2$= 93.25, p$<$0.001). Among Independents, about 19\% expect the number to be 10,000 or fewer, about 21\% to be between 20,000 and 30,000, another 19\% to be 50,000, and about 18\% to be 100,000 or more.}.

\begin{figure}[H]
	\caption{Share of respondents believing that the annual COVID-19 death rate in 2020 could be considered a success by political party and expected death rate.}
	\label{fig:forecast_death}
	\begin{center}
		\includegraphics[height=8cm]{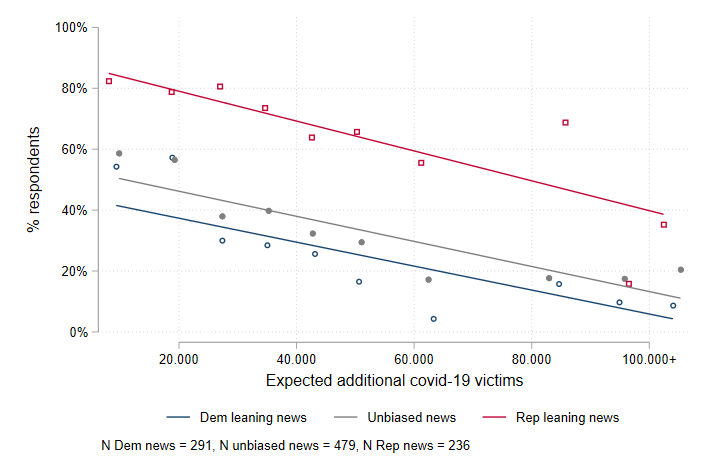}
	\end{center}
\begin{minipage}{1\linewidth \setstretch{0.75}}
	{\scriptsize{\textit{Notes}:} 
		\scriptsize The figure shows a binned scatterplot in which the x-axis variable (estimated deaths) is grouped into equal-sized bins, and the means of the x- and y-axis within each bin are computed. The plot controls for a set of variables.}
\end{minipage}	
\end{figure}

In figure \ref{fig:forecast_death}, we plot the correlation between the expected additional deaths and whether respondents considered this figure a success\footnote{Following \cite{chetty2014measuring}, we report a binscatter, controlling for a set of variables, and using a restricted model in which each covariate has the same coefficient in each by-value sample. Binscatter is a binned scatterplot, in which the x-axis variable (estimated deaths) is grouped into equal-sized bins, and the means of the x- and y-axis within each bin are computed. This allows us to visualize the expected number of respondents considering the estimated death rate as a success, conditional on the value that they had assigned}\footnote{We also repeated the same exercise by plotting the residuals of a regression with a dummy variable indicating whether the additional expected deaths were a success, as the dependent variable, and a set of controls as explanatory variables. This way, we control for the demographic characteristics that might be correlated with both our outcome (success) and our explanatory variable (forecast deaths). Results are robust also to this specification.}. A simple linear probability regression confirms again how the partisan gap is further exacerbated by respondents' source of news (see Table \ref{tab:expected_death}): Democrats consuming Democratic-leaning news estimated, on average, about 11,500 more deaths than those consuming unbiased sources; Republicans consuming Republican-leaning media reported about 11,000 deaths less. Similarly, consuming Democratic-leaning news is correlated with a decrease in the probability of considering the death rate as a success of 11.5 percentage points, whereas consuming Republican-leaning news with an increase of 18 percentage points. The effects of party and media are mostly robust to the inclusion of the expected number of deaths as a control, as shown in Column (3) of table \ref{tab:expected_death}. 


These results, however, also show that a convergence in beliefs (i.e., expected deaths) between Democrats and Republicans is associated with a convergence in judgment on how public authorities handled the crisis, further supporting the hypothesis that polarization might be driven by biased media consumption rather than a cheerleading effect \citep{gaines2007}\footnote{A debated issue with survey-based measures is whether some answers are biased by a cheerleading effect - that is, survey respondents' inclination to respond to questions to signal support for their political party rather than what they actually believe in. Recent studies, however, show that cheerleading effects might be less of a concern and that respondents do engage in motivated reasoning even in financially incentivized contexts \citep{peterson2021partisan}.} Motivated by these findings, we experimentally test whether exposure to the same information can overcome political divergence. \\ 

\subsubsection{Survey experiment} 

In the sixth wave of the survey (week of June 22), we asked all respondents to report the latest COVID-19 death rate in the U.S. and in their state of residence\footnote{The questions were as follows: \textit{How many people have died in your state because of coronavirus from the first death until today?} and \textit{How many people have died in the U.S. because of coronavirus from the first death until today?}. To avoid any survey fatigue effects, we asked these questions within the first block of ten questions of the survey.}. Immediately before seeing these questions, half of the respondents, the treatment group\footnote{see table \ref{tab:balancetable_deathexp} in the Appendix for the balance tables)}, was shown a blank page with the following text: \textit{Please answer the following questions carefully. If you wish to do so, you can look up the answer on the official CDC website at the following link: https://www.cdc.gov/coronavirus/2019-ncov/cases-updates/cases-in-us.html}. The URL link was spelled out so respondents could see it was a CDC webpage and could choose not to click on it and move to the next question. If they clicked on the link, a separate browser windows would open, redirecting them to the CDC webpage with the up to date death rate in the country, and a map that showed the same statistics in each state by simply hovering the mouse over the interested area (see figure \ref{fig:wte} in the Appendix). 

\par Through a hidden timer embedded in the survey question, we see that respondents in the treatment group spent significantly more time in answering the question, and in particular Republicans, suggesting that respondents did not avoid this new information even if they were not incentivized to consult the link\footnote{Due to privacy regulation, we could not check whether respondents clicked on the link, but we can track the time they spent answering the questions, which we use as a proxy for engagement with the website. We see that treated respondents spent an average of 40 seconds to answer the first question on the total number of deaths in their state, compared to a lower 26.5 seconds in the control group (Adj.Wald test with survey weights: F(1,189)=14.49; \textit{p}$<$0.001), but about the same time to answer the second question on the number of deaths in the U.S. (25.5 seconds in the control group and 25.6 in the treatment group; Adj. Wald test with survey weights: F(1,189)=0.00; \textit{p}=0.973). These estimates are confirmed in a linear regression. We also find differences across political lines, with the treatment being effective at increasing the time Republicans (50.8 seconds for the control group and 89.1 for the treated one, Adj. Wald test F(1,189)=5.59; \textit{p}=0.015)) and Independents (42.2 seconds for the control group and 52.4 in the treated one, Adj. Wald test F(1,189)=4.72; \textit{p}=0.033)) spent answering the questions, but we do not notice significant effects between Democrats in the control and treatment group. In other words, Republicans did not discard or avoid the new information, even if they might have anticipated the objective of the question asked \citep{saccardo2020}.}. The treatment also significantly increased the share of respondents who reported the state death rate according to the CDC, especially among Republicans (from 41\% to 60.4\% in the treated group, F(1,189)=6.319, \textit{p}=0.013) than the Democrats (from 51.5\% to 61.2\% in the treated group, F(1,189)=2.903, \textit{p}=0.09)\footnote{We analyzed whether the treatment had a stronger impact on respondents who expected a low number (i.e., below the median) of additional deaths in wave 5. Results show a positive but not significant effect}. These effects are confirmed in a series of regressions showing that treatment significantly increased the likelihood of reporting the correct death rate both at the State and the country level. 

\par Leveraging the wording of the previous survey wave, all respondents were then asked to judge how they thought public authorities handled the crisis\footnote{The question asked: \textit{Looking again at your estimated number of total coronavirus deaths in your state and in the US so far, and considering how public authorities in the country have been managing the pandemic crisis, do you think the current death rate can be defined as a: Great success; success; failure; or great failure}}. Among Democrats, already 88\% consider the outcome a failure or a great failure, but having answered the death rate questions according to CDC figures further increases the likelihood of stating so (from 85\% among those who didn't answer it correctly to 92\%, F(1,190)=3.187, \textit{p}=0.076). Among Republicans, a lower 40\% overall considered the death rate a failure or great failure of how public authorities managed the crisis, but answering the death rate question correctly reduced this, although not significantly, from 40\% among those who didn't answer it correctly to 29\%, F(1,189)=20.026, \textit{p}=0.156). Figure \autoref{f:deaths__success_f} provides a visual summary of the experimental results. Importantly, we do not observe a backfiring effect of information exposure among Republicans, suggesting that respondents might not have engaged in motivated reasoning \citep{nyhan2021}. 

\par As estimating the number of deaths according to the CDC might be endogenous to a person's political beliefs, we exploit the exogenous variation in the likelihood of correctly estimating the number of deaths caused by our treatment, which was randomly assigned, as follows:

$$ Pr(Success_{ic}) = \alpha + \beta Shock_{i} + \gamma Shock_{c} + \theta_1 Rep_i + \theta_2 Dem_i + $$ 
$$ \phi Treat_i + \delta X_{ic} + \epsilon_{ic} $$

The dependent variable in our regression is the probability of considering the current deaths as a ``success''; $Shock_i$ and $Shock_c$ indicated whether the respondent incurred a direct or indirect shock\footnote{The indirect economic shock in this regression is the variation in consumer spending between the time of the survey wave and the baseline of January 2020.}, and $X_{ic}$ captures a set of demographic variables. In Table \ref{tab:death_exp_main}, we show the results of this OLS regressions. In the first two columns, we show that the treatment succeeded in increasing the chances of stating the death rate as per CDC figures, both at the federal and the national level, while in the remaining columns, we report the effect of the treatment on the likelihood of declaring the number of deaths a success. In Table \ref{tab:death_tr_effect} in the Appendix, we show that the treatment effectively increased the time respondents spent answering the questions. In columns (3)-(6), we further break down the outcomes of the experiment, separating between those who under, over, or correctly estimated the number of deaths at the State or the US level. We see that Republicans were significantly more likely to underestimate the number of State and US deaths, while Democrats were less so (35\% of the Republicans under-reported both the number of US and State deaths, while the Democrats doing so were 18\% and 26\%, respectively; 35\% of Democrats overestimated the number of deaths in the US compared to 27\% of the Republicans). These results suggest that exposure to the same information can correct for the partisan gap in estimating the gravity of a crisis, in line with recent studies \citep{haaland2019}. We also find a directional, although not significant, change in the way respondents judged the gravity of the crisis and the success of the response by public authorities as a result of this intervention. 

To test whether respondents have a preference for consistency in their (motivated) response \citep{falk2018information}. To do this, we replicate the same above regressions and add a dummy for whether the respondent stated in the previous wave that the expected additional death rate could be deemed a success. We find that this dummy significantly increases the probability that respondents regarded the actual death rate as a success\footnote{We also replicate the same analysis by looking at whether the treatment had heterogeneous effects depending on the size of the gap between the forecast in wave 5 and the actual measure in wave 6. We find that the treatment had a similar effect regardless of how `'far'' a person's forecast was.}. The inclusion of this dummy does not change the statistical significance of the treatment effect in the first stage regression, nor the significance of party identity and biased media consumption variables in the second stage. As an additional check, we instrument ``correctly estimated the state and country number of deaths'' with the treatment assignment, which provides an exogenous variation, and the results confirm our findings (see Table \ref{tab:covid_iv} in Appendix).

\par We further investigate whether this light-touch non-incentivized treatment had lasting effects. To do this, in wave 7 (end of October), more than 3 months later than the survey experiment, we asked respondents to compare the U.S. COVID-19 death rate to the rest of the world\footnote{The possible answers ranged from ``\textit{The highest in the world}'' to ``\textit{The lowest in the world}'', on a four-point scale}. In column 6 of Table \ref{tab:death_tr_effect} we see that the treatment had a persistent, significant, and large effect in changing respondents' beliefs about the gravity of the crisis in the long run. Further, we see from the interaction terms that the treatment counterbalanced the effect of consuming biased media (see figure \ref{f:death_rate_pol})\footnote{When this survey wave was administered, the U.S. cumulative death rate was the 10th highest in the world, with 685 deaths per million inhabitants. We consider the cumulative death rate per million inhabitants reported by the website ``Our World In Data'' on October, the 26th 2020 (url: \url{https://ourworldindata.org/covid-deaths})}.\\ 


The final research question we investigate is the relationship between negative personal experiences and exposure to new information in a crisis, accounting for media consumption. We do this by first testing for heterogenous treatment effects, employing a causal forest methodology \citep{athey2019estimating}\footnote{While there is not a clear consensus on causal forest validation, one approach suggested by \citet{athey2019estimating} is the use of a “best linear predictor” method, which fits the conditional average treatment effects (CATE) as a linear function of out-of-bag causal forest estimates.}. This approach allows us to construct non-biased partitions of the data ex-ante from which a valid CATE may be achieved. To improve precision, we first train a pilot random forest on all baseline characteristics included in the OLS regression to identify relative variable importance to the model. We then train a second forest using only the subset of covariates that score above mean importance to eliminate possible confounding effects \citep{basu2018iterative}. We then run tests to detect any heterogeneity in our primary outcomes of interest: (1) correctly identifying state and national COVID-19 death rates; and (2) evaluating these rates as a success. Additionally, we test for heterogeneity in sustained informational effects, measured through a question in the next wave evaluating if respondents correctly identify the relative US death rate to other countries. 

We find strong evidence of association between causal forest estimates and heterogeneity in treatment effect for correct estimation of state and national COVID death rates, but not for other outcomes, consistent with the non-significance of our OLS estimates. This suggests we might not be powered to detect variation in lasting treatment effects. We then employ a series of tests suggested by \citep{davis2020rethinking} to verify that out-of-bag predictions and actual treatment effects are related and find that the results for correct estimation of COVID death rates are consistent with our calibration test (see Appendix Table \ref{tab:bestlinearfit}). Together, these tests suggest there is a meaningful source of heterogeneity in treatment effectiveness that is worthy of further examination. Using a quartile breakout by predicted treatment effects for correct estimation of state and national US death rates, we find having incurred a direct shock does not increase responsiveness to the information treatment, and does not increase the probability of answering the estimate questions correctly (see Table \ref{tab:causalforestquartiles} for summary statistics by quartile for our baseline characteristics, as well as the mean CATE prediction). Perhaps unsurprisingly, instead, we find that a higher level of education reinforces the treatment effect: respondents with a bachelor’s degree or higher display significantly higher treatment effects, representing over 60\% of the highest quartile; in contrast, respondents with a high school education experience a constant diminishing representation in each subsequent quartile. Democratic respondents who consumed more Democratic-leaning news were also more responsive to the treatment than other political sub-groups.

\section{Conclusions}

In this study, we used a longitudinal multi-wave survey on a representative sample of Americans and find that large-scale crises such as COVID-19 can induce changes in policy preferences, trust in institutions, and beliefs. Contrary to previous studies that suggested that such changes only occur over long periods of time, we show that they can actually happen quickly, sometimes in a matter of weeks. We also offer several novel insights in the mechanisms behind these shifts by showing that direct negative experiences explain most of these changes. A direct economic or health shock during the COVID-19 pandemic, increased Americans' preferences for greater government spending, particularly on welfare assistance, and decreased trust in most public institutions. Political polarization on these outcomes is instead explained by a gap in understanding the gravity of the crisis that can be largely explained by the consumption of partisan-leaning media. In a light-touch experiment, we show that exposing respondents to the same source of information reduces this gap with long-lasting effects. Our results contribute to a growing literature on how crises transform societies, pointing to the importance of tracking preferences frequently and disentangling the different channels at play. 

\newpage

\bibliographystyle{chicago}
\bibliography{polarization}

\newpage

\appendix

\section{Appendix}

\subsection{Survey design and methodology}

\begin{figure}[H]
	\caption{Timing of the longitudinal survey waves against health indicators of the COVID-19 pandemic}
	\label{f:timeline}
	\begin{center}
		\includegraphics[height=12cm]{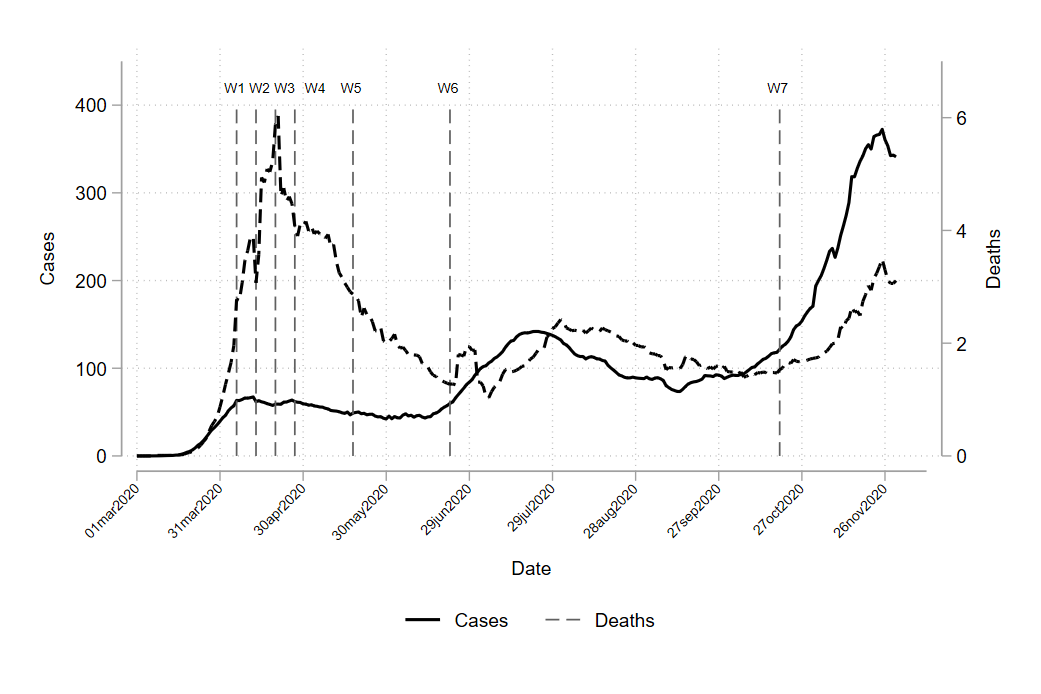}
	\end{center}
\begin{minipage}{1\linewidth \setstretch{0.75}}
	{\scriptsize{\textit{Notes}:} 
		\scriptsize The figure shows the timing of the seven survey waves implemented between early April and end of October 2020 against the curves of the 7-day rate of COVID-19 cases and deaths per 100,000 inhabitants, allowing for comparisons between areas with different population sizes in the United States. The figure is based on the U.S. Centers for Disease Control and Prevention publicly available data.}
\end{minipage}	
\end{figure}

\newpage

\textbf{Sample characteristics}
\begin{table}[H]
\centering
\caption{Summary statistics of the key demographics in each survey wave, applying survey weights.}
\label{tab:summary_stats}
\resizebox{\textwidth}{!}{%
\begin{tabular}{lccccccc} \hline \hline
& Wave1 & Wave2 & Wave3 & Wave4 & Wave5 & Wave6 & Wave7 \\ \hline
Republican & 27.37 & 25.33 & 26.24 & 25.94 & 25.55 & 24.53 & 26.47 \\
Democrat & 37.93 & 37.19 & 35.87 & 37.41 & 36.53 & 36.2 & 35.94 \\
Independent \& other & 34.71 & 37.48 & 37.88 & 36.64 & 37.92 & 39.27 & 37.59 \\
Woman & 51.69 & 51.68 & 51.68 & 51.69 & 51.68 & 51.68 & 51.7 \\
Age: 18-29 & 20.51 & 20.5 & 20.5 & 20.5 & 20.5 & 20.5 & 20.53 \\
Age: 30-44 & 24.71 & 24.83 & 25.52 & 25.97 & 25.36 & 25.45 & 25.5 \\
Age: 45-59 & 25.02 & 24.9 & 24.21 & 23.76 & 24.38 & 24.29 & 24.22 \\
Age: 60+ & 29.76 & 29.77 & 29.77 & 29.77 & 29.77 & 29.77 & 29.75 \\
Less than HS & 9.77 & 9.77 & 9.77 & 9.77 & 9.77 & 9.77 & 9.77 \\
High school & 28.26 & 28.26 & 28.26 & 28.26 & 28.26 & 28.26 & 28.25 \\
Some college & 27.71 & 27.7 & 27.7 & 27.7 & 27.69 & 27.7 & 27.73 \\
Bachelor + & 34.27 & 34.27 & 34.27 & 34.27 & 34.28 & 34.27 & 34.26 \\
I income q & 22.63 & 22.89 & 23.89 & 24.37 & 23.5 & 24.41 & 26.35 \\
II income q & 21.35 & 21.15 & 20.47 & 20.4 & 21.48 & 19.99 & 17.86 \\
III income q & 17.99 & 17.28 & 17.23 & 17.65 & 16.39 & 17.64 & 17.9 \\
IV income q & 18.98 & 19.37 & 19.6 & 18.66 & 19.7 & 19.29 & 18.69 \\
V income q & 19.05 & 19.3 & 18.81 & 18.93 & 18.93 & 18.66 & 19.21 \\
\begin{tabular}[c]{@{}l@{}}Financial hardship\\ pre-COVID-19\end{tabular} & 31.52 & 31.04 & 31.89 & 31.33 & 31.14 & 31.75 & 31.38 \\
African American & 11.93 & 11.93 & 11.93 & 11.93 & 11.93 & 11.93 & 11.93 \\
Hispanic & 16.67 & 16.67 & 16.67 & 16.67 & 16.67 & 16.67 & 16.66 \\
Other Race & 8.59 & 8.58 & 8.58 & 8.58 & 8.58 & 8.58 & 8.62 \\
White & 62.81 & 62.82 & 62.82 & 62.82 & 62.82 & 62.82 & 62.79 \\
Cohabitating & 54.58 & 57.9 & 57.36 & 57.98 & 58.55 & 62.43 & 62.68 \\
Parent of minor & 27.87 & 26.64 & 27.57 & 27.13 & 26.68 & 26.61 & 27.07 \\
Caring responsibilities & 16.48 & 15.95 & 16.65 & 16.21 & 15.87 & 16.42 & 16.37 \\
Not in the labor force & 27.68 & 27.14 & 28.18 & 27.75 & 26.91 & 27.24 & 26.99 \\
Unemployed in Feb & 9.81 & 8.96 & 10.8 & 9.73 & 9.06 & 8.83 & 9.56 \\
North-East & 17.45 & 17.45 & 17.45 & 17.45 & 17.45 & 17.45 & 17.44 \\
Midwest & 20.74 & 20.74 & 20.74 & 20.74 & 20.74 & 20.74 & 20.73 \\
South & 37.98 & 37.97 & 37.97 & 37.97 & 37.97 & 37.97 & 38 \\
West & 23.84 & 23.84 & 23.84 & 23.84 & 23.84 & 23.84 & 23.83 \\
Metropolitan area & 85.49 & 85.1 & 86.07 & 84.78 & 85.29 & 86.07 & 85.49 \\
No health insurance & 6.67 & 6.39 & 6.54 & 5.89 & 6.6 & 6.29 & 6.24 \\
Population density (ZCTA) & 371350.9 & 368034.6 & 379836.2 & 380862.6 & 380720.3 & 373379.7 & 383178.7 \\
\textit{N} & 27.69 & 27.15 & 28.19 & 27.76 & 26.92 & 27.25 & 26.99 \\\hline
\end{tabular}%
}
\end{table}

\newpage

\textbf{List of questions and outcomes}. 

\begin{table}[H]
\centering
\caption{Summary table of questions by waves}
\label{tab:questions}
\resizebox{\textwidth}{!}{%
\begin{tabular}{llll}
\hline
\textbf{Outcomes} &
 \textbf{Questions} &
 \textbf{Scale} &
 \textbf{Waves} \\
 &
  &
  &
  \\
Welfare policy preferences &
 \begin{tabular}[c]{@{}l@{}}Do you favor or oppose a universal health care system covered \\ by the government so that every American can have equal access \\ to health care, even if this means that you will have to pay higher taxes?\end{tabular} &
 \begin{tabular}[c]{@{}l@{}}1 (strongly oppose) \\ to 5 (strongly favor)\end{tabular} &
 \multicolumn{1}{c}{1, 4, 7} \\
 &
 \begin{tabular}[c]{@{}l@{}}Do you think the following should or should not be the government’s \\ responsibility to:\end{tabular} &
 \begin{tabular}[c]{@{}l@{}}1 (definitely should)\\ to 4 (definitely should not be)\end{tabular} &
 \multicolumn{1}{c}{1, 4, 7} \\
 &
 Provide mental health care for persons with mental illnesses &
  &
  \\
 &
 Help individuals affected by natural disasters &
  &
  \\
 &
 Keep prices under control &
  &
  \\
 &
 Provide a decent standard of living for the old &
  &
  \\
 &
 Provide a decent standard of living for the unemployed &
  &
  \\
 &
 Provide everyone with a guaranteed basic income &
  &
  \\
 &
 Provide industry with the help it needs to grow &
  &
  \\
 &
 Reduce income differences between the rich and the poor &
  &
  \\
 &
 Give financial help to university students from low-income families &
  &
  \\
 &
  &
  &
  \\
Temporary relief policies &
 To what extent do you agree or disagree with the following statements? &
 \begin{tabular}[c]{@{}l@{}}1 (strongly disagree) \\ to 5 (strongly agree)\end{tabular} &
 \multicolumn{1}{c}{4, 7} \\
 &
 \begin{tabular}[c]{@{}l@{}}The government should transfer money directly to families and businesses until \\ the US economy can fully return to its pre-crisis levels\end{tabular} &
  &
  \\
 &
 \begin{tabular}[c]{@{}l@{}}The government should do more to protect essential workers from \\ contracting the virus\end{tabular} &
  &
  \\
 &
 \begin{tabular}[c]{@{}l@{}}The government should spend more on public healthcare to reduce the \\ number of preventable deaths\end{tabular} &
  &
  \\
 &
  &
  &
  \\
Trust in institutions &
 \begin{tabular}[c]{@{}l@{}}How much confidence do you have in the people running \\ the following institutions?\end{tabular} &
 \begin{tabular}[c]{@{}l@{}}1 (complete confidence) to \\ 5 (no confidence at all)\end{tabular} &
 \multicolumn{1}{c}{1, 4, 7} \\
 &
 U.S. Congress and Senate &
  &
  \\
 &
 The White House &
  &
  \\
 &
 Scientific community &
  &
  \\
 &
 Banks and financial institutions &
  &
  \\
 &
 The private sector &
  &
  \\
 &
 Hospitals and healthcare professionals &
  &
  \\
 &
 Health insurance companies &
  &
  \\
 &
  &
  &
  \\
\begin{tabular}[c]{@{}l@{}}Information processing \\ and interpretation of reality\end{tabular} &
 \begin{tabular}[c]{@{}l@{}}By May 17, the U.S. Centers for Disease Control and Prevention \\ (CDC) stated that about 90,000 Americans have so far died from \\ COVID-19 (coronavirus). In addition to this, how many more\\ Americans do you think will die by the end of this year due to coronavirus?\end{tabular} &
 \begin{tabular}[c]{@{}l@{}}10 options, from ''10,000 or fewer'' \\ to ''100,000 or more''\end{tabular} &
 \multicolumn{1}{c}{5} \\
 &
  &
  &
  \\
 &
 \begin{tabular}[c]{@{}l@{}}Looking again at your estimated number of total coronavirus deaths \\ in the U.S. by the end of the year, and considering how public authorities \\ in the country have been managing the pandemic crisis, do you think the \\ estimate you expect can be defined as a:\end{tabular} &
 \begin{tabular}[c]{@{}l@{}}1 (great success) to \\ 4 (great failure)\end{tabular} &
 \multicolumn{1}{c}{5} \\
 &
  &
  &
 \multicolumn{1}{c}{} \\
 &
 \begin{tabular}[c]{@{}l@{}}How many people have died in your state because of coronavirus \\ from the first death until today?\end{tabular} &
 \begin{tabular}[c]{@{}l@{}}8 options, from ''less than 500''\\ to ''more than 30,000''\end{tabular} &
 \multicolumn{1}{c}{6} \\
 &
 \begin{tabular}[c]{@{}l@{}}How many people have died in the U.S. because of coronavirus \\ from the first death until today?\end{tabular} &
 \begin{tabular}[c]{@{}l@{}}Slider from 0 to 200,000 with \\ intervals of 20,000\end{tabular} &
 \multicolumn{1}{c}{6} \\
 &
  &
  &
 \multicolumn{1}{c}{} \\
 &
 Do you believe the current COVID-19 death rate per capita in the U.S. is: &
 \begin{tabular}[c]{@{}l@{}}1 (the highest in the world) to \\ 4 (the lowest in the world)\end{tabular} &
 \multicolumn{1}{c}{7} \\ \hline
\end{tabular}%
}
\end{table}

\begin{table}[H]
\caption{Welfare policy pre-trends prior to household-level shocks}
\label{tab:policy_pretrends}
\resizebox{\textwidth}{!}{%
\begin{tabular}{lcc}
\hline
\hline
 & (1) & (2) \\
 & \begin{tabular}[c]{@{}c@{}}Lost 20\%\\ income\end{tabular} & \begin{tabular}[c]{@{}c@{}}Knows\\ hospitalized\end{tabular} \\ \hline
 & & \\
 Provide universal health care & 0.076 & 0.146* \\
 & (0.075) & (0.077) \\
 Guarantee basic income & -0.004 & 0.001 \\
 & (0.060) & (0.061) \\
 Reduce income inequality & 0.031 & 0.016 \\
 & (0.059) & (0.060) \\
 Help the unemployed & 0.069 & 0.030 \\
 & (0.052) & (0.053) \\
 Provide mental health care & 0.049 & 0.098** \\
 & (0.046) & (0.047) \\
 Help the elderly & 0.030 & 0.087* \\
 & (0.047) & (0.047) \\
 Help those affected by natural disasters & 0.006 & 0.012 \\
 & (0.039) & (0.039) \\
 Give financial help to low-income students & 0.007 & 0.030 \\
 & (0.050) & (0.051) \\
 Help industry grow & 0.106** & -0.007 \\
 & (0.053) & (0.054) \\
 Keep prices under control & -0.004 & -0.026 \\
 & (0.054) & (0.055) \\
 & & \\ \hline
\multicolumn{3}{l} {
\begin{minipage}{1.33\columnwidth}%
\small \textit{Notes}. Standard errors in parentheses. *** p\textless{}0.01, ** p\textless{}0.05, * p\textless{}0.1. Pre-trends on policy preferences are established using wave 1 responses. All regressions are OLS regressions that take into account population survey weights and the sampling procedure. The dependent variable ranges on a Likert scale from 1 to 5, with a 1 indicating respondent expressed a negative preference for the policies listed above at wave 1. The control variables include: gender, race, age, education, parental status, caring responsibilities for an elderly or a person with a disability, baseline income in February 2020, cohabitation with a partner, labor force participation and employment status in February 2020, health insurance provider, if the respondent had financial difficulties before the pandemic, macro-region, metro vs. rural, the population density at the zip code, two dummy variables indicating if they consume at least 30min a week of international news and if they consume news from social media, reported political party, and two dummy variables indicating consumption of Republican or Democrat-leaning news. We also control for whether respondents completed the survey in a shorter time than the 99$^{th}$ percentile as well as ceiling effects.\end{minipage}
}
\end{tabular}%
}
\end{table}

\begin{table}[H]
\caption{Institutional trust pre-trends prior to household-level shocks}
\label{tab:institution_pretrends}
\resizebox{\textwidth}{!}{%
\begin{tabular}{lcc}
\hline
\hline
 & (1) & (2) \\
 & \begin{tabular}[c]{@{}c@{}}Lost 20\%\\ income\end{tabular} & \begin{tabular}[c]{@{}c@{}}Knows\\ hospitalized\end{tabular} \\ \hline
 & & \\
 Congress \& Senate & 0.050 & 0.025 \\
 & (0.056) & (0.056) \\
 White House & -0.049 & -0.070 \\
 & (0.065) & (0.066) \\
 Scientific community & 0.104* & 0.021 \\
 & (0.057) & (0.058) \\
 Financial institutions \& banks & 0.002 & 0.012 \\
 & (0.058) & (0.059) \\
 Private sector & -0.007 & -0.003 \\
 & (0.055) & (0.056) \\
 Hospitals & 0.015 & 0.008 \\
 & (0.056) & (0.057) \\
 Health insurance companies & -0.079 & -0.069 \\
 & (0.061) & (0.0.062) \\
 & & \\ \hline
\multicolumn{3}{l} {
\begin{minipage}{1.33\columnwidth}%
\small \textit{Notes}. Standard errors in parentheses. *** p\textless{}0.01, ** p\textless{}0.05, * p\textless{}0.1. Pre-trends on institutional trust are established using wave 1 responses. All regressions are OLS regressions that take into account population survey weights and the sampling procedure. The dependent variable ranges on a Likert scale from 1 to 5, with a 1 indicating respondent expressed a negative preference for the policies listed above at wave 1. The control variables include: gender, race, age, education, parental status, caring responsibilities for an elderly or a person with a disability, baseline income in February 2020, cohabitation with a partner, labor force participation and employment status in February 2020, health insurance provider, if the respondent had financial difficulties before the pandemic, macro-region, metro vs. rural, the population density at the zip code, two dummy variables indicating if they consume at least 30min a week of international news and if they consume news from social media, reported political party, and two dummy variables indicating consumption of Republican or Democrat-leaning news. We also control for whether respondents completed the survey in a shorter time than the 99$^{th}$ percentile as well as ceiling effects.\end{minipage}
}
\end{tabular}%
}
\end{table}

\newpage

\subsection{Previous GSS waves}

\begin{table}[H]
\centering
\caption{Policy support across GSS waves between Democrats and Republicans}
\label{tab:GSS_policies}
\resizebox{\textwidth}{!}{%
\begin{tabular}{lcccccccccc}
\cline{2-11}
 & \multicolumn{2}{c}{1996} & \multicolumn{2}{c}{2006} & \multicolumn{2}{c}{2016} & \multicolumn{2}{c}{Apr-20} & \multicolumn{2}{c}{Oct-20} \\ \cline{2-11} 
                     & Dem & Rep & Dem & Rep & Dem & Rep & Dem & Rep & Dem & Rep \\ \hline
Give fin. Help to low income students  & 90\% & 79\% & 95\% & 86\% & 95\% & 79\% & 93\% & 61\% & 92\% & 59\% \\
Help industry grow            & 71\% & 60\% & 77\% & 68\% & 79\% & 62\% & 71\% & 66\% & 73\% & 68\% \\
Help the elderly             & 92\% & 79\% & 95\% & 80\% & 94\% & 80\% & 96\% & 72\% & 91\% & 68\% \\
Help the unemployed           & 57\% & 35\% & 62\% & 33\% & 69\% & 37\% & 88\% & 49\% & 80\% & 33\% \\
Help those affacted by natural disasters & N/A & N/A & 94\% & 83\% & N/A & N/A & 98\% & 96\% & 95\% & 92\% \\
Keep prices under control        & 76\% & 58\% & 82\% & 63\% & 79\% & 61\% & 87\% & 69\% & 91\% & 69\% \\
Provide mental health care        & 87\% & 69\% & 88\% & 74\% & N/A & N/A & 96\% & 75\% & 95\% & 74\% \\
Reduce income inequality         & 59\% & 33\% & 64\% & 31\% & 72\% & 31\% & 83\% & 29\% & 84\% & 32\% \\ \hline
\end{tabular}%
}
\end{table}

\begin{table}[H]
\centering
\caption{Trust in institutions across GSS waves between Democrats and Republicans}
\label{tab:GSS_trust1}
\resizebox{\textwidth}{!}{%
\begin{tabular}{lcccccccccc}
\cline{2-11}
 & \multicolumn{2}{c}{1996} & \multicolumn{2}{c}{2006} & \multicolumn{2}{c}{2016} & \multicolumn{2}{c}{Apr-20} & \multicolumn{2}{c}{Oct-20} \\ \cline{2-11} 
                 & Dem & Rep & Dem & Rep & Dem & Rep & Dem & Rep & Dem & Rep \\ \hline
Congress \&  Senate       & 7\% & 9\% & 8\% & 15\% & 5\% & 6\% & 7\% & 12\% & 3\% & 12\% \\
Financial institutions and banks & 26\% & 26\% & 27\% & 38\% & 11\% & 17\% & 18\% & 32\% & 11\% & 19\% \\
Scientific community       & 40\% & 42\% & 41\% & 44\% & 46\% & 36\% & 68\% & 51\% & 70\% & 36\% \\ \hline
\end{tabular}%
}
\end{table}

\begin{table}[H]
\centering
\caption{Trust in institutions across GSS waves between Democrats and Republicans}
\label{tab:GSS_trust2}
\resizebox{0.75\textwidth}{!}{%
\begin{tabular}{lccccc} \hline \hline
Year & Party & \begin{tabular}[c]{@{}c@{}}Banks\\ and financial\\ institutions\end{tabular} & \begin{tabular}[c]{@{}c@{}}Scientific \\ community\end{tabular} & \begin{tabular}[c]{@{}c@{}}U.S. Congress\\ \& Senate\end{tabular} & \begin{tabular}[c]{@{}c@{}}Private\\ sector\end{tabular} \\ \hline
\multirow{2}{*}{1994} & Democrats & 18\% & 38\% & 9\% & N/A \\
 & Republican & 19\% & 42\% & 6\% & N/A \\
\multirow{2}{*}{1996} & Democrats & 26\% & 40\% & 7\% & N/A \\
 & Republican & 26\% & 42\% & 9\% & N/A \\
\multirow{2}{*}{1998} & Democrats & 25\% & 39\% & 11\% & 9\% \\
 & Republican & 29\% & 42\% & 9\% & 15\% \\
\multirow{2}{*}{2000} & Democrats & 27\% & 43\% & 12\% & N/A \\
 & Republican & 34\% & 44\% & 14\% & N/A \\
\multirow{2}{*}{2002} & Democrats & 21\% & 39\% & 12\% & N/A \\
 & Republican & 26\% & 39\% & 15\% & N/A \\
\multirow{2}{*}{2004} & Democrats & 22\% & 40\% & 12\% & N/A \\
 & Republican & 36\% & 44\% & 16\% & N/A \\
\multirow{2}{*}{2006} & Democrats & 27\% & 41\% & 8\% & N/A \\
 & Republican & 38\% & 44\% & 15\% & N/A \\
\multirow{2}{*}{2008} & Democrats & 20\% & 40\% & 10\% & 9\% \\
 & Republican & 19\% & 38\% & 9\% & 15\% \\
\multirow{2}{*}{2010} & Democrats & 10\% & 45\% & 12\% & N/A \\
 & Republican & 11\% & 38\% & 6\% & N/A \\
\multirow{2}{*}{2012} & Democrats & 10\% & 44\% & 7\% & N/A \\
 & Republican & 13\% & 35\% & 5\% & N/A \\
\multirow{2}{*}{2014} & Democrats & 12\% & 45\% & 7\% & N/A \\
 & Republican & 15\% & 36\% & 3\% & N/A \\
\multirow{2}{*}{2016} & Democrats & 11\% & 46\% & 5\% & N/A \\
 & Republican & 17\% & 36\% & 6\% & N/A \\
\multirow{2}{*}{2018} & Democrats & 15\% & 50\% & 4\% & 9\% \\
 & Republican & 25\% & 41\% & 6\% & 15\% \\ \hline
\end{tabular}%
}
\end{table}

\subsection{Policy preference regression specifications}

\begin{table}[H]
\caption{The effect of shocks and media on welfare policy preferences - A}
\label{tab:media_policies_A}
\resizebox{\textwidth}{!}{%
\begin{tabular}{lcccccc}
\hline
\hline
 & \multicolumn{6}{c}{\begin{tabular}[c]{@{}c@{}}Stronger belief between Apr and Oct 2020 that\\ it should be the role of government to:\end{tabular}} \\
 & (1) & (2) & (3) & (4) & (5) & (6) \\
 & \begin{tabular}[c]{@{}c@{}}Provide \\ universal \\ health care\end{tabular} & \begin{tabular}[c]{@{}c@{}}Provide \\ universal \\ health care\end{tabular} & \begin{tabular}[c]{@{}c@{}}Guarantee \\ basic \\ income\end{tabular} & \begin{tabular}[c]{@{}c@{}}Guarantee \\ basic\\ income\end{tabular} & \begin{tabular}[c]{@{}c@{}}Reduce \\ income\\ inequality\end{tabular} & \begin{tabular}[c]{@{}c@{}}Reduce \\ income\\ inequality\end{tabular} \\ \hline
 & & & & & & \\
Republican & -0.0356 & -0.00854 & -0.0256 & -0.00701 & -0.0349 & -0.0114 \\
 & (0.0452) & (0.0458) & (0.0407) & (0.0408) & (0.0426) & (0.0423) \\
Democrat & 0.0515 & 0.0198 & 0.0502 & 0.0255 & 0.114*** & 0.0955** \\
 & (0.0319) & (0.0322) & (0.0355) & (0.0376) & (0.0350) & (0.0371) \\
Lost 20\% income & 0.0264 & 0.0335 & 0.0463* & 0.0492* & 0.0336 & 0.0374 \\
 & (0.0354) & (0.0365) & (0.0270) & (0.0277) & (0.0253) & (0.0252) \\
Knows hospitalized & 0.0182 & 0.0241 & -0.00935 & -0.00656 & -0.0104 & -0.00873 \\
 & (0.0289) & (0.0298) & (0.0391) & (0.0388) & (0.0327) & (0.0328) \\
Var consumer expenditures & -0.0397** & -0.0495** & 0.00297 & -0.00589 & 0.0333** & 0.0280* \\
 & (0.0196) & (0.0208) & (0.0190) & (0.0193) & (0.0158) & (0.0161) \\
Incr COVID-19 cases & -0.00735 & -0.00626 & 0.0221 & 0.0189 & -0.0139 & -0.0144 \\
 & (0.0196) & (0.0194) & (0.0193) & (0.0187) & (0.0143) & (0.0147) \\
Rep leaning news & & -0.136*** & & -0.0892*** & & -0.0878** \\
 & & (0.0436) & & (0.0289) & & (0.0380) \\
Dem leaning news & & 0.0226 & & 0.0125 & & 0.0190 \\
 & & (0.0382) & & (0.0328) & & (0.0408) \\
Constant & 0.0808 & 0.155 & -2.32e-06 & 0.0563 & 0.229 & 0.242* \\
 & (0.130) & (0.132) & (0.123) & (0.124) & (0.151) & (0.141) \\
 & & & & & & \\
Controls & Yes & Yes & Yes & Yes & Yes & Yes \\ 
Observations & 1,010 & 1,010 & 999 & 999 & 1,006 & 1,006 \\
R-squared & 0.173 & 0.188 & 0.182 & 0.194 & 0.231 & 0.240 \\
Avg increase support & 0.265 & 0.265 & 0.220 & 0.220 & 0.225 & 0.225 \\
Avg decrease support & 0.210 & 0.210 & 0.233 & 0.233 & 0.231 & 0.231 \\ \hline
\multicolumn{7}{l}{%
 \begin{minipage}{1.15\columnwidth}%
\small \textit{Notes}. Standard errors in parentheses. *** p\textless{}0.01, ** p\textless{}0.05, * p\textless{}0.1. The percentages in the first row report the share of respondents who increased their Likert-based score between the first and last wave of the survey. All regressions are OLS regressions that take into account population survey wights and the sampling procedure. The dependent variable is a dummy=1 if the respondent increased their belief that it should be the government's responsibility to provide the following policies. The control variables include: gender, race, age, education, parental status, caring responsibilities for an elderly or a person with a disability, baseline income in February 2020, cohabitation with a partner, labor force participation and employment status in February 2020, health insurance provider, if the respondent had financial difficulties before the pandemic, macro-region, metro vs. rural, the population density at the zip code, and two dummy variables indicating if they consume at least 30min a week of international news and if they have at least one social media account. We also control for whether respondents completed the survey in a shorter time than the 99$^{th}$ percentile as well as ceiling effects.\end{minipage}
}
\end{tabular}%
}
\end{table}

\begin{table}[H]
\caption{The effect of shocks and media on welfare policy preferences - B}
\label{tab:media_policies_B}
\resizebox{\textwidth}{!}{%
\begin{tabular}{lccccccc}
\hline
\hline
 & \multicolumn{6}{c}{\begin{tabular}[c]{@{}c@{}}Stronger belief between Apr and Oct 2020 that\\ it should be the role of government to:\end{tabular}} \\
 & (1) & (2) & (3) & (4) & (5) & (6) \\
 & \begin{tabular}[c]{@{}c@{}}Help the \\ unemployed\end{tabular} & \begin{tabular}[c]{@{}c@{}}Help the \\ unemployed\end{tabular} & \begin{tabular}[c]{@{}c@{}}Provide \\ mental \\ health care\end{tabular} & 
 \begin{tabular}[c]{@{}c@{}}Provide \\ mental \\ health care\end{tabular} &
 \begin{tabular}[c]{@{}c@{}}Help the \\ elderly\end{tabular} & \begin{tabular}[c]{@{}c@{}}Help the \\ elderly\end{tabular} & \\ \hline
 & & & & & & \\
Republican & -0.0443 & -0.0426 & -0.00545 & 0.0121 & 0.00698 & 0.0268 \\
 & (0.0310) & (0.0337) & (0.0354) & (0.0350) & (0.0389) & (0.0401) \\
Democrat & 0.0220 & 0.0268 & 0.0392 & 0.0340 & 0.0141 & 0.00544 \\
 & (0.0312) & (0.0308) & (0.0281) & (0.0306) & (0.0289) & (0.0299) \\
Lost 20\% income & 0.0253 & 0.0263 & -0.0181 & -0.0171 & 0.0440 & 0.0482* \\
 & (0.0308) & (0.0304) & (0.0200) & (0.0199) & (0.0284) & (0.0271) \\
Knows hospitalized & -0.0195 & -0.0141 & 0.0181 & 0.0276 & 0.0439 & 0.0592** \\
 & (0.0246) & (0.0234) & (0.0260) & (0.0262) & (0.0269) & (0.0253) \\
Var consumer expenditures & -0.00159 & -0.00233 & -0.0140 & -0.0179 & 0.0135 & 0.00793 \\
 & (0.0168) & (0.0165) & (0.0173) & (0.0169) & (0.0222) & (0.0213) \\
Incr COVID-19 cases & 0.0136 & 0.0112 & 0.00542 & 0.00334 & -0.000729 & -0.00302 \\
 & (0.0158) & (0.0158) & (0.0152) & (0.0155) & (0.0158) & (0.0156) \\
Rep leaning news & & -0.0244 & & -0.109*** & & -0.123*** \\
 & & (0.0343) & & (0.0347) & & (0.0237) \\
Dem leaning news & & -0.0122 & & -0.0515** & & -0.0273 \\
 & & (0.0263) & & (0.0255) & & (0.0279) \\
Constant & 0.0506 & 0.0783 & 0.344*** & 0.407*** & 0.267** & 0.354*** \\
 & (0.0971) & (0.0966) & (0.0989) & (0.103) & (0.122) & (0.122) \\
 & & & & & & \\
Controls & Yes & Yes & Yes & Yes & Yes & Yes \\
Observations & 1,002 & 1,002 & 1,010 & 1,010 & 1,004 & 1,004 \\
R-squared & 0.150 & 0.161 & 0.276 & 0.291 & 0.248 & 0.267 \\
Avg increase support & 0.144 & 0.144 & 0.158 & 0.158 & 0.168 & 0.168 \\
Avg decrease support & 0.325 & 0.325 & 0.251 & 0.251 & 0.254 & 0.254 \\ \hline
\multicolumn{7}{l}{ \begin{minipage}{1.15\columnwidth}%
\small \textit{Notes}. Standard errors in parentheses. *** p\textless{}0.01, ** p\textless{}0.05, * p\textless{}0.1. The percentages in the first row report the share of respondents who increased their Likert-based score between the first and last wave of the survey. All regressions are OLS regressions that take into account population survey wights and the sampling procedure. The dependent variable is a dummy=1 if the respondent increased their belief that it should be a government's responsibility to provide the following policies. The control variables include: gender, race, age, education, parental status, caring responsibilities for an elderly or a person with a disability, baseline income in February 2020, cohabitation with a partner, labor force participation and employment status in February 2020, health insurance provider, if the respondent had financial difficulties before the pandemic, macro-region, metro vs. rural, the population density at the zip code, and two dummy variables indicating if they consume at least 30min a week of international news and if they have at least one social media account. We also control for whether respondents completed the survey in a shorter time than the 99$^{th}$ percentile as well as ceiling effects.\end{minipage}
}
\end{tabular}%
}
\end{table}

\begin{table}[H]
\caption{The effect of shocks and media on welfare policy preferences - C}
\label{tab:media_policies_C}
\resizebox{\textwidth}{!}{%
\begin{tabular}{lcccccccc}
\hline
\hline
 & \multicolumn{8}{c}{\begin{tabular}[c]{@{}c@{}}Stronger belief between Apr and Oct 2020 that\\ it should be the role of government to:\end{tabular}} \\
 & (1) & (2) & (3) & (4) & (5) & (6) & (7) & (8) \\
 & \begin{tabular}[c]{@{}c@{}}Help \\ those affected \\ by natural \\ disasters\end{tabular} & \begin{tabular}[c]{@{}c@{}}Help \\ those affected \\ by natural \\ disasters\end{tabular} & \begin{tabular}[c]{@{}c@{}}Give \\ financial help \\ to low income \\ students\end{tabular} & \begin{tabular}[c]{@{}c@{}}Give \\ financial help \\ to low income \\ students\end{tabular} & \begin{tabular}[c]{@{}c@{}}Help \\ industry \\ grow\end{tabular} & \begin{tabular}[c]{@{}c@{}}Help \\ industry \\ grow\end{tabular} & \begin{tabular}[c]{@{}c@{}}Keep \\ prices \\ under \\ control\end{tabular} & \begin{tabular}[c]{@{}c@{}}Keep \\ prices \\ under \\ control\end{tabular} \\ \hline
 & \multicolumn{1}{l}{} & \multicolumn{1}{l}{} & \multicolumn{1}{l}{} & \multicolumn{1}{l}{} & \multicolumn{1}{l}{} & \multicolumn{1}{l}{} & \multicolumn{1}{l}{} & \multicolumn{1}{l}{} \\
Republican & 0.0124 & 0.0137 & -0.0103 & 0.00701 & 0.0291 & 0.0446 & -0.0313 & -0.0249 \\
 & (0.0292) & (0.0310) & (0.0302) & (0.0315) & (0.0386) & (0.0378) & (0.0348) & (0.0349) \\
Democrat & -0.0109 & -0.0151 & 0.0494 & 0.0447 & 0.109*** & 0.0894** & 0.0455 & 0.0360 \\
 & (0.0202) & (0.0181) & (0.0335) & (0.0343) & (0.0368) & (0.0354) & (0.0331) & (0.0327) \\
Lost 20\% income & 0.0151 & 0.0181 & 0.00654 & 0.00945 & -0.0780*** & -0.0745*** & -0.00655 & -0.00456 \\
 & (0.0278) & (0.0289) & (0.0281) & (0.0275) & (0.0295) & (0.0283) & (0.0294) & (0.0294) \\
Knows hospitalized & 0.0130 & 0.0121 & 0.0329 & 0.0422* & -0.00536 & -0.0128 & 0.0531* & 0.0526* \\
 & (0.0248) & (0.0252) & (0.0248) & (0.0252) & (0.0322) & (0.0323) & (0.0279) & (0.0295) \\
Var consumer expenditures & 0.0348** & 0.0362** & 0.00413 & 0.00213 & -0.0198 & -0.0235 & 0.0345* & 0.0329* \\
 & (0.0163) & (0.0155) & (0.0190) & (0.0176) & (0.0232) & (0.0230) & (0.0179) & (0.0188) \\
Incr COVID-19 cases & -0.00658 & -0.00517 & 0.00458 & 0.00439 & -0.00964 & -0.00937 & -0.0179 & -0.0181 \\
 & (0.0103) & (0.0104) & (0.0144) & (0.0140) & (0.0165) & (0.0158) & (0.0141) & (0.0142) \\
Rep leaning news & & 0.00136 & & -0.0904** & & -0.0368 & & -0.0157 \\
 & & (0.0276) & & (0.0441) & & (0.0406) & & (0.0426) \\
Dem leaning news & & 0.0171 & & -0.0258 & & 0.0388 & & 0.0222 \\
 & & (0.0271) & & (0.0360) & & (0.0411) & & (0.0365) \\
Constant & 0.407*** & 0.388*** & 0.217 & 0.255* & 0.209 & 0.186 & 0.180 & 0.173 \\
 & (0.125) & (0.134) & (0.151) & (0.153) & (0.175) & (0.184) & (0.121) & (0.124) \\
 & & & & & & & & \\
Controls & Yes & Yes & Yes & Yes & Yes & Yes & Yes & Yes \\
Observations & 1,007 & 1,007 & 1,003 & 1,003 & 1,002 & 1,002 & 1,005 & 1,005 \\
R-squared & 0.381 & 0.387 & 0.225 & 0.235 & 0.192 & 0.200 & 0.316 & 0.318 \\
Avg increase support & 0.126 & 0.126 & 0.187 & 0.187 & 0.262 & 0.262 & 0.244 & 0.244 \\
Avg decrease support & 0.224 & 0.224 & 0.215 & 0.215 & 0.244 & 0.244 & 0.233 & 0.233 \\ \hline
\multicolumn{9}{l}{ \begin{minipage}{1.5\columnwidth}%
\small \textit{Notes}. Standard errors in parentheses. *** p\textless{}0.01, ** p\textless{}0.05, * p\textless{}0.1. The percentages in the first row report the share of respondents who increased their Likert-based score between the first and last wave of the survey. All regressions are OLS regressions that take into account population survey wights and the sampling procedure. The dependent variable is a dummy=1 if the respondent increased their belief that it should be the government's responsibility to provide the following policies. The control variables include: gender, race, age, education, parental status, caring responsibilities for an elderly or a person with a disability, baseline income in February 2020, cohabitation with a partner, labor force participation and employment status in February 2020, health insurance provider, if the respondent had financial difficulties before the pandemic, macro-region, metro vs. rural, the population density at the zip code, and two dummy variables indicating if they consume at least 30min a week of international news and if they have at least one social media account. We also control for whether respondents completed the survey in a shorter time than the 99$^{th}$ percentile as well as ceiling effects.\end{minipage}
}
\end{tabular}%
}
\end{table}

\begin{table}[H]
\caption{The effect of shocks and media on temporary relief policies}
\label{tab:covid_policies}
\resizebox{\textwidth}{!}{%
\begin{tabular}{lcccccc}
\hline
\hline
 & \multicolumn{6}{c}{\begin{tabular}[c]{@{}c@{}}Stronger belief between Apr and Oct 2020 that\\ it should be the role of government to:\end{tabular}} \\
 & (1) & (2) & (3) & (4) & (5) & (6) \\
 & \begin{tabular}[c]{@{}c@{}}Spend more on \\ health care \\ to reduce \\ deaths\end{tabular} & \begin{tabular}[c]{@{}c@{}}Spend more on \\ health care \\ to reduce \\ deaths\end{tabular} & \begin{tabular}[c]{@{}c@{}}Do more \\ to protect \\ essential \\ workers\end{tabular} & \begin{tabular}[c]{@{}c@{}}Do more \\ to protect \\ essential \\ workers \end{tabular} & \begin{tabular}[c]{@{}c@{}}Transfer money\\ directly to \\ families \\ \& businesses\end{tabular} & \begin{tabular}[c]{@{}c@{}}Transfer money\\ directly to \\ families \\ \& businesses\end{tabular} \\ \hline
 & & & & & & \\
Republican & -0.0459 & -0.0456 & 0.0233 & 0.0497 & -0.0574 & -0.0425 \\
 & (0.0444) & (0.0437) & (0.0476) & (0.0446) & (0.0369) & (0.0364) \\
Democrat & 0.129*** & 0.133*** & 0.104*** & 0.0765** & 0.0119 & 0.00229 \\
 & (0.0407) & (0.0379) & (0.0318) & (0.0303) & (0.0334) & (0.0348) \\
Lost 20\% income & 0.0606* & 0.0629* & 0.0663** & 0.0727** & 0.0527* & 0.0536* \\
 & (0.0335) & (0.0320) & (0.0317) & (0.0330) & (0.0309) & (0.0314) \\
Knows hospitalized & 0.0131 & 0.0131 & -0.00242 & -0.0118 & -0.0131 & -0.0173 \\
 & (0.0290) & (0.0296) & (0.0221) & (0.0222) & (0.0246) & (0.0259) \\
Var consumer expenditures & 0.0136** & 0.0151*** & 0.0112 & 0.00972 & 0.0145** & 0.0139** \\
 & (0.00628) & (0.00534) & (0.0110) & (0.0103) & (0.00626) & (0.00642) \\
Incr COVID-19 cases & 0.0140 & 0.0168 & 0.00663 & 0.00723 & -0.00921 & -0.00851 \\
 & (0.0147) & (0.0144) & (0.0160) & (0.0155) & (0.0138) & (0.0137) \\
Rep leaning news & & -0.00428 & & -0.0929** & & -0.0338 \\
 & & (0.0379) & & (0.0421) & & (0.0370) \\
Dem leaning news & & -0.0157 & & 0.00439 & & -0.0116 \\
 & & (0.0328) & & (0.0357) & & (0.0381) \\
Constant & 0.228 & 0.197 & 0.172 & 0.135 & 0.183 & 0.167* \\
 & (0.147) & (0.154) & (0.161) & (0.136) & (0.113) & (0.0958) \\
 & & & & & & \\
Controls & Yes & Yes & Yes & Yes & Yes & Yes \\
Observations & 937 & 937 & 938 & 938 & 942 & 942 \\
R-squared & 0.197 & 0.206 & 0.208 & 0.238 & 0.127 & 0.135 \\
Average increase & 0.177 & 0.177 & 0.188 & 0.188 & 0.181 & 0.181 \\
Average decrease & 0.295 & 0.295 & 0.317 & 0.317 & 0.369 & 0.369 \\ \hline
\multicolumn{7}{l}{ \begin{minipage}{1.33\columnwidth}%
\small \textit{Notes}. Standard errors in parentheses. *** p\textless{}0.01, ** p\textless{}0.05, * p\textless{}0.1. The percentages in the first row report the share of respondents who increased their Likert-based score between the first and last wave of the survey. All regressions are OLS regressions that take into account population survey wights and the sampling procedure. The dependent variable is a dummy=1 if the respondent increased their belief that it should be the government's responsibility to provide the following policies. The control variables include: gender, race, age, education, parental status, caring responsibilities for an elderly or a person with a disability, baseline income in February 2020, cohabitation with a partner, labor force participation and employment status in February 2020, health insurance provider, if the respondent had financial difficulties before the pandemic, macro-region, metro vs. rural, the population density at the zip code, and two dummy variables indicating if they consume at least 30min a week of international news and if they have at least one social media account. We also control for whether respondents completed the survey in a shorter time than the 99$^{th}$ percentile as well as ceiling effects.\end{minipage}
}
\end{tabular}%
}
\end{table}

\begin{table}[H]
\caption{The effect of shocks and media on trust in institutions - A}
\label{tab:trust_a}
\resizebox{\textwidth}{!}{%
\begin{tabular}{lcccccccc}
\hline
\hline
 & \multicolumn{8}{c}{Decreased confidence in people running the following institutions:} \\
 & (1) & (2) & (3) & (4) & (5) & (6) & (7) & (8) \\
 & \begin{tabular}[c]{@{}c@{}}Congress \\ \& Senate\end{tabular} & \begin{tabular}[c]{@{}c@{}}Congress \\ \& Senate\end{tabular} & \begin{tabular}[c]{@{}c@{}}White\\ House\end{tabular} & \begin{tabular}[c]{@{}c@{}}White\\ House\end{tabular} & \begin{tabular}[c]{@{}c@{}}Financial\\ institutions\\ \& banks\end{tabular} & \begin{tabular}[c]{@{}c@{}}Financial\\ institutions\\ \& banks\end{tabular} & \begin{tabular}[c]{@{}c@{}}Private\\ sector\end{tabular} & \multicolumn{1}{l}{\begin{tabular}[c]{@{}l@{}}Private\\ sector\end{tabular}} \\ \hline
 & & & & & & & & \multicolumn{1}{l}{} \\
Republican & -0.0959** & -0.0797* & -0.137*** & -0.117*** & 0.0202 & 0.0473 & 0.0159 & 0.0305 \\
 & (0.0419) & (0.0453) & (0.0398) & (0.0365) & (0.0386) & (0.0418) & (0.0469) & (0.0477) \\
Democrat & 0.0611* & 0.0382 & 0.134*** & 0.102*** & 0.0544 & 0.0406 & 0.00710 & 0.00737 \\
 & (0.0342) & (0.0355) & (0.0322) & (0.0305) & (0.0444) & (0.0471) & (0.0404) & (0.0409) \\
Lost 20\% income & 0.0605 & 0.0679* & 0.0110 & 0.0196 & 0.0743** & 0.0734** & 0.0577* & 0.0578* \\
 & (0.0381) & (0.0377) & (0.0304) & (0.0302) & (0.0330) & (0.0338) & (0.0302) & (0.0301) \\
Knows hospitalized & 0.0302 & 0.0392 & 0.0353 & 0.0401 & 0.0317 & 0.0296 & 0.0173 & 0.0262 \\
 & (0.0352) & (0.0342) & (0.0347) & (0.0338) & (0.0382) & (0.0367) & (0.0412) & (0.0414) \\
Var consumer expenditures & 0.00543 & -3.66e-05 & 0.0506** & 0.0384* & -0.000902 & -0.00634 & -0.0121 & -0.0138 \\
 & (0.0332) & (0.0319) & (0.0216) & (0.0213) & (0.0303) & (0.0300) & (0.0212) & (0.0211) \\
Incr COVID-19 cases & -0.00686 & -0.00450 & -0.00984 & -0.0118 & 0.00477 & 0.00286 & 0.0289 & 0.0287 \\
 & (0.0210) & (0.0206) & (0.0173) & (0.0163) & (0.0192) & (0.0199) & (0.0197) & (0.0199) \\
Rep leaning news & & -0.108** & & -0.140*** & & -0.0869* & & -0.0678* \\
 & & (0.0479) & & (0.0517) & & (0.0496) & & (0.0404) \\
Dem leaning news & & -0.00928 & & 0.0397 & & -0.0130 & & -0.0373 \\
 & & (0.0418) & & (0.0304) & & (0.0442) & & (0.0384) \\
Constant & 0.455** & 0.527*** & 0.631*** & 0.719*** & 0.468*** & 0.475*** & -0.0529 & -0.0264 \\
 & (0.211) & (0.195) & (0.131) & (0.136) & (0.159) & (0.170) & (0.0959) & (0.105) \\
 & & & & & & & & \\
 Controls & Yes & Yes & Yes & Yes & Yes & Yes & Yes & Yes \\
Observations & 1,009 & 1,009 & 1,003 & 1,003 & 1,007 & 1,007 & 1,006 & 1,006 \\
R-squared & 0.167 & 0.179 & 0.357 & 0.376 & 0.113 & 0.126 & 0.124 & 0.130 \\
Avg increase trust & 0.159 & 0.159 & 0.142 & 0.142 & 0.142 & 0.142 & 0.185 & 0.185 \\
Avg decrease trust & 0.351 & 0.351 & 0.312 & 0.312 & 0.299 & 0.299 & 0.247 & 0.247 \\ \hline
\multicolumn{9}{l} {
\begin{minipage}{1.33\columnwidth}%
\small \textit{Notes}. Standard errors in parentheses. *** p\textless{}0.01, ** p\textless{}0.05, * p\textless{}0.1. The percentages in the first row report the share of respondents who increased their Likert-based score between the fourth and last wave of the survey. All regressions are OLS regressions that take into account population survey weights and the sampling procedure. The dependent variable is a dummy=1 if the respondent reduced their confidence in the people running the following institutions. The control variables include: gender, race, age, education, parental status, caring responsibilities for an elderly or a person with a disability, baseline income in February 2020, cohabitation with a partner, labor force participation and employment status in February 2020, health insurance provider, if the respondent had financial difficulties before the pandemic, macro-region, metro vs. rural, the population density at the zip code, and two dummy variables indicating if they consume at least 30min a week of international news and if they consume news from social media. We also control for whether respondents completed the survey in a shorter time than the 99$^{th}$ percentile as well as ceiling effects.\end{minipage}
}
\end{tabular}%
}
\end{table}

\begin{table}[H]
\caption{The effect of shocks and media on trust in institutions - B}
\label{tab:trust_b}
\resizebox{\textwidth}{!}{%
\begin{tabular}{lcccccc}
\hline
\hline
 & \multicolumn{6}{c}{Decreased confidence in people running the following institutions:} \\
 & (1) & (2) & (3) & (4) & (5) & (6) \\
 & \begin{tabular}[c]{@{}c@{}}Scientific\\ community\end{tabular} & \begin{tabular}[c]{@{}c@{}}Scientific\\ community\end{tabular} & \begin{tabular}[c]{@{}c@{}}Health \\ insurance\\ companies\end{tabular} & \begin{tabular}[c]{@{}c@{}}Health \\ insurance\\ companies\end{tabular} & Hospitals & Hospitals \\ \hline
 & & & & & & \\
Republican & 0.0370 & 0.0214 & -0.0319 & -0.0349 & 0.0589 & 0.0392 \\
 & (0.0439) & (0.0447) & (0.0437) & (0.0440) & (0.0357) & (0.0373) \\
 Democrat & -0.117*** & -0.0983*** & -0.0359 & -0.0353 & -0.0575 & -0.0399 \\
 & (0.0345) & (0.0357) & (0.0395) & (0.0388) & (0.0401) & (0.0385) \\
Lost 20\% income & -0.0218 & -0.0249 & 0.0367 & 0.0402 & 0.0579* & 0.0546* \\
 & (0.0314) & (0.0302) & (0.0367) & (0.0359) & (0.0319) & (0.0328) \\
Knows hospitalized & 0.0472 & 0.0486 & 0.0204 & 0.0193 & 0.0419 & 0.0393 \\
 & (0.0438) & (0.0446) & (0.0347) & (0.0333) & (0.0375) & (0.0366) \\
Var consumer expenditures & 0.00691 & 0.0116 & -0.0647*** & -0.0622*** & -0.0560** & -0.0519** \\
 & (0.0339) & (0.0335) & (0.0218) & (0.0209) & (0.0235) & (0.0236) \\
Incr COVID-19 cases & 0.0326 & 0.0310 & -0.00768 & -0.00480 & -0.0165 & -0.0197 \\
 & (0.0198) & (0.0204) & (0.0191) & (0.0191) & (0.0235) & (0.0230) \\
Rep leaning news & & 0.0609 & & 0.0433 & & 0.102** \\
 & & (0.0586) & & (0.0581) & & (0.0489) \\
Dem leaning news & & -0.0145 & & 0.0564 & & 0.0253 \\
 & & (0.0488) & & (0.0446) & & (0.0452) \\
Constant & 0.318* & 0.300* & 0.178 & 0.136 & 0.403*** & 0.361** \\
 & (0.165) & (0.168) & (0.167) & (0.163) & (0.152) & (0.149) \\
 & & & & & & \\
 Controls & Yes & Yes & Yes & Yes & Yes & Yes \\
Observations & 1,002 & 1,002 & 1,006 & 1,006 & 1,007 & 1,007 \\
R-squared & 0.088 & 0.095 & 0.123 & 0.129 & 0.096 & 0.109 \\
Avg increase trust & 0.185 & 0.185 & 0.172 & 0.172 & 0.164 & 0.164 \\
Avg decrease trust & 0.286 & 0.286 & 0.284 & 0.284 & 0.306 & 0.306 \\ \hline 
\multicolumn{7}{l} { 
\begin{minipage}{1.15\columnwidth}%
\small \textit{Notes}. Standard errors in parentheses. *** p\textless{}0.01, ** p\textless{}0.05, * p\textless{}0.1. The percentages in the first row report the share of respondents who increased their Likert-based score between the fourth and last wave of the survey. All regressions are OLS regressions that take into account population survey weights and the sampling procedure. The dependent variable is a dummy=1 if the respondent reduced their confidence in the people running the following institutions. The control variables include: gender, race, age, education, parental status, caring responsibilities for an elderly or a person with a disability, baseline income in February 2020, cohabitation with a partner, labor force participation and employment status in February 2020, health insurance provider, if the respondent had financial difficulties before the pandemic, macro-region, metro vs. rural, the population density at the zip code, and two dummy variables indicating if they consume at least 30min a week of international news and if they consume news from social media. We also control for whether respondents completed the survey in a shorter time than the 99$^{th}$ percentile as well as ceiling effects.\end{minipage}
}
\end{tabular}%
}
\end{table}


\subsection{Survey experiment}

\begin{figure}[H]
	\caption{Judgment as a function of accurate information }
	\label{f:deaths__success_f}
	\begin{center}
		\includegraphics[height=11cm]{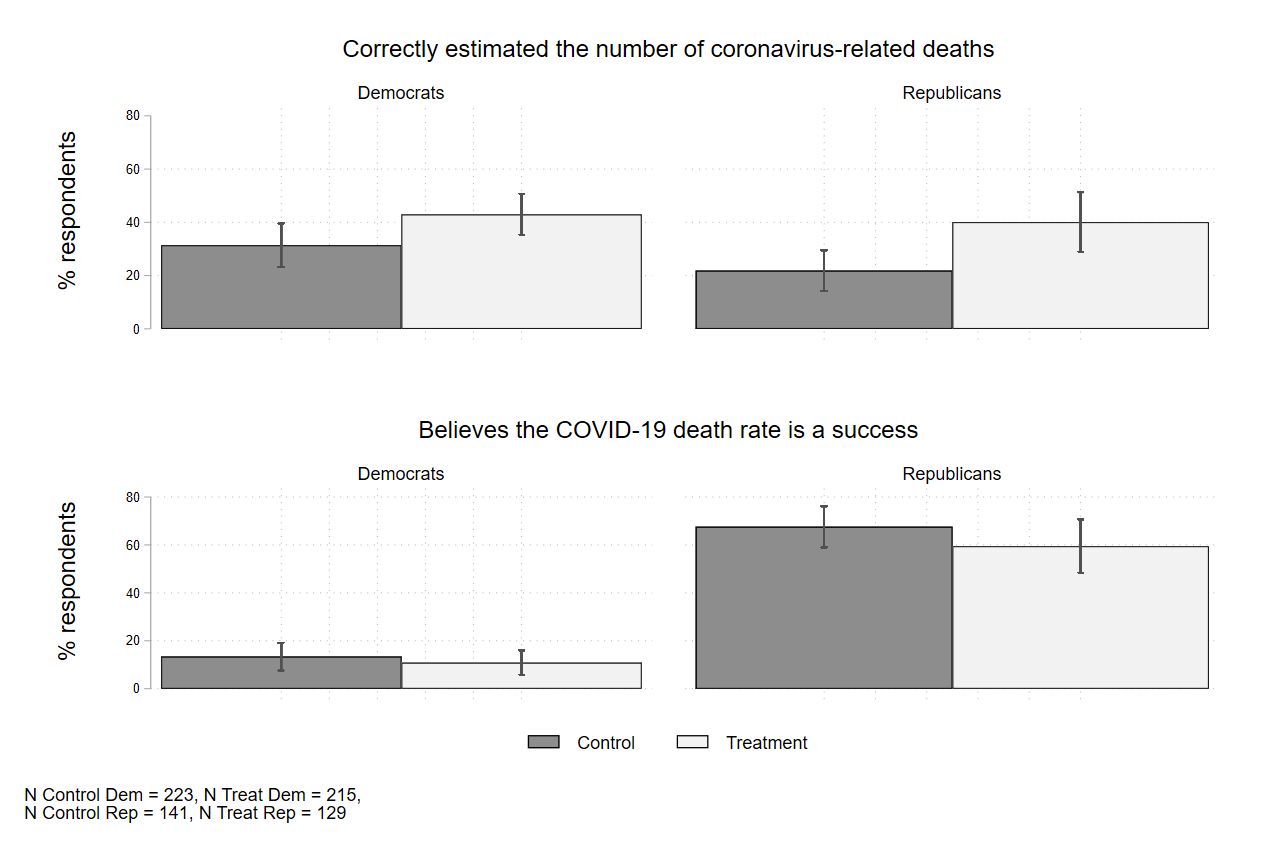}
	\end{center}
\begin{minipage}{1\linewidth \setstretch{0.75}}
	{\scriptsize{\textit{Notes}:} 
		\scriptsize The figure on top shows the share of respondents who correctly estimated the number of COVID-19 deaths in both their state and the U.S. by party and treatment group. The figure at the bottom shows the share of respondents who believed the COVID-19 death rate could be considered a success by party and by whether they were in the treatment or the control group. Error bars are 95\% confidence intervals.}
\end{minipage}	
\end{figure}

\begin{table}[H]
\centering
\caption{Information processing, expectations, and judgment}
\label{tab:expected_death}
\resizebox{0.78\textwidth}{!}{%
\begin{tabular}{lccc}
\hline
\hline
 & (1) & (2) & (3) \\
 &  \begin{tabular}[c]{@{}c@{}}Expected \\ death rate \\ (in 1K) \end{tabular} &
 \begin{tabular}[c]{@{}c@{}}Expected \\ death rate \\ is a success\end{tabular} &
 \begin{tabular}[c]{@{}c@{}}Expected \\ death rate \\ is a success\end{tabular} \\ \hline
Democrat & 855.3*** & -0.234*** & -0.214*** \\
 & (215.4) & (0.0391) & (0.0409) \\
Republican & -748.7*** & 0.220*** & 0.206*** \\
 & (206.6) & (0.0472) & (0.0462) \\
Lost 20\% income & 519.7* & -0.0155 & 0.00287 \\
 & (302.7) & (0.0440) & (0.0427) \\
Knows hospitalized & 409.9 & 0.0231 & 0.0333 \\
 & (361.9) & (0.0514) & (0.0508) \\
Consumer exp - May & 1,482* & -0.182 & -0.210 \\
 & (815.5) & (0.151) & (0.174) \\
ln COVID-19 cases & -264.5*** & 0.0143 & 0.00497 \\
 & (95.99) & (0.0132) & (0.0141) \\
Democratic leaning news & 1,164*** & -0.115*** & -0.0671 \\
 & (281.3) & (0.0422) & (0.0430) \\
Republican leaning news & -1,043*** & 0.179*** & 0.148*** \\
 & (272.2) & (0.0441) & (0.0418) \\
 Expected additional deaths (1k) & & & -0.0284*** \\
 & & & (0.00652) \\
Constant & 3,544*** & 0.641*** & 0.774*** \\
 & (879.1) & (0.130) & (0.145) \\
 & & & \\
 Controls & Yes & Yes & Yes \\
Observations & 1,184 & 1,182 & 1,176 \\
R-squared & 0.238 & 0.306 & 0.340 \\
Mean dep. var. & 4983 & 0.413 & 0.413\\ \hline
\multicolumn{4}{l}{ %
\begin{minipage}{0.85\columnwidth}%
 \small \textit{Notes}: Standard errors in parentheses. *** p\textless{}0.01, ** p\textless{}0.05, * p\textless{}0.1. \\
All regressions are OLS regressions that take into account population survey wights and the sampling procedure. The dependent variable is a continuous value of the expected COVID-19 deaths by the end of the year in columns (1), and a dummy=1 if the respondent believed the expected death rate can be considered as a success when judging the work done by public authorities in columns (2) to (3). The control variables include: gender, race, age, education, parental status, caring responsibilities for an elderly or a person with a disability, baseline income in February 2020, cohabitation with a partner, labor force participation and employment status in February 2020, health insurance provider, if the respondent had financial difficulties before the pandemic, macro-region, metro vs. rural, and the population density at the zip code. We also control for whether respondents completed the survey in a shorter time than the 99$^{th}$ percentile. We also consider the media diet and control for social media usage and the amount of international news consumed. \\ \end{minipage}
}%
\end{tabular}
}
\end{table}

\begin{table}[H]
\centering
\caption{Balance table across the treatment and the control group for the experiment on death estimation.}
\label{tab:balancetable_deathexp}
\resizebox*{!}{0.9\textheight}{%
\begin{tabular}{lccc} \hline \hline
 & (1) & (2) & (3) \\
& Mean controls & Mean treated & Difference \\ \hline
&&&\\
Republican & 0.240 & 0.241 & 0.002 \\
 & (0.427) & (0.428) & (0.025) \\
Democrat & 0.377 & 0.381 & 0.004 \\
 & (0.485) & (0.486) & (0.028) \\
Independent/non-voter & 0.383 & 0.378 & -0.006 \\
 & (0.487) & (0.485) & (0.028) \\
Woman & 0.483 & 0.479 & -0.004 \\
 & (0.500) & (0.500) & (0.029) \\
Age: 18-29 & 0.207 & 0.216 & 0.008 \\
 & (0.406) & (0.412) & (0.024) \\
Age: 30-44 & 0.282 & 0.286 & 0.004 \\
 & (0.450) & (0.452) & (0.026) \\
Age: 45-59 & 0.219 & 0.224 & 0.006 \\
 & (0.414) & (0.417) & (0.024) \\
High school & 0.150 & 0.159 & 0.009 \\
 & (0.357) & (0.366) & (0.021) \\
Some college & 0.392 & 0.400 & 0.008 \\
 & (0.488) & (0.490) & (0.028) \\
Bachelor + & 0.424 & 0.414 & -0.010 \\
 & (0.495) & (0.493) & (0.029) \\
\$10-23k & 0.188 & 0.190 & 0.002 \\
 & (0.391) & (0.392) & (0.023) \\
\$23-37k & 0.189 & 0.209 & 0.019 \\
 & (0.392) & (0.407) & (0.023) \\
\$37-62 & 0.194 & 0.219 & 0.025 \\
 & (0.396) & (0.414) & (0.023) \\
Over \$62k & 0.215 & 0.191 & -0.024 \\
 & (0.411) & (0.394) & (0.023) \\
\multirow{2}{*}{\begin{tabular}[c]{@{}l@{}}Financial hardship\\ pre-COVID-19\end{tabular}} & 0.302 & 0.274 & -0.028 \\
 & (0.459) & (0.446) & (0.027) \\
African American & 0.100 & 0.095 & -0.005 \\
 & (0.300) & (0.293) & (0.017) \\
Hispanic & 0.150 & 0.152 & 0.002 \\
 & (0.357) & (0.359) & (0.021) \\
Other Race & 0.116 & 0.128 & 0.012 \\
 & (0.320) & (0.334) & (0.019) \\
Coabitating & 0.626 & 0.614 & -0.013 \\
 & (0.484) & (0.487) & (0.028) \\
Parent of minor & 0.263 & 0.286 & 0.024 \\
 & (0.440) & (0.452) & (0.026) \\
Caring responsibilities & 0.152 & 0.169 & 0.017 \\
 & (0.359) & (0.375) & (0.021) \\
Not in the labor force & 0.256 & 0.249 & -0.007 \\
 & (0.437) & (0.433) & (0.025) \\
Unemployed in Feb & 0.059 & 0.054 & -0.005 \\
 & (0.235) & (0.225) & (0.013) \\
Midwest & 0.268 & 0.257 & -0.011 \\
 & (0.443) & (0.437) & (0.026) \\
South & 0.357 & 0.352 & -0.006 \\
 & (0.480) & (0.478) & (0.028) \\
West & 0.235 & 0.234 & -0.000 \\
 & (0.424) & (0.424) & (0.025) \\
Metropolitan area & 0.856 & 0.876 & 0.019 \\
 & (0.351) & (0.330) & (0.020) \\
No health insurance & 0.074 & 0.078 & 0.005 \\
 & (0.261) & (0.269) & (0.015) \\
Population density in ZCTA & 3,921.175 & 3,771.923 & -149.253 \\
 & (9,715.106) & (8,380.174) & (527.568) \\
Dem leaning news & 0.288 & 0.310 & 0.023 \\
 & (0.453) & (0.463) & (0.029) \\
Rep leaning news & 0.237 & 0.224 & -0.013 \\
 & (0.426) & (0.418) & (0.027) \\
\multirow{2}{*}{\begin{tabular}[c]{@{}l@{}}30+ mins/day\\ international news\end{tabular}} & 0.235 & 0.264 & 0.029 \\
 & (0.424) & (0.441) & (0.026) \\
News from social media & 0.368 & 0.404 & 0.036 \\
 & (0.483) & (0.491) & (0.028) \\ 
\textit{N} & 613 & 580 & 1,193 \\ \hline
\end{tabular}%
}
\end{table}

\begin{table}[H]
\centering
\caption{The effect of providing factual information in changing misunderstanding and assessment of the gravity of the crisis.}
\label{tab:death_exp_main}
\resizebox{\textwidth}{!}{%
\begin{tabular}{lcccccc} \hline \hline
 & (1) & (2) & (3) & (4) & (5) & (6) \\
 & \begin{tabular}[c]{@{}c@{}}Correctly \\ estimated\\ US \& State\\ deaths\end{tabular} & \begin{tabular}[c]{@{}c@{}}Correctly\\ estimated\\ US \& State\\ deaths\end{tabular} & \begin{tabular}[c]{@{}c@{}}US \& State \\ deaths\\ are a\\ success\end{tabular} & \begin{tabular}[c]{@{}c@{}}US \& State \\ deaths\\ are a\\ success\end{tabular} & \begin{tabular}[c]{@{}c@{}}US \& State \\ deaths\\ are a\\ success\end{tabular} & \begin{tabular}[c]{@{}c@{}}Correctly \\ stated\\ the US deaths\\ vs. the world\end{tabular} \\ \hline
 & & & & & & \\
CDC Tx & 0.118*** & 0.149*** & -0.0415 & -0.0198 & -0.0404 & 0.0125 \\
 & (0.0305) & (0.0341) & (0.0313) & (0.0370) & (0.0368) & (0.0423) \\
CDC Tx*Rep news & & -0.0370 & & -0.0996 & -0.0613 & 0.236** \\
 & & (0.0643) & & (0.0719) & (0.0729) & (0.0922) \\
CDC Tx*Dem news & & -0.0905 & & -0.000831 & -0.00802 & -0.0417 \\
 & & (0.0624) & & (0.0617) & (0.0571) & (0.0671) \\
Democrat & 0.0615 & 0.0596 & -0.130*** & -0.130*** & -0.0533* & -0.111*** \\
 & (0.0402) & (0.0404) & (0.0307) & (0.0306) & (0.0291) & (0.0416) \\
Republican & -0.0330 & -0.0331 & 0.230*** & 0.230*** & 0.143*** & -0.0336 \\
 & (0.0369) & (0.0366) & (0.0431) & (0.0432) & (0.0425) & (0.0386) \\
Lost 20\% income & -0.0307 & -0.0313 & -0.0109 & -0.0108 & 0.00956 & -0.0441 \\
 & (0.0395) & (0.0398) & (0.0383) & (0.0380) & (0.0351) & (0.0429) \\
Knows hospitalized & -0.0730* & -0.0746* & -0.0135 & -0.0145 & -0.0164 & -0.0166 \\
 & (0.0422) & (0.0418) & (0.0329) & (0.0333) & (0.0324) & (0.0389) \\
ln COVID-19 cases & -0.0178 & -0.0187 & -0.00475 & -0.00522 & -0.0150 & 0.0191 \\
 & (0.0178) & (0.0179) & (0.0191) & (0.0195) & (0.0199) & (0.0251) \\
Consumer exp - June & 0.158 & 0.153 & -0.204* & -0.228** & -0.179 & 0.0111 \\
 & (0.124) & (0.123) & (0.115) & (0.114) & (0.111) & (0.119) \\
Dem leaning news & 0.0188 & 0.0635 & -0.0419 & -0.0420 & -0.0175 & 0.0280 \\
 & (0.0369) & (0.0478) & (0.0380) & (0.0526) & (0.0503) & (0.0577) \\
Rep leaning news & -0.0267 & -0.00890 & 0.267*** & 0.311*** & 0.214*** & -0.284*** \\
 & (0.0514) & (0.0565) & (0.0420) & (0.0508) & (0.0562) & (0.0726) \\
\multirow{2}{*}{\begin{tabular}[c]{@{}l@{}}Expected additional death\\ rate is a success (w5)\end{tabular}} & & & & & 0.390*** & \\
 & & & & & (0.0390) & \\
Constant & 0.300** & 0.297** & 0.396*** & 0.395*** & 0.174 & 0.954*** \\
 & (0.140) & (0.140) & (0.139) & (0.140) & (0.134) & (0.249) \\
 & & & & & & \\
Controls & Yes & Yes & Yes & Yes & Yes & Yes \\
Observations & 1,141 & 1,141 & 1,137 & 1,137 & 1,137 & 948 \\
R-squared & 0.158 & 0.160 & 0.285 & 0.287 & 0.390 & 0.102 \\
Mean dep. var. & 0.330 & 0.330 & 0.335 & 0.335 & 0.335 & 0.552 \\ \hline
\multicolumn{7}{l}{%
 \begin{minipage}{1.25\columnwidth}%
  \small \textit{Notes}: Standard errors in parentheses. *** p\textless{}0.01, ** p\textless{}0.05, * p\textless{}0.1. The dep. var. in Col (1) and (2) is a dummy=1 if the respondent provided the correct death rate, while col (2), (3), and (4) it is a dummy=1 if the respondents believed the COVID-19 death rate at the National and State level was a success. Col. (6) reports a regression predicting whether the respondent correctly stated that the US death rate was higher than in most countries in the world in wave 7. The control variables include: gender, race, age, education, parental status, caring responsibilities for an elderly or a person with a disability, baseline income in February 2020, cohabitation with a partner, labor force participation and employment status in February 2020, health insurance provider, if the respondent had financial difficulties before the pandemic, macro-region, metro vs. rural, and the population density at the zip code. We also control for whether respondents completed the survey in a shorter time than the 99$^{th}$ percentile. Finally, we consider social media usage and the amount of international news consumed.
 \end{minipage}%
}
\end{tabular}%
}
\end{table}

\begin{figure}[H]
	\caption{CDC webpage}
	\label{fig:wte}
	\begin{center}
		\includegraphics[height=12cm]{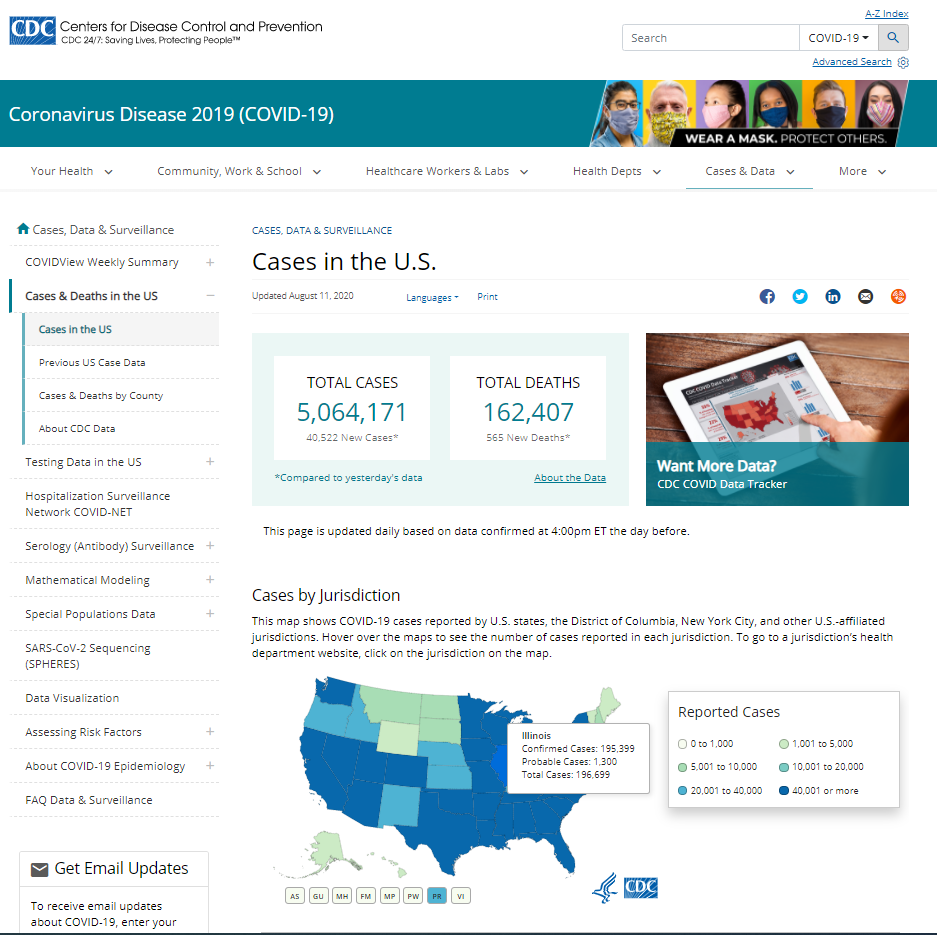}
	\end{center}
\begin{minipage}{1\linewidth \setstretch{0.75}}
	{\scriptsize{\textit{Notes}: The figure shows the landing webpage that respondents in the treatment group saw when clicking on the link in the experiment. This snapshot was taken on August 12 as an example. The webpage has not changed in design or layout through the year}}. 
\end{minipage}	
\end{figure}

\begin{table}[H]
\centering
\caption{Additional analyses for the death experiment}
\label{tab:death_tr_effect}
\resizebox{\textwidth}{!}{%
\begin{tabular}{lcccccc} \hline \hline
 & (1) & (2) & (3) & (4) & (5) & (6) \\
 & \begin{tabular}[c]{@{}c@{}}Time\\ to select\\ State deaths\end{tabular} & \begin{tabular}[c]{@{}c@{}}Time\\ to select\\ US deaths\end{tabular} & \begin{tabular}[c]{@{}c@{}}Underestimated \\ State deaths\end{tabular} & \begin{tabular}[c]{@{}c@{}}Overestimated \\ State deaths\end{tabular} & \begin{tabular}[c]{@{}c@{}}Underestimated\\ US deaths\end{tabular} & \begin{tabular}[c]{@{}c@{}}Overestimated\\ US deaths\end{tabular} \\ \hline
 & & & & & & \\
Death Tx & 11.18** & 1.082 & -0.0735* & -0.0220 & -0.0599 & 0.0456 \\
 & (4.868) & (2.891) & (0.0409) & (0.0432) & (0.0556) & (0.0502) \\
Death Tx* Democrat & -3.111 & -4.299 & -0.0389 & -0.00755 & 0.0628 & -0.144** \\
 & (8.647) & (6.064) & (0.0560) & (0.0590) & (0.0737) & (0.0696) \\
Death Tx* Republican & 18.29 & 6.951 & -0.0909 & -0.0613 & -0.00450 & -0.123 \\
 & (12.79) & (9.582) & (0.0629) & (0.0624) & (0.0830) & (0.0829) \\
Democrat & 12.85* & 5.528 & -0.0678 & -0.00681 & -0.0824 & 0.137*** \\
 & (7.111) & (5.422) & (0.0475) & (0.0447) & (0.0554) & (0.0514) \\
Republican & 3.692 & 3.696 & 0.125*** & -0.0340 & 0.0739 & 0.0715 \\
 & (4.804) & (3.912) & (0.0479) & (0.0429) & (0.0657) & (0.0715) \\
Lost 20\% income & 4.019 & -0.733 & 0.0311 & 0.00278 & 0.0215 & -0.0466 \\
 & (5.878) & (3.299) & (0.0305) & (0.0364) & (0.0352) & (0.0405) \\
Knows hospitalized & -7.397 & 5.118 & 0.0862** & -0.0246 & 0.00202 & 0.0661 \\
 & (5.560) & (6.049) & (0.0395) & (0.0389) & (0.0447) & (0.0552) \\
ln COVID-19 cases & -0.788 & -2.379 & 0.0299** & -0.0249* & 0.0131 & -0.0284 \\
 & (2.284) & (1.894) & (0.0138) & (0.0140) & (0.0141) & (0.0216) \\
Consumer exp - June & 23.23 & 20.93 & -0.348** & -0.00105 & 0.197* & -0.218 \\
 & (26.42) & (15.10) & (0.140) & (0.152) & (0.116) & (0.204) \\
Democratic leaning news & -1.431 & -2.691 & 0.0217 & -0.0493 & -0.0996** & -0.0180 \\
 & (6.032) & (4.575) & (0.0307) & (0.0349) & (0.0409) & (0.0315) \\
Republican leaning news & -6.467 & -8.875 & -0.0405 & 0.00633 & 0.00953 & -0.0233 \\
 & (7.452) & (5.684) & (0.0380) & (0.0550) & (0.0551) & (0.0448) \\
Constant & 53.68** & 56.68*** & 0.106 & 0.419*** & 0.350** & 0.486*** \\
 & (23.07) & (16.73) & (0.141) & (0.142) & (0.143) & (0.165) \\
 & & & & & & \\
Observations & 1,128 & 1,128 & 1,141 & 1,141 & 1,141 & 1,141 \\
R-squared & 0.113 & 0.088 & 0.192 & 0.136 & 0.157 & 0.086 \\
Controls & Yes & Yes & Yes & Yes & Yes & Yes \\
Mean dep. var. & 34.13 & 26.27 & 0.237 & 0.244 & 0.251 & 0.262 \\ \hline
\multicolumn{7}{l}{%
 \begin{minipage}{1.33\textwidth}%
  \small \textit{Notes}: Standard errors in parentheses. *** p\textless{}0.01, ** p\textless{}0.05, * p\textless{}0.1. \\ Column 1 reports an OLS regression, with the time taken to respond to the two questions on the number of deaths as the dependent variable.
Col 2-5 contain the results of two multinomial logistic regressions. The dependent variables are categorical variables reporting whether the respondent under- or over-estimated the number of deaths at the State or the US level. The excluded category is the correct estimation. All regressions take into account population survey weights and the sampling procedure. The dependent variable is a dummy=1 if the respondent increased their belief that it should be a government's responsibility to provide the following policies. The control variables include: gender, race, age, education, parental status, caring responsibilities for an elderly or a person with a disability, baseline income in February 2020, cohabitation with a partner, labor force participation and employment status in February 2020, health insurance provider, if the respondent had financial difficulties before the pandemic, macro-region, metro vs. rural, the population density at the zip code, and two dummy variables indicating if they consume at least 30min a week of international news and if they have consume news from social media. We also control for whether respondents completed the survey in a shorter time than the 99$^{th}$ percentile as well as ceiling effects.
 \end{minipage}%
}
\end{tabular}%
}
\end{table}

In the following table, we report the results of an IV regression. We instrument having correctly estimated the number of casualties both at the State and the federal level (i.e. a more stringent outcome) with our treatment, and we study whether the number of deaths affects respondents' judgement, controlling for a set of demographic characteristics, media consumption, and exposure to the virus:
 $$ Pr(Success_{ic}) = \alpha + \beta Shock_{i} + \gamma Shock_{c} + \theta_1 Rep_i + \theta_2 Dem_i + $$ 
 $$ \phi \left(Correct deaths_i = Treat_i \right) + \delta X_{ic} + \epsilon_{ic} $$

We see that our instrument (i.e. assignment to the treatment) is strong when considering both a traditional and a more stringer F-test threshold \citep{stock2002survey, lee2020valid}\footnote{\cite{lee2020valid} suggest that when maintaining the F-statistics of 10, the t-test associated with the instrumental variable should be above 3.43, which is the case in our experiment}.

\begin{table}[H]
\centering
\caption{IV regression}
\label{tab:covid_iv}
\resizebox{0.7\textwidth}{!}{%
\begin{tabular}{lcccc}
 & (1) & (2) & (3) & (4) \\ \hline \hline
& \begin{tabular}[c]{@{}l@{}}First stage\\ Correct \\ death rate\end{tabular} & \begin{tabular}[c]{@{}l@{}}Second stage\\ Death number\\ is a success\end{tabular} & \begin{tabular}[c]{@{}l@{}}First stage\\ Correct \\ death rate\end{tabular} & \begin{tabular}[c]{@{}l@{}}Second stage\\ Death number\\ is a success\end{tabular} \\ \hline
 & & & & \\
Death Tx & 0.118*** & & 0.149*** & \\
 & (0.0303) & & (0.0338) & \\
Correct deaths & & -0.347 & & 0.114 \\
 & & (0.296) & & (0.340) \\
Correct deaths* Rep news & & & & -1.102 \\
 & & & & (0.915) \\
Correct deaths* Dem news & & & & -0.290 \\
 & & & & (0.926) \\
Death Tx*Rep leaning news & & & -0.0386 & \\
 & & & (0.0644) & \\
Death Tx*Dem leaning news & & & -0.0906 & \\
 & & & (0.0625) & \\
Democrat & 0.0617 & -0.107** & 0.0599 & -0.133*** \\
 & (0.0398) & (0.0417) & (0.0400) & (0.0365) \\
Republican & -0.0364 & 0.216*** & -0.0366 & 0.181*** \\
 & (0.0364) & (0.0448) & (0.0360) & (0.0572) \\
Lost 20\% income & -0.0309 & -0.0229 & -0.0315 & 0.00609 \\
 & (0.0394) & (0.0498) & (0.0398) & (0.0657) \\
Knows hospitalized & -0.0736* & -0.0407 & -0.0752* & -0.0580 \\
 & (0.0423) & (0.0468) & (0.0418) & (0.0619) \\
ln COVID-19 cases & -0.0179 & -0.0111 & -0.0188 & -0.00546 \\
 & (0.0178) & (0.0230) & (0.0179) & (0.0317) \\
Consumer exp - June & 0.161 & -0.140 & 0.155 & -0.150 \\
 & (0.122) & (0.142) & (0.122) & (0.218) \\
Dem leaning news & 0.0180 & -0.0366 & 0.0628 & 0.0568 \\
 & (0.0369) & (0.0423) & (0.0478) & (0.352) \\
Rep leaning news & -0.0246 & 0.259*** & -0.00598 & 0.616** \\
 & (0.0511) & (0.0535) & (0.0563) & (0.281) \\
Constant & 0.306** & 0.502** & 0.303** & 0.224 \\
 & (0.139) & (0.202) & (0.139) & (0.240) \\
 & & & & \\
Observations & 1,146 & 1,142 & 1,146 & 977 \\
R-squared & 0.159 & 0.167 & 0.160 & 0.032 \\
Controls & Yes & Yes & Yes & Yes \\
F test model & 21.28 & 30.36 & 23.38 & 19.54 \\
Death Tx t-test & 3.900 & & 4.417 & \\
Mean dep. var. & 0.330 & 0.335 & & \\
Death Tx*Dem t-test & & & -0.599 & \\
Death Tx*Rep t-test & & & -1.450 & \\ \hline
\multicolumn{5}{l}{%
 \begin{minipage}{1\columnwidth}%
  \small \textit{Notes}: Standard errors in parentheses. *** p\textless{}0.01, ** p\textless{}0.05, * p\textless{}0.1.\\
  Col. (1) and (3) report the first stage results, col. (2) and (4), the second stage. In the first stage, the dependent variable is a dummy=1 if the respondent correctly estimated the deaths both at the national and federal level, while in the second stage, it's a dummy=1 if they deem such figures a success. In the second stage, we instrument ``correctly estimating the death rates'' with the treatment status. The control variables include: gender, race, age, education, parental status, caring responsibilities for an elderly or a person with a disability, baseline income in February 2020, cohabitation with a partner, labor force participation and employment status in February 2020, health insurance provider, if the respondent had financial difficulties before the pandemic, macro-region, metro vs. rural, and the population density at the zip code. We also control for whether respondents completed the survey in a shorter time than the 99$^{th}$ percentile.
 \end{minipage}%
}\\ 
\end{tabular}%
}
\end{table}

\begin{table}[H]
\centering
\caption{Death experiment controlling for whether the death rate predicted in wave 5 was considered a success.}
\label{tab:death_exp2}
\resizebox{0.5\textwidth}{!}{%
\begin{tabular}{lcc} \hline \hline
 & (1) & (2) \\
 & First stage & Second stage \\
& \begin{tabular}[c]{@{}c@{}}Correct \\ death rate\end{tabular} & \begin{tabular}[c]{@{}c@{}}Death rate \\ is a success\end{tabular} \\ \hline
 & & \\
Treated & 0.120*** & \\
 & (0.0302) & \\
\multirow{2}{*}{\begin{tabular}[c]{@{}l@{}}Expected death rate \\ in May is a success\end{tabular}} & -0.0538 & 0.369*** \\
 & (0.0363) & (0.0446) \\
Correct deaths & & -0.472 \\
 & & (0.315) \\
Democrat & 0.0487 & -0.0290 \\
 & (0.0390) & (0.0415) \\
Republican & -0.0251 & 0.130*** \\
 & (0.0378) & (0.0449) \\
Lost \textgreater{}20\% income & -0.0347 & -0.00874 \\
 & (0.0383) & (0.0502) \\
Knows hospitalized & -0.0720* & -0.0529 \\
 & (0.0412) & (0.0496) \\
log county cases & -0.0166 & -0.0215 \\
 & (0.0178) & (0.0240) \\
Var consumer spending & 0.155 & -0.0871 \\
 & (0.122) & (0.150) \\
Dem leaning news & 0.0163 & -0.0138 \\
 & (0.0371) & (0.0429) \\
Rep leaning news & -0.0158 & 0.179*** \\
 & (0.0512) & (0.0586) \\
International news & -0.0324 & -0.0182 \\
 & (0.0399) & (0.0359) \\
Constant & 0.343** & 0.318 \\
 & (0.141) & (0.214) \\
 & & \\
 Controls & Yes & Yes \\
Observations & 1,146 & 1,142 \\
R-squared & 0.160 & 0.166 \\
F test model & 22.05 & 47.60 \\
Treatment t-test & 3.991 & \\ \hline
\multicolumn{3}{l}{%
 \begin{minipage}{0.7\columnwidth}%
  \small \textit{Notes}: Standard errors in parentheses. *** p\textless{}0.01, ** p\textless{}0.05, * p\textless{}0.1.\\
  Col. (1) reports the first stage results, col. (2) the second stage. In the first stage, the dependent variable is a dummy=1 if the respondent correctly estimated the deaths both at the national and federal level, while in the second stage, it's a dummy=1 if they deem such figures a success. In the second stage, we instrument ``correctly estimating the death rates'' with the treatment status. The control variables include: gender, race, age, education, parental status, caring responsibilities for an elderly or a person with a disability, baseline income in February 2020, cohabitation with a partner, labor force participation and employment status in February 2020, health insurance provider, if the respondent had financial difficulties before the pandemic, macro-region, metro vs. rural, and the population density at the zip code. We also control for whether respondents completed the survey in a shorter time than the 99$^{th}$ percentile. Finally, we consider whether respondents use social media and they consumer international news.
 \end{minipage}%
}\\ \end{tabular}
}
\end{table}

\begin{figure}[H]
	\caption{The long-term effect of information treatment on beliefs.}	\label{f:death_rate_pol}
	\begin{center}
		\includegraphics[height=11cm]{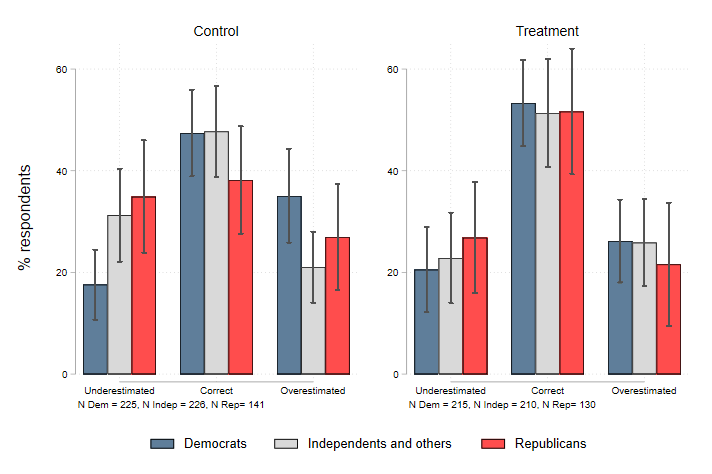}
	\end{center}
\begin{minipage}{1\linewidth \setstretch{0.75}}
	{\scriptsize{\textit{Notes}:} 
		\scriptsize The figure shows the share of respondents who correctly estimated the death rate of the U.S. compared to the rest of the world by political party and by whether they were in the treatment group in the previous similar question we asked more than 3 months earlier. Error bars show 95\% confidence intervals.}
\end{minipage}	
\end{figure}

\begin{table}[H]
\centering
\caption{Comparing the US coronavirus death rate per capita with the rest of the world.}
\label{tab:us_compare}
\resizebox{\textwidth}{!}{%
\begin{tabular}{lcccccc} \hline \hline
 & (1) & (2) & (3) & (4) & (5) & (6) \\
 & \begin{tabular}[c]{@{}c@{}}Underestimated \\ US death rate\\ relative to the \\ rest of the world\end{tabular} & \begin{tabular}[c]{@{}c@{}}Correctly \\ estimated \\ US death rate\\ relative to the \\ rest of the world\end{tabular} & \begin{tabular}[c]{@{}c@{}}Overestimated \\ US death rate\\ relative to the \\ rest of the world\end{tabular} & \begin{tabular}[c]{@{}c@{}}Underestimated \\ US death rate\\ relative to the \\ rest of the world\end{tabular} & \begin{tabular}[c]{@{}c@{}}Correctly \\ estimated \\ US death rate\\ relative to the \\ rest of the world\end{tabular} & \begin{tabular}[c]{@{}c@{}}Overestimated \\ US death rate\\ relative to the \\ rest of the world\end{tabular} \\ \hline
 & & & & & & \\
Republican & 0.0960** & -0.0492 & -0.0469 & 0.0976** & -0.0550 & -0.0426 \\
 & (0.0469) & (0.0442) & (0.0297) & (0.0433) & (0.0419) & (0.0311) \\
Democrat & -0.0663** & -0.0972*** & 0.163*** & -0.0664** & -0.101*** & 0.167*** \\
 & (0.0291) & (0.0348) & (0.0323) & (0.0289) & (0.0362) & (0.0326) \\
Death Tx & & & & -0.0709** & 0.0151 & 0.0558 \\
 & & & & (0.0338) & (0.0440) & (0.0361) \\
Death Tx*Dem news & & & & 0.0733* & -0.0791 & 0.00577 \\
 & & & & (0.0372) & (0.0714) & (0.0674) \\
Death Tx*Rep news & & & & -0.139* & 0.250*** & -0.111** \\
 & & & & (0.0810) & (0.0858) & (0.0522) \\
Lost 20\% income & -0.00924 & 0.0110 & -0.00173 & -0.00951 & 0.0130 & -0.00353 \\
 & (0.0313) & (0.0403) & (0.0318) & (0.0310) & (0.0390) & (0.0319) \\
Knows hospitalized & -0.0214 & 0.0326 & -0.0111 & -0.0224 & 0.0330 & -0.0105 \\
 & (0.0361) & (0.0378) & (0.0319) & (0.0364) & (0.0377) & (0.0313) \\
Consumer exp - Oct & 0.136 & -0.148 & 0.0120 & 0.0928 & -0.101 & 0.00770 \\
 & (0.0869) & (0.108) & (0.0832) & (0.0892) & (0.105) & (0.0851) \\
ln COVID-19 cases & 0.0507* & -0.0213 & -0.0294 & 0.0538* & -0.0228 & -0.0310 \\
 & (0.0296) & (0.0481) & (0.0428) & (0.0292) & (0.0484) & (0.0426) \\
Dem leaning news & -0.0182 & -0.00776 & 0.0260 & -0.0541 & 0.0351 & 0.0190 \\
 & (0.0289) & (0.0447) & (0.0503) & (0.0381) & (0.0545) & (0.0556) \\
Rep leaning news & 0.253*** & -0.191*** & -0.0626 & 0.311*** & -0.295*** & -0.0166 \\
 & (0.0618) & (0.0640) & (0.0416) & (0.0681) & (0.0745) & (0.0470) \\
Constant & -0.271 & 1.128*** & 0.143 & -0.287 & 1.151*** & 0.136 \\
 & (0.264) & (0.390) & (0.330) & (0.262) & (0.397) & (0.328) \\
 & & & & & & \\
Observations & 1,061 & 1,061 & 1,061 & 1,061 & 1,061 & 1,061 \\
R-squared & 0.223 & 0.080 & 0.126 & 0.239 & 0.097 & 0.132 \\
Mean dep. var. & 0.202 & 0.550 & 0.249 & 0.202 & 0.550 & 0.249\\ \hline
\multicolumn{7}{l}{%
 \begin{minipage}{1.45\columnwidth}%
  \small \textit{Notes}: Standard errors in parentheses. *** p\textless{}0.01, ** p\textless{}0.05, * p\textless{}0.1.\\
  All regressions are OLS regressions that take into account population survey wights and the sampling procedure. The dependent variable is a dummy=1 if the respondent stated that they believed the U.S. COVID-19 death rate was the among the lowest or the lowest (Underestimated) - col. (1) and (4), higher than most countries (Correct) - col. (2) and (5), or the highest in the world (Overestimated) - col. (3) and (6). The control variables include: gender, race, age, education, parental status, caring responsibilities for an elderly or a person with a disability, baseline income in February 2020, cohabitation with a partner, labor force participation and employment status in February 2020, health insurance provider, if the respondent had financial difficulties before the pandemic, macro-region, metro vs. rural, and the population density at the zip code. We also control for whether respondents completed the survey in a shorter time than the 99$^{th}$ percentile.
 \end{minipage}%
}\\                  
\end{tabular}%
}
\end{table}

\subsubsection{Causal forest estimation}

\begin{table}[H]
\centering
\caption{Best linear fit using forest predictions for heterogeneity calibration test.}
\label{tab:bestlinearfit}
\resizebox{0.6\textwidth}{!}{%
\begin{tabular}{lccc} \hline \hline
 & (1) & (2) & (3) \\
 & \begin{tabular}[c]{@{}c@{}}Correctly \\ estimated\\ US and\\ State deaths\end{tabular} &
  \begin{tabular}[c]{@{}c@{}}US and\\State deaths\\ are a\\ success\end{tabular} & 
  \begin{tabular}[c]{@{}c@{}}Correctly\\ stated\\ US deaths\\vs. the world\end{tabular} \\
 \hline
 & & & \\
 Mean forest prediction & 0.9907*** & 1.3815 & 0.3251 \\
 & (0.1988) & (1.8756) & (16.0403) \\
 Differential forest prediction & 0.94199*** & -2.5150 & -3.4927 \\
 & (0.29131) & (1.2971) & (1.9099) \\
 & & & \\
 \hline
\multicolumn{4}{l}{%
 \begin{minipage}{0.75\columnwidth}%
  \small \textit{Notes}: Standard errors in parentheses. *** p\textless{}0.01, ** p\textless{}0.05, * p\textless{}0.1.\\
  Best linear fit mean and different forest predictions are run on the final model, which excludes potentially confounding or otherwise non-essential variables through a pre-fitting step identifying covariates with relatively high importance to the model. Pre-fitting occurs independently for each outcome, meaning each model includes a unique set of 7 to 9 predictors. Model (1) includes covariates indicating: county log COVID cases, county consumer expenditure, information on hours of international media consumed, education level, income, financial hardship pre-COVID, any care responsibilities, whether respondent has public health insurance, and ZCTA population density.
 \end{minipage}}                  
\end{tabular}}
\end{table}

\begin{figure}[H]
	\caption{Causal forest out-of-bad predictions vs. actual treatment effect by quintile}
	\label{fig:causalforesttest}
	\begin{center}
		\includegraphics[height=8cm]{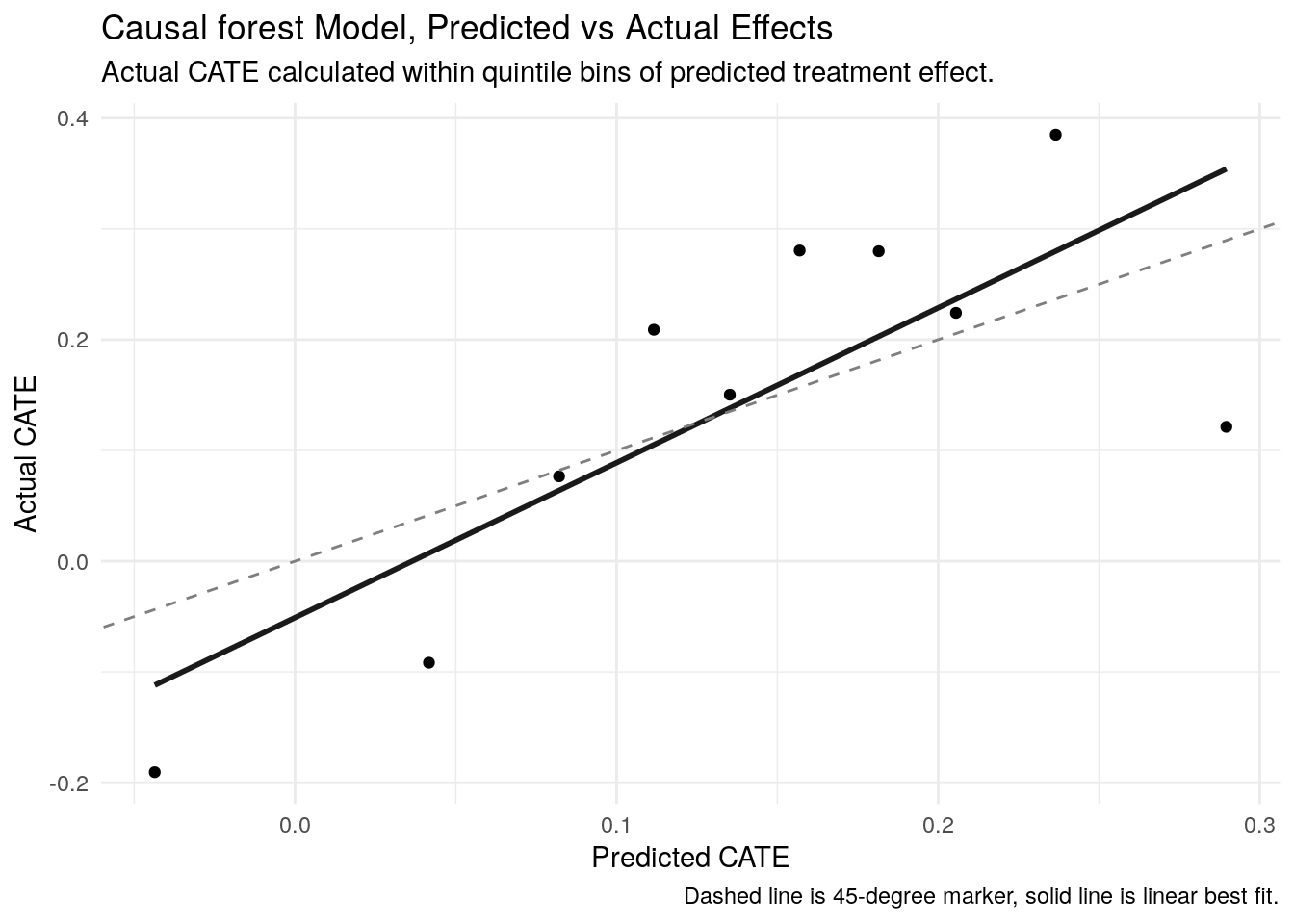}
	\end{center}
\begin{minipage}{1\linewidth \setstretch{0.75}}
	{\scriptsize{\textit{Notes}:} 
		\scriptsize This figure summarizes tests suggested by Davis \& Heller (2020) to verify a relation between heterogeneous predictions and actual CATE.}
\end{minipage}	
\end{figure}

\begin{table}[H]
\centering
\caption{Covariate means by CATE quartile, obtained from casual forest on whether respondent correctly stated US and State death rates following wave 6 information treatment.}
\label{tab:causalforestquartiles}
\resizebox{0.75\textwidth}{!}{%
\begin{tabular}{lcccc} \hline \hline
 & (1) & (2) & (3) & (4) \\
 & Q1 & Q2 & Q3 & Q4 \\
 \hline
 & & & \\
 Mean CATE & 0.014 & 0.117 & 0.175 & 0.253 \\
 & & & \\
 Incr COVID-19 cases & 6.056 & 6.017 & 6.141 & 6.309 \\
 Var consumer expenditures & -0.081 & -0.114 & -0.119 & -0.093 \\
 Hours international news & 0.202 & 0.234 & 0.226 & 0.247 \\
 Hours int. news missing & 0.108 & 0.091 & 0.080 & 0.056 \\
 High school degree & 0.178 & 0.136 & 0.188 & 0.087 \\
 Bachelor's degree or higher & 0.383 & 0.385 & 0.341 & 0.603 \\
 Income over \$62k & 0.258 & 0.196 & 0.157 & 0.226 \\
 Finances weak pre- COVID-19 & 0.226 & 0.269 & 0.303 & 0.282 \\
 Caring responsibilities & 0.139 & 0.175 & 0.195 & 0.125 \\
 Population density (ZCTA) & 2,937.1 & 4,152.0 & 4,913.2 & 3,416.8 \\
 & & & \\
 Republican & 0.226 & 0.231 & 0.258 & 0.230 \\
 Democrat & 0.355 & 0.388 & 0.369 & 0.422 \\
 Independent & 0.418 & 0.381 & 0.373 & 0.348 \\
 Lost 20\% income & 0.230 & 0.213 & 0.223 & 0.254 \\
 Lost 20\% income missing & 0.174 & 0.150 & 0.185 & 0.143 \\
 Knows hospitalized & 0.136 & 0.161 & 0.132 & 0.178 \\
 Knows hospitalized missing & 0.178 & 0.150 & 0.188 & 0.146 \\
 Fast Response & 0.021 & 0.052 & 0.042 & 0.024 \\
 Democrat leaning news & 0.230 & 0.269 & 0.216 & 0.314 \\
 Democrat leaning news missing & 0.125 & 0.178 & 0.167 & 0.111 \\
 Republican leaning news & 0.206 & 0.196 & 0.192 & 0.195 \\
 Has social media & 0.397 & 0.402 & 0.411 & 0.359 \\
 Has social media missing & 0.000 & 0.003 & 0.000 & 0.010 \\
 Female & 0.491 & 0.465 & 0.523 & 0.429 \\
 Some college education & 0.411 & 0.448 & 0.446 & 0.289 \\
 Age 18-29 & 0.195 & 0.248 & 0.220 & 0.195 \\
 Age 30-44 & 0.279 & 0.280 & 0.265 & 0.314 \\
 Age 45-59 & 0.213 & 0.210 & 0.237 & 0.230 \\
 Black/African American & 0.087 & 0.084 & 0.105 & 0.115 \\
 Hispanic/Latinx & 0.143 & 0.178 & 0.185 & 0.101 \\
 Other race & 0.091 & 0.136 & 0.111 & 0.139 \\
 Cohabiting partner & 0.627 & 0.605 & 0.564 & 0.659 \\
 Parent (under 18) & 0.244 & 0.287 & 0.289 & 0.293 \\
 Not always working & 0.261 & 0.227 & 0.000 & 0.226 \\
 Unemployed pre-COVID-19 & 0.028 & 0.073 & 0.077 & 0.049 \\
 Midwest & 0.296 & 0.266 & 0.223 & 0.251 \\
 South & 0.328 & 0.339 & 0.397 & 0.355 \\
 West & 0.209 & 0.283 & 0.226 & 0.230 \\
 In metropolitan area & 0.808 & 0.871 & 0.875 & 0.923 \\
 No insurance & 0.073 & 0.073 & 0.101 & 0.056 \\
 & & & \\
 \hline
\multicolumn{5}{l}{%
 \begin{minipage}{0.80\columnwidth}%
  \small \textit{Notes}: The first row presents the mean conditional average treatment effect estimate for the quartile in question. The first set of covariates are those determined to have high importance to the model through the pre-fit step, and as such were part of the final causal forest prediction; all other covariates are listed beneath this set.
 \end{minipage}%
}\\                  
\end{tabular}%
}
\end{table}

\section{Robustness checks} \label{robust}

\textbf{Alternative measures of shocks}. In the results presented in the main section of the paper, we considered as a direct economic shock whether respondents lost at least 20\% of their household income between any two months between the baseline survey wave and the last survey wave. We replicate the same model specifications using two different assumptions of direct economic shock: (a) whether respondents lost at least 20\% of their household income between February and October - that is, they incurred a more permanent loss in income, thus excluding those who eventually recovered from their loss by our last wave, and (b) the percentage decrease in income between the baseline and the outcome month, to account for possible different magnitudes of the level of shock. The two measures are, respectively:

$$shock_2 = \begin{cases} 1, & \mbox{if } \frac{income_{final}-income_{baseline}}{income_{baseline}} \leq -0.20 \\ 0, & \mbox{otherwise } \end{cases} $$

and

 $$shock_3 = \begin{cases} \frac{income_{final}-income_{baseline}}{income_{baseline}}, & \mbox{if } < 0 \\ 0, & \mbox{otherwise} \end{cases} $$ \\

Among respondents in our sample who participated in the first and the last survey waves (i.e., \textit{n}=1,076), about 27\% of our respondents lost at least 20\% of their income in a permanent way between February and October, compared to 38\% who lost it between any two months but potentially recovered. When looking at the continuous measure of shock, we find that between February and October, about 4\% of respondents reported having lost all of their household income, while about 17\% lost up to half of their household income. In tables \ref{tab:policy_inc2}, \ref{tab:policy_inc3} and \ref{tab:covid_policy_inc23} in the Online Appendix, we report the results of the regressions on policy preferences and trust in institutions using these two alternative measures of shocks. The magnitude and the coefficient signs are consistent with our main specification: direct income shocks increased support for most government interventions, with the exclusion of providing mental healthcare and universal healthcare, whose associated coefficients are not significant, and help the industry grow. Support for the latter significantly decreased among respondents who incurred a shock, regardless of how it was measured, in line with our main results. Regarding temporary relief policies, we witness even stronger support, both in terms of outcomes and significance, among respondents who incurred an income shock and had not recovered by October, suggesting that support for welfare policies increased with the severity of a person's income loss. We also report results related to institutional trust in Tables \ref{tab:trust_inc2} and \ref{tab:trust_inc3}. Also in this case, the coefficients are consistent with our main specification: incurring an economic shock is associated with an increase in the likelihood of having lost confidence in institutions, particularly so in the U.S. Congress and Senate and in the private sector. \\

\textbf{Alternative measures of outcomes and regression models}. We focused on analyzing an increase in support for policies and government interventions and a decrease in institutional trust. However, we also considered the opposite direction - that is, a decrease in support for welfare and an increase in institutional trust. We report these results in Tables \ref{tab:policy_inc2}, \ref{tab:policy_inc3} \ref{tab:covid_policy_inc23}, \ref{tab:trust_inc2}, and \ref{tab:trust_inc3} in the Online Appendix and show that they are in line with what presented above: Democrats are significantly less likely to have decreased their support for most of the government interventions, while Republicans are more likely to do so, and the biased media diet further increased this trend. On the other side, Democrats are less likely to have increased their trust in President Trump and in the U.S. Congress and Senate but have significantly increased their confidence in people running the scientific community, whereas the opposite is true for respondents supporting the Republican party.

\par Lastly, since most of our outcomes are binary variables, for completeness, we also show that our results hold when using a logistic regression instead of OLS, as shown in the Online Appendix, in Tables \ref{tab:policy_logit}, \ref{tab:covid_policy_logit}, and \ref{tab:trust_logit}. \\

\textbf{Average effect sizes.} Another robustness check we perform is testing whether our results hold when considered as a bundle, which allows for making more general claims. To do so, we replicate the analyses using Average Effect Sizes, as in \cite{kling2004moving, clingingsmith2009estimating, heller2017thinking}. To perform such an analysis, one needs to make several assumptions about the nature of the outcomes being studied since an AES estimation requires stacking multiple outcomes. As we have seen in the main specifications of our results, support for policies and trust in institutions change in different directions according to a person's political beliefs and depending on the nature of the shock they incurred. As such, this requires grouping dependent variables into sub-groups using a more subjective judgment. In the Appendix, we propose one plausible stacking approach and show that the results remain qualitatively similar to those presented in the previous sections. We group the variables according to the type of institutions or policies considered. When analyzing policies, we separate between questions related to whether it's a government's responsibility to provide a set of services and those concerning coronavirus relief. Within the first ones, we further split the variables into two groups: one considering traditional macroeconomic policies (keep prices under control and help the industry grow), and one focused on welfare issues (reduce inequality, provide for the unemployed, provide help to university students from a disadvantaged background, and provide a basic income, universal healthcare, provide mental health care services to people with mental illnesses, provide for the elderly and help those affected by natural disasters). For what concerns institutional trust, we separate between government-related institutions (the U.S. Congress and Senate and the White House), science-related ones (scientific community, hospitals and health care professionals, and health insurance companies), and the ones related to the economy (banks and financial institutions, and the private sector). Again, we see that our results remain qualitatively identical to the main specifications presented in the body of the paper.

In order to assess the overall impact of such shocks on preferences, we also compute the Average Effect Sizes (AES), following several other authors \citep{kling2004moving, clingingsmith2009estimating, heller2017thinking}.

Let $\beta_k$ indicate the estimated shock $s$ coefficient for the outcome variable $k$, and let $\sigma_k$ denote the standard deviation of such coefficient. The AES for shock $s$ across all $K$ outcomes is equal to: 
$$ \frac{1}{K} \sum_{k=1}^K \frac{\beta_k}{\sigma_k} $$

In order to calculate the AES standard errors, the regressions are estimated simultaneously in a Seemingly Unrelated Regression (SUR) framework. We stack the K outcomes and use our shock $s$ effects regression fully interacted with dummy variables for each outcome as the right-hand side. The coefficients $\beta_k$ are the same as those estimated in the outcome-by-outcome regressions, but the stacked regression provides the correct covariance matrix to form a test of significance for the AES. We compute our estimates, following \cite{clingingsmith2009estimating}. \par

Further, we group institutions and policies in sub-groups, according to their topics or their area of expertise, to reduce the heterogeneity. For what concerns the institutional trust, we group: health-related institutions (health insurance companies, hospitals and healthcare professionals, and the scientific community); political institutions (President Trump and the U.S. Congress and Senate); economic institutions (banks and financial institutions, and the private sector). With regard to the support for government interventions, we form the following sets: Welfare policies (Support for universal healthcare, Support unemployed, Proved a basic income, Reduce inequality, Help those affected by natural disasters, Provide mental healthcare, and Provide for the elderly); Economic interventions (Help the industry grow, and Keep prices under control); COVID-19 response policies (w4-w7 / protect essential workers, transfer money to families and businesses, and increase spending on public health to reduce deaths). 

We further split policies according to the type of welfare in wealth-related policies (Support unemployed, Proved a basic income, Reduce inequality) and health-related ones (Support for universal and Provide mental healthcare), and institutions according to the area of interest in health-related institutions (Scientific community, Hospitals and healthcare professionals, and Health insurance companies) and economic/financial ones (Banks and financial institutions, and Private sector).

\begin{table}[H]
\centering
\caption{AES - Government's responsibilities}
\label{tab:AES_policy}
\resizebox{\textwidth}{!}{%
\begin{tabular}{lcccccc} \hline \hline
 & \multicolumn{6}{c}{Increase in belief that it's a government's responsibility to provide...} \\
 & (1) & (2) & (3) & (4) & (5) & (6) \\
 & All policies & All policies & \begin{tabular}[c]{@{}c@{}}Welfare \\ policies\end{tabular} & \begin{tabular}[c]{@{}c@{}}Welfare \\ policies\end{tabular} & \begin{tabular}[c]{@{}c@{}}Economic\\ interventions\end{tabular} & \begin{tabular}[c]{@{}c@{}}Economic\\ interventions\end{tabular} \\ \hline
 & & & & & & \\
Lost 20\% income & 0.00994 & & 0.0253 & & -0.0515 & \\
 & (0.0298) & & (0.0323) & & (0.0480) & \\
Knows hospitalized & & 0.00980 & & -0.00927 & & 0.0861 \\
 & & (0.0314) & & (0.0340) & & (0.0533) \\
 & & & & & & \\
 Controls & Yes & Yes & Yes & Yes & Yes & Yes \\
Observations & 971 & 971 & 971 & 971 & 971 & 971\\ \hline
\multicolumn{7}{l}{%
 \begin{minipage}{1.05\columnwidth}%
  \small \textit{Notes}: Standard errors in parentheses. *** p\textless{}0.01, ** p\textless{}0.05, * p\textless{}0.1.\\
  The dependent variable is a dummy=1 if the respondent increased their belief that it should be the government's responsibility to provide a set of policies. The shock coefficients are then combined according to the type of intervention. Welfare policies include support for the unemployed, basic income, decrease inequality, provide for the elderly, support financially university students from poor households, provide mental health services to those affected by mental health diseases, provide support to those affected by natural disasters and support for universal healthcare, while economic interventions refer to keep prices under control and help industry grow. The control variables include: gender, race, age, education, parental status, and caring responsibilities for an elderly or a person with a disability, income in February 2020, housing, labor force participation and employment status in February 2020, health insurance provider, and whether respondents had financial difficulties before the pandemic, the area in which the respondents live, whether it's a metropolitan or rural area, and the population density in the zip code. We also control for whether respondents completed the related surveys in a shorter time than the 99$^{th}$ percentile, and ceiling effects, along with indirect economic and health-related shocks. Finally, we also include whether the sources of news consulted lean politically, the amount of international news consumed and social media usage.
 \end{minipage}%
}\\ 
\end{tabular}%
}
\end{table}

\begin{table}[H]
\centering
\caption{AES - COVID-19 relief policies}
\label{tab:AES_covid_policy}
\resizebox{0.8\textwidth}{!}{%
\begin{tabular}{lcc} \hline \hline
 & (1) & (2) \\
 & \multicolumn{2}{l}{Increased support for COVID-19 relief policies} \\ \hline
 & &  \\
Lost 20\% income & 0.0350 &  \\
 & (0.0521) & \\
Know hospitalized & & -0.0161 \\
 & & (0.0527) \\
 & &  \\ \hline
 Controls & Yes & Yes\\
Observations & 935 & 933 \\ \hline \\
\multicolumn{3}{l}{%
 \begin{minipage}{0.8\columnwidth}%
  \small \textit{Notes}: Standard errors in parentheses. *** p\textless{}0.01, ** p\textless{}0.05, * p\textless{}0.1.\\
  The dependent variable is a dummy=1 if the respondent increased their support for coronavirus relief policies: invest more in healthcare to reduce preventable deaths, protect essential workers and provide financial support to families and businesses. The shock coefficients are then combined. The control variables include: gender, race, age, education, parental status, and caring responsibilities for an elderly or a person with a disability, income in February 2020, housing, labor force participation and employment status in February 2020, health insurance provider, and whether respondents had financial difficulties before the pandemic, the area in which the respondents live, whether it's a metropolitan or rural area, and the population density in the zip code. We also control for whether respondents completed the related surveys in a shorter time than the 99$^{th}$ percentile, and ceiling effects, along with indirect economic and health-related shocks. Finally, we also include whether the sources of news consulted lean politically, the amount of international news consumed and social media usage.
 \end{minipage}%
}\\ 
\end{tabular}%
}
\end{table}

\begin{table}[H]
\centering
\caption{AES - Institutional trust}
\label{tab:AES_trust}
\resizebox{\textwidth}{!}{%
\begin{tabular}{lcccccccc} \hline \hline
& \multicolumn{8}{c}{Decrease in trust in people running the following institutions...} \\
 & (1) & (2) & (3) & (4) & (5) & (6) & (7) & (8) \\
 \\
 & \begin{tabular}[c]{@{}c@{}}All\\ institutions\end{tabular} & \begin{tabular}[c]{@{}c@{}}All\\ institutions\end{tabular} & \begin{tabular}[c]{@{}c@{}}Economic\\ institutions\end{tabular} & \begin{tabular}[c]{@{}c@{}}Economic\\ institutions\end{tabular} & \begin{tabular}[c]{@{}c@{}}Scientific\\ institutions\end{tabular} & \begin{tabular}[c]{@{}c@{}}Scientific\\ institutions\end{tabular} & \begin{tabular}[c]{@{}c@{}}Government\\ institutions\end{tabular} & \begin{tabular}[c]{@{}c@{}}Government\\ institutions\end{tabular} \\ \hline
 & & & & & & & & \\
Lost 20\% income & 0.0603* & & 0.108** & & 0.0699 & & 0.0559 & \\
 & (0.0353) & & (0.0544) & & (0.0528) & & (0.0540) & \\
Knows hospitalized & & 0.0596 & & 0.0322 & & 0.0495 & & 0.0926 \\
 & & (0.0366) & & (0.0551) & & (0.0560) & & (0.0578) \\ 
 & & & & & & & & \\
Controls & Yes & Yes & Yes & Yes & Yes & Yes & Yes & Yes\\
Observations & 984 & 984 & 984 & 984 & 984 & 984 & 984 & 984 \\ \hline
\multicolumn{9}{l}{%
 \begin{minipage}{1.45\columnwidth}%
  \small \textit{Notes}: Standard errors in parentheses. *** p\textless{}0.01, ** p\textless{}0.05, * p\textless{}0.1.\\
  The dependent variable is a dummy=1 if the respondent decreased their trust in an institution. The shock coefficients are then combined according to the type of intervention. Economic institutions include banks and financial institutions and the private sector, whereas Scientific institutions include the scientific community, health insurance companies and hospitals and healthcare professionals. Government institutions are the U.S. Senate and Congress and the White House. The control variables include: gender, race, age, education, parental status, and caring responsibilities for an elderly or a person with a disability, income in February 2020, housing, labor force participation and employment status in February 2020, health insurance provider, and whether respondents had financial difficulties before the pandemic, the area in which the respondents live, whether it's a metropolitan or rural area, and the population density in the zip code. We also control for whether respondents completed the related surveys in a shorter time than the 99$^{th}$ percentile, and ceiling effects, along with indirect economic and health-related shocks. Finally, we also include whether the sources of news consulted lean politically, the amount of international news consumed and social media usage.
 \end{minipage}%
}\\ 
\end{tabular}%
}
\end{table}
\end{spacing}

\textbf{Entropy weights}. The COVID-19 pandemic affected communities and citizens differently, also depending on their income levels. As such, some shocks, such as incurring an income loss, are correlated with several demographic characteristics, including income, gender, and race. Even though we consider variations at the individual level, which reduces concerns related to endogeneity, we cannot entirely exclude that those who have been affected by a shock were systematically different from those who did not, and that their preferences and opinions would have varied in a different way. In order to minimize this potential source of endogeneity, we repeat our analyses with entropy balancing weights. The entropy balancing technique re-weights the observations in order to reduce the differences with respect to a set of balance conditions across treated and control units (in our case, those who incurred a shock vs. those who did not)\footnote{See \cite{hainmueller2013ebalance} for the Stata package and \cite{hainmueller2012entropy} for the theory behind this approach. We opt for applying entropy balancing weights, instead of performing any matching technique, in order to avoid excluding any observation.}. These survey weights still take into account the population weights, so the resulting weights still reflect the whole population. In Tables \ref{tab:policy_entropy}, \ref{tab:covid_policy_entropy}, and \ref{tab:trust_entropy} in the Online Appendix, we report the regression results using entropy balancing weights. Coefficients do not vary in a substantial way with regard to the magnitude and the signs, suggesting that the level of endogeneity is not of particular concern in the interpretation of our results. \\

\textbf{Voting intentions}. The COVID-19 crisis occurred at a time of great political polarization in the U.S., also due to the Presidential elections. The months just before the elections of November 2020 saw greater division among the public, with some voters not necessarily reflecting themselves in one of the two main parties but rather in the Presidential nominees. To account for different political identity effects, we replicate our analysis considering voting intentions, which we collected from our respondents in the middle of May. Results are presented in the Online Appendix, in Tables \ref{tab:policy_vote}, \ref{tab:covid_policy_vote}, and \ref{tab:trust_vote}. Again, the sign and the magnitude of the coefficients associated with the political parties are consistent across specifications. The only marginal differences we note are that Trump voters are significantly less likely to have increased their belief that it's a government responsibility to provide for the unemployed, to provide a basic income, or to reduce inequality, while Republicans, in general, were not. However, Biden voters, unlike Democrats, have not significantly increased their support for coronavirus-related policies or for other government interventions. Yet, such differences are minor, and the coefficient signs are consistent with our main specifications. \\

\textbf{Fixed effects}. We also perform similar analyses to those presented above, but considering a model with longitudinal data and controlling for fixed effects at the individual level.
 $$ y_{ict}= \alpha_i + wave_t + shock_{it} +shock_{ct} + \epsilon_{ict}$$

with $y_{ict}$ being one outcome of interest for individual $i$, in county $c$, in time $t$; $shock_{it}$ and $shock_{ct}$ being a shock for individual $i$ or county $c$, in time $t$; $\alpha_i$ the individual fixed effects, and $wave_t$ the survey wave. Variables referring to direct shocks are dummy variables flagging if the respondent incurred a shock at any time preceding the current wave, so if the event occurred in a certain month, the shock variable will be equal to one for all the subsequent observations. In this way, we track the impact of having had an income loss or knowing someone hospitalized at least once in our time frame, similarly to what was measured in the regression in differences. 

Since the individual effects absorb all time-invariant variables, from the main specification, we cannot assess whether respondents' political views affected their opinions and preferences in time. Thus, we repeat the same analysis but in subgroups, considering a sample of Republicans and one of Democrats. The results of the analysis concerning institutional trust are presented in the Online Appendix, in Tables \ref{tab:fe_policy_1}-\ref{tab:fe_trust_2}. Again, we can see that the results don't change drastically. The fixed effect model allows us to assess how support for government interventions and institutional trust have varied over time. Since the beginning of April, respondents have decreased their belief that the government should keep prices under control, and this seems to be driven by the Republicans, and we observe a similar pattern for two other welfare policies: support for the unemployed and for the elderly. For what concerns trust, the Democrats increased their confidence in the U.S. Senate and Congress between the first and the last week of April, but by mid-May, the level of trust had dropped back to the baseline levels. On the contrary, confidence in President Trump dropped significantly both in May and October, and the coefficients remain negative for both sub-samples of Democrats and Republicans, although they are not significant for the latter ones. Trust in financial institutions and in the private sector has oscillated in time, while confidence in scientific institutions has dropped in time across all parties, reaching the lowest point in June.

\newpage 

\section{Online Appendix}

\subsection{Partisan gap during the crisis} \label{sec: partisan_gap}

\textbf{Partisan gap by media over time}.

\begin{figure}[H]
	\caption{Welfare policy partisan gap, by media consumption, between April and October 2020}
	\label{fig:policies_1}
	\begin{center}
		\includegraphics[height=9cm]{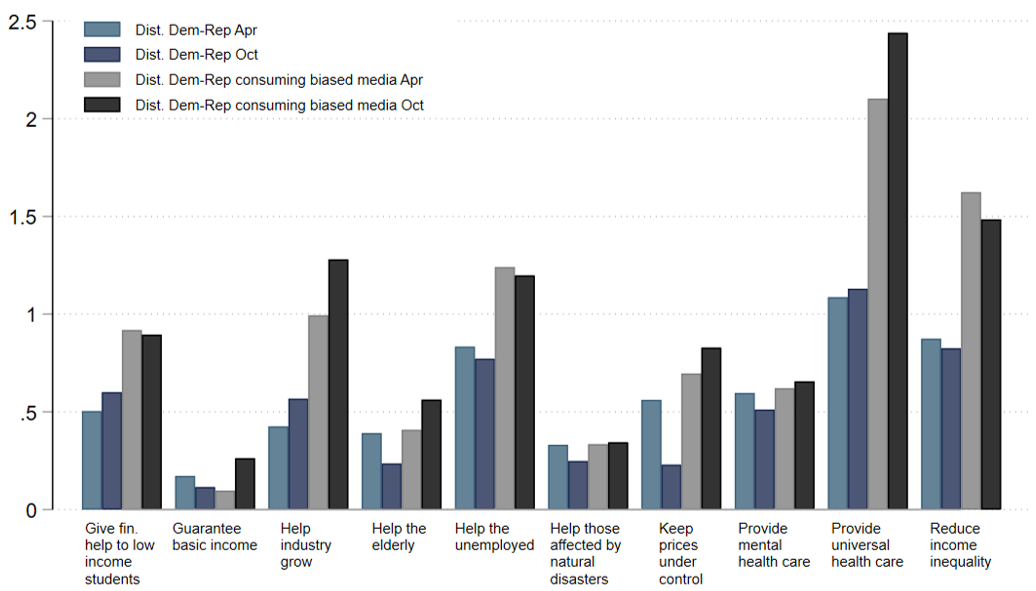}
	\end{center}
\begin{minipage}{1\linewidth \setstretch{0.75}}
	{\scriptsize{\textit{Notes}:} 
		\scriptsize The figure shows the difference of the differences in average Likert scale scores between Democrats and Republicans, and between Democrats and Republicans consuming biased media, between April and October 2020.}
\end{minipage}	
\end{figure}

\begin{figure}[H]
	\caption{Partisan gap in temporary relief policy, by media consumption, between April and October 2020}
	\label{fig:policies_2}
	\begin{center}
		\includegraphics[height=9cm]{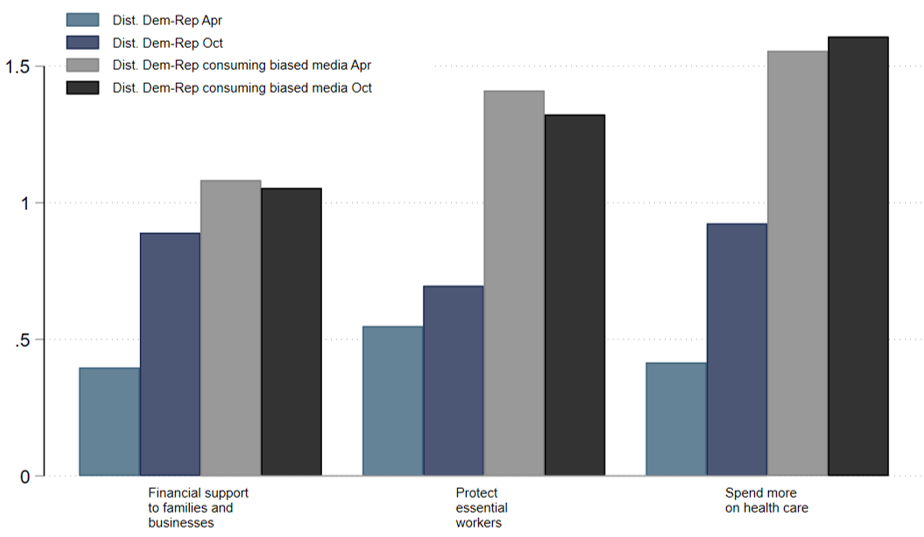}
	\end{center}
\begin{minipage}{1\linewidth \setstretch{0.75}}
	{\scriptsize{\textit{Notes}:} 
		\scriptsize The figure shows the difference of the differences in average Likert scale scores between Democrats and Republicans, and between Democrats and Republicans consuming biased media, between April and October 2020. }
\end{minipage}	
\end{figure}

\begin{figure}[H]
	\caption{Partisan gap in institutional trust, by media consumption, between April and October 2020}
	\label{fig:confid}
	\begin{center}
		\includegraphics[height=9cm]{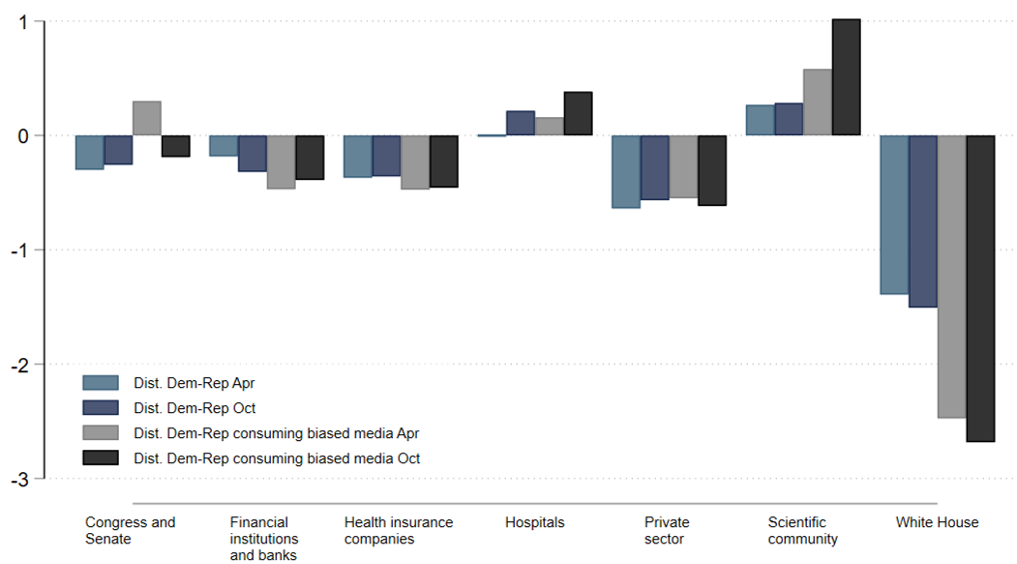}
	\end{center}
\begin{minipage}{1\linewidth \setstretch{0.75}}
	{\scriptsize{\textit{Notes}:} 
		\scriptsize The figure shows the difference of the differences in average Likert scale scores between Democrats and Republicans, and between Democrats and Republicans consuming biased media, between April and October 2020.}
\end{minipage}	
\end{figure}

\newpage 

\textbf{Partisan gap by shocks over time}.

\begin{figure}[H]
	\caption{Welfare policy partisan gap, by economic shock, between April and October 2020}
	\label{fig:policies_3}
	\begin{center}
		\includegraphics[height=10cm]{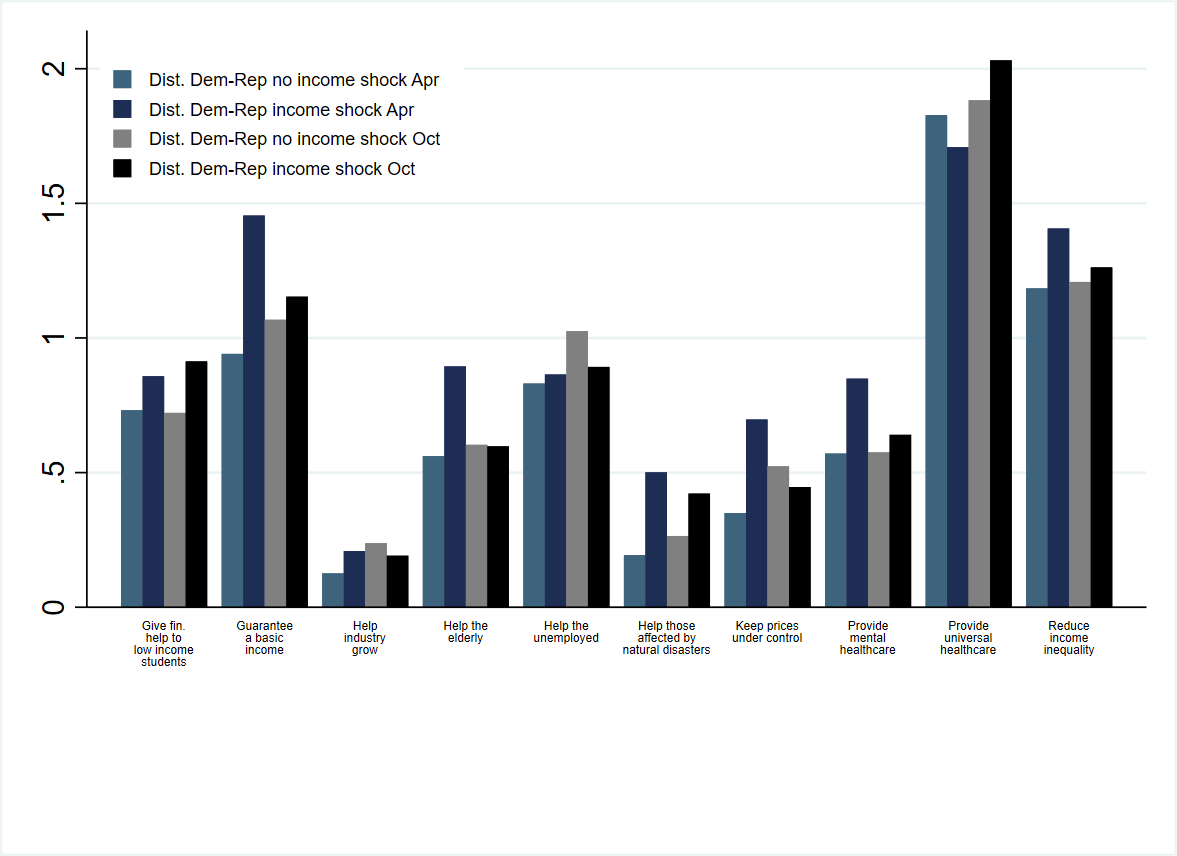}
	\end{center}
\begin{minipage}{1\linewidth \setstretch{0.75}}
	{\scriptsize{\textit{Notes}:} 
		\scriptsize The figure shows the difference of the differences in average Likert scale scores between Democrats and Republicans, and between Democrats and Republicans, by shock, between April and October 2020.}
\end{minipage}	
\end{figure}

\begin{figure}[H]
	\caption{Welfare policy partisan gap, by health shock, between April and October 2020}
	\label{fig:policies_4}
	\begin{center}
		\includegraphics[height=10cm]{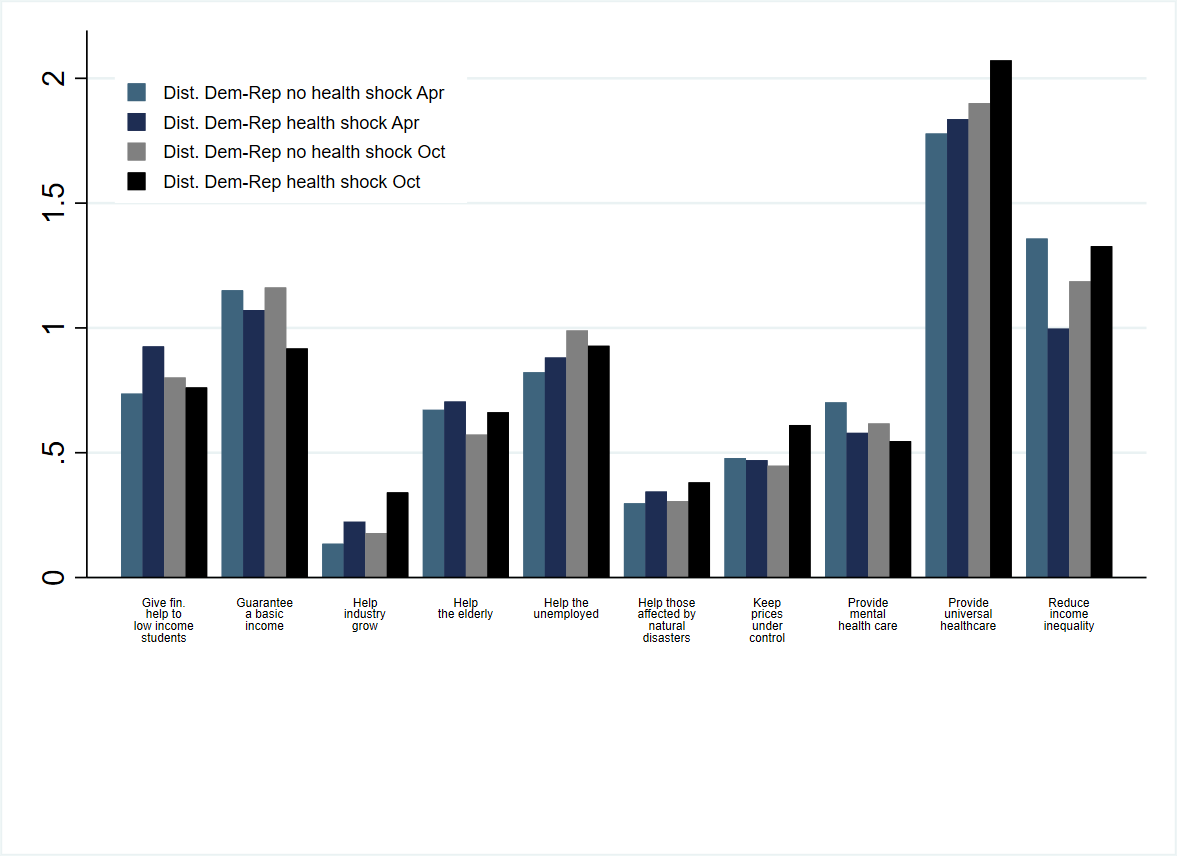}
	\end{center}
\begin{minipage}{1\linewidth \setstretch{0.75}}
	{\scriptsize{\textit{Notes}:} 
		\scriptsize The figure shows the difference of the differences in average Likert scale scores between Democrats and Republicans, and between Democrats and Republicans, by shock, between April and October 2020.}
\end{minipage}	
\end{figure}

\begin{figure}[H]
	\caption{Partisan gap in temporary relief policy, by economic shock, between April and October 2020}
	\label{fig:policies_5}
	\begin{center}
		\includegraphics[height=10cm]{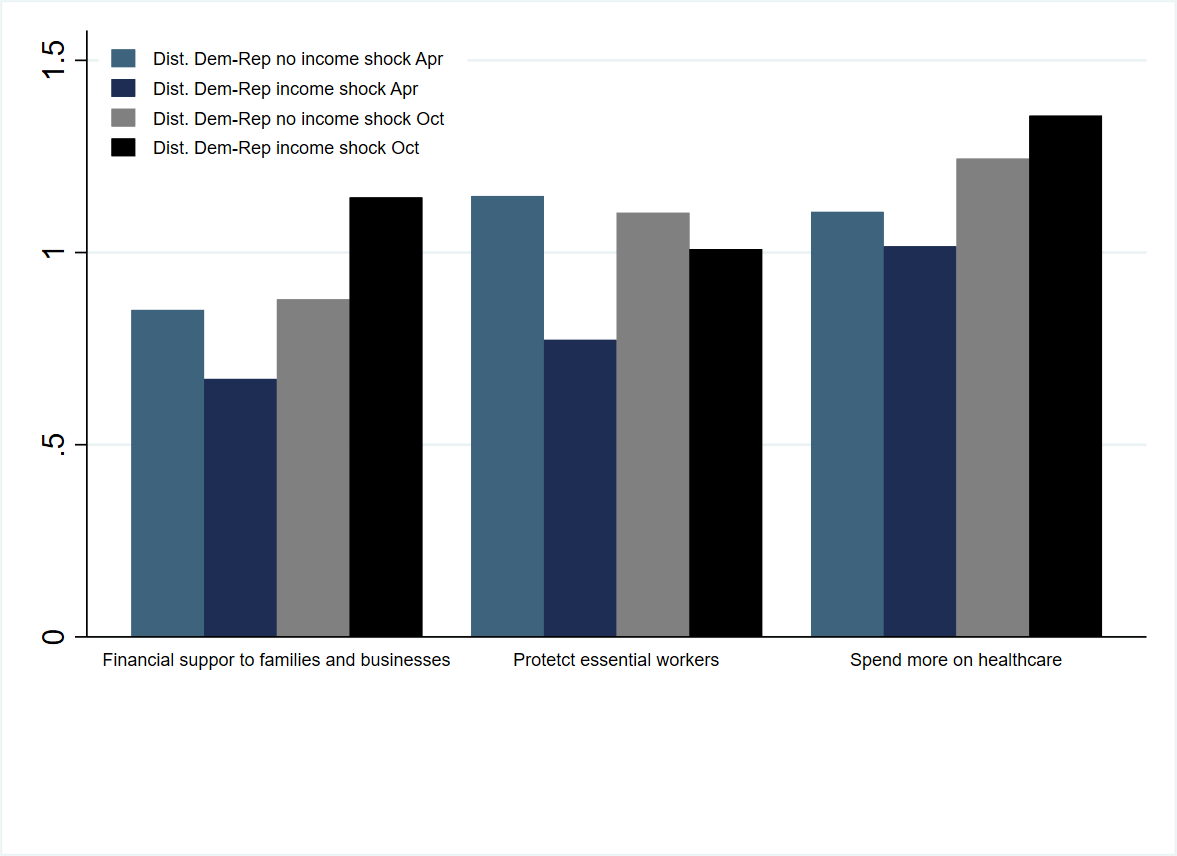}
	\end{center}
\begin{minipage}{1\linewidth \setstretch{0.75}}
	{\scriptsize{\textit{Notes}:} 
		\scriptsize The figure shows the difference of the differences in average Likert scale scores between Democrats and Republicans, and between Democrats and Republicans, by shock, between April and October 2020.}
\end{minipage}	
\end{figure}

\begin{figure}[H]
	\caption{Partisan gap in temporary relief policy, by health shock, between April and October 2020}
	\label{fig:policies_6}
	\begin{center}
		\includegraphics[height=10cm]{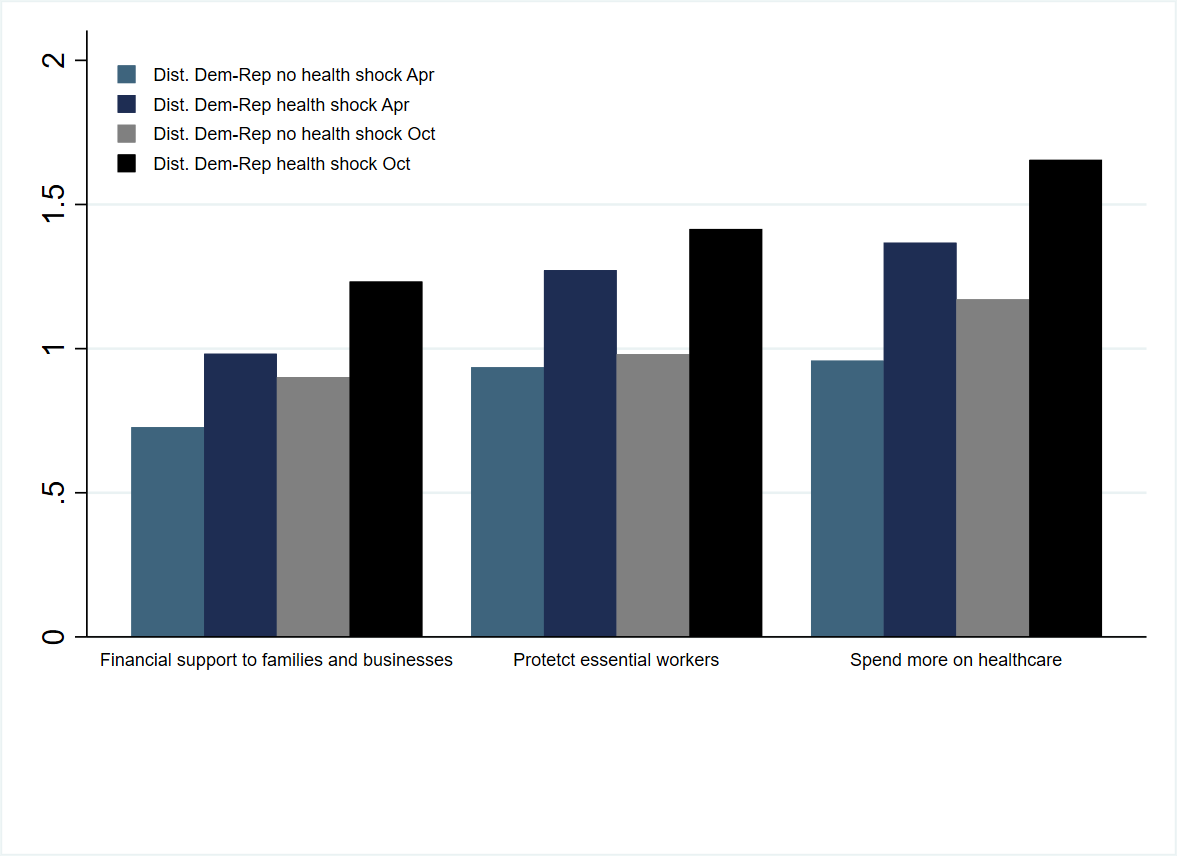}
	\end{center}
\begin{minipage}{1\linewidth \setstretch{0.75}}
	{\scriptsize{\textit{Notes}:} 
		\scriptsize The figure shows the difference of the differences in average Likert scale scores between Democrats and Republicans, and between Democrats and Republicans, by shock, between April and October 2020.}
\end{minipage}	
\end{figure}

\begin{figure}[H]
	\caption{Partisan gap in institutional trust, by economic shock, between April and October 2020}
	\label{fig:confid_2}
	\begin{center}
		\includegraphics[height=10cm]{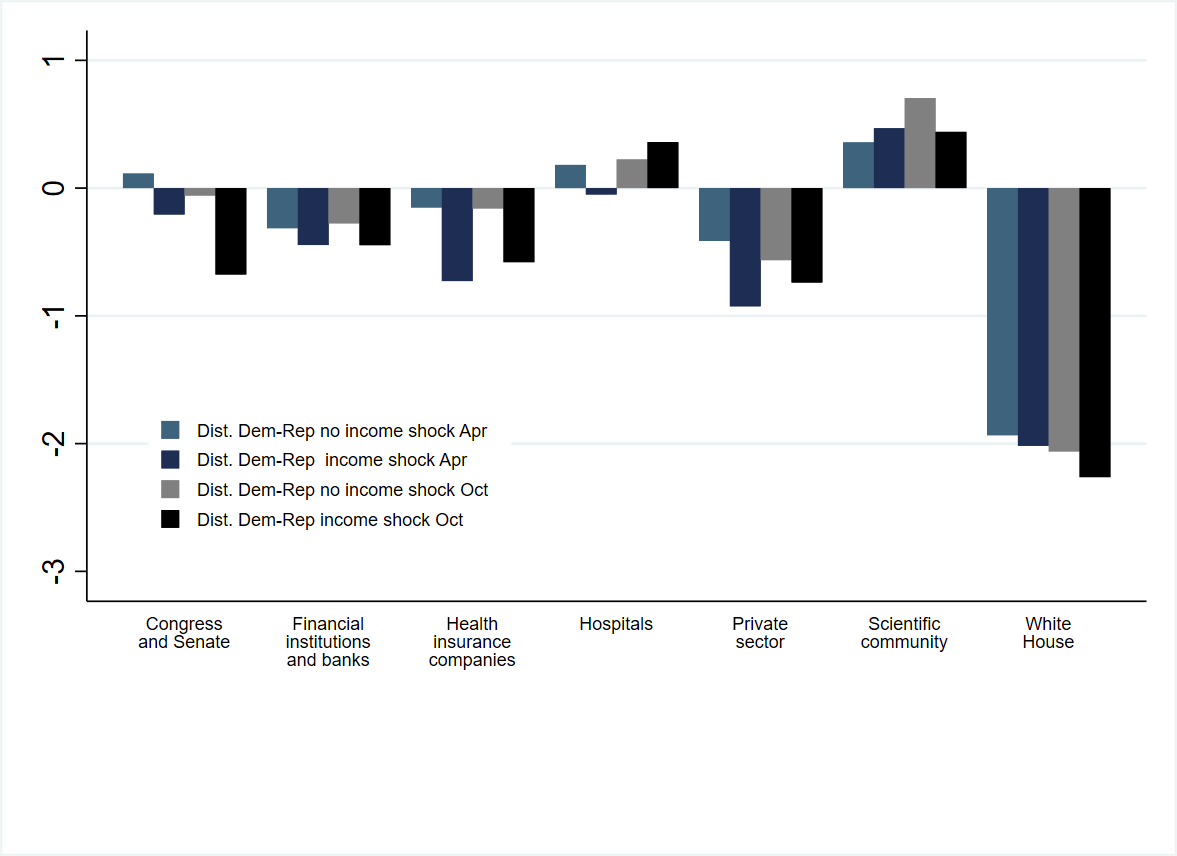}
	\end{center}
\begin{minipage}{1\linewidth \setstretch{0.75}}
	{\scriptsize{\textit{Notes}:} 
		\scriptsize The figure shows the difference of the differences in average Likert scale scores between Democrats and Republicans, and between Democrats and Republicans, by shock, between April and October 2020.}
\end{minipage}	
\end{figure}

\begin{figure}[H]
	\caption{Partisan gap in institutional trust, by health shock, between April and October 2020}
	\label{fig:confid_3}
	\begin{center}
		\includegraphics[height=10cm]{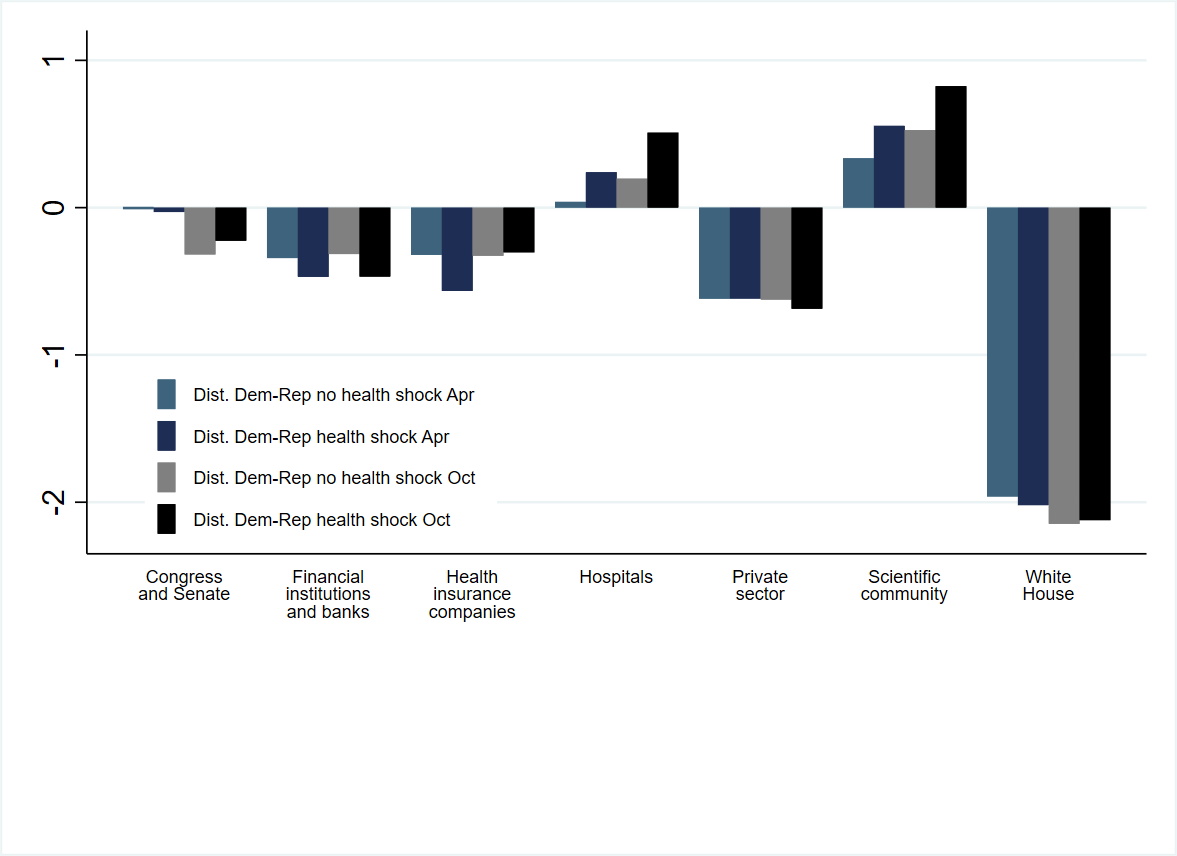}
	\end{center}
\begin{minipage}{1\linewidth \setstretch{0.75}}
	{\scriptsize{\textit{Notes}:} 
		\scriptsize The figure shows the difference of the differences in average Likert scale scores between Democrats and Republicans, and between Democrats and Republicans, by shock, between April and October 2020.}
\end{minipage}	
\end{figure}

\newpage

\textbf{Attrition}. Our study population comprises the US Adults who are at least 18 years old. The research institution NORC submitted our first survey to 11263 individuals from their AmeriSpeak panel, a representative sample of the US population, and obtained a response rate of 12.8\% (1442 respondents). Survey weights have been applied to take in account the sampling strategy and to re-weights individuals belonging to specific groups who have been over or under sampled. In each subsequent wave, we re-interviewed the same 1,442 individuals, experiencing different attrition rates. In our last wave, 1076 individuals (or 74.6\% of the sample) completed our survey, with income being one of the best predictors for attrition, while 814 respondents completed all the seven waves. In table \ref{tab:summary_stats}, we report the average demographic characteristics of our sample in each wave, once the survey weights have been applied. 

\begin{table}[H]
\centering
\caption{Differences in demographics across groups who are affected by attrition and those who are not.}
\label{tab:balancetable_attr}
\resizebox*{!}{0.97\textheight}{%
\begin{tabular}{lccccccc} \hline \hline
 & (1) & (2) & (3) & (4) & (5) & (6) & (7) \\
 & \begin{tabular}[c]{@{}c@{}}Mean\\ baseline\end{tabular} & \begin{tabular}[c]{@{}c@{}}Mean \\ no attrition \\ w4-w7\end{tabular} & \begin{tabular}[c]{@{}c@{}}Mean\\ attrition \\ w4-w7\end{tabular} & \begin{tabular}[c]{@{}c@{}}Diff \\ w7-w4\end{tabular} & \begin{tabular}[c]{@{}c@{}}Mean \\ no attrition\\ w1-w7\end{tabular} & \begin{tabular}[c]{@{}c@{}}Mean \\ attrition \\ w1-w7\end{tabular} & \begin{tabular}[c]{@{}c@{}}Diff \\ w7-w1\end{tabular} \\ \hline
Republican & 0.265 & 0.257 & 0.265 & 0.008 & 0.259 & 0.283 & 0.024 \\
 & (0.441) & (0.437) & (0.442) & (0.033) & (0.438) & (0.451) & (0.027) \\
Democrat & 0.397 & 0.379 & 0.452 & 0.073** & 0.379 & 0.451 & 0.072** \\
 & (0.489) & (0.485) & (0.499) & (0.036) & (0.485) & (0.498) & (0.030) \\
Independent/other & 0.338 & 0.364 & 0.283 & -0.081** & 0.362 & 0.266 & -0.096*** \\
 & (0.473) & (0.481) & (0.452) & (0.036) & (0.481) & (0.443) & (0.029) \\
Woman & 0.476 & 0.458 & 0.523 & 0.064* & 0.460 & 0.523 & 0.063** \\
 & (0.500) & (0.499) & (0.501) & (0.037) & (0.499) & (0.500) & (0.030) \\
Age: 18-29 & 0.223 & 0.190 & 0.286 & 0.096*** & 0.201 & 0.290 & 0.090*** \\
 & (0.417) & (0.393) & (0.453) & (0.030) & (0.401) & (0.455) & (0.025) \\
Age: 30-44 & 0.278 & 0.292 & 0.205 & -0.088*** & 0.288 & 0.247 & -0.042 \\
 & (0.448) & (0.455) & (0.404) & (0.033) & (0.453) & (0.432) & (0.027) \\
Age: 45-59 & 0.223 & 0.223 & 0.205 & -0.019 & 0.223 & 0.222 & -0.001 \\
 & (0.416) & (0.417) & (0.404) & (0.031) & (0.416) & (0.416) & (0.025) \\
Age: 60+ & 0.276 & 0.294 & 0.305 & 0.010 & 0.288 & 0.241 & -0.047* \\
 & (0.447) & (0.456) & (0.461) & (0.034) & (0.453) & (0.428) & (0.027) \\
Less than HS & 0.033 & 0.026 & 0.050 & 0.024* & 0.028 & 0.049 & 0.021** \\
 & (0.180) & (0.159) & (0.218) & (0.013) & (0.165) & (0.217) & (0.011) \\
High school & 0.162 & 0.159 & 0.127 & -0.032 & 0.164 & 0.156 & -0.007 \\
 & (0.368) & (0.366) & (0.334) & (0.027) & (0.370) & (0.364) & (0.022) \\
Some college & 0.412 & 0.379 & 0.495 & 0.116*** & 0.384 & 0.496 & 0.112*** \\
 & (0.492) & (0.485) & (0.501) & (0.036) & (0.487) & (0.501) & (0.030) \\
Bachelor + & 0.393 & 0.435 & 0.327 & -0.108*** & 0.425 & 0.299 & -0.126*** \\
 & (0.489) & (0.496) & (0.470) & (0.037) & (0.495) & (0.458) & (0.029) \\
I income q & 0.203 & 0.220 & 0.100 & -0.120*** & 0.227 & 0.132 & -0.095*** \\
 & (0.402) & (0.415) & (0.301) & (0.030) & (0.419) & (0.338) & (0.024) \\
II income q & 0.198 & 0.171 & 0.286 & 0.115*** & 0.172 & 0.277 & 0.105*** \\
 & (0.399) & (0.377) & (0.453) & (0.029) & (0.377) & (0.448) & (0.024) \\
III income q & 0.199 & 0.190 & 0.241 & 0.051* & 0.189 & 0.230 & 0.041* \\
 & (0.400) & (0.393) & (0.429) & (0.030) & (0.391) & (0.421) & (0.024) \\
IV income q & 0.201 & 0.204 & 0.205 & 0.000 & 0.204 & 0.192 & -0.012 \\
 & (0.401) & (0.403) & (0.404) & (0.030) & (0.403) & (0.394) & (0.024) \\
V income q & 0.199 & 0.214 & 0.168 & -0.046 & 0.209 & 0.170 & -0.039 \\
 & (0.400) & (0.410) & (0.375) & (0.030) & (0.407) & (0.376) & (0.024) \\
Financial hardship pre-COVID & 0.301 & 0.277 & 0.324 & 0.046 & 0.283 & 0.354 & 0.071** \\
 & (0.459) & (0.448) & (0.469) & (0.035) & (0.451) & (0.479) & (0.029) \\
African American & 0.113 & 0.094 & 0.141 & 0.047** & 0.101 & 0.148 & 0.047** \\
 & (0.317) & (0.292) & (0.349) & (0.023) & (0.302) & (0.356) & (0.019) \\
Hispanic & 0.162 & 0.133 & 0.209 & 0.076*** & 0.138 & 0.230 & 0.092*** \\
 & (0.368) & (0.340) & (0.408) & (0.026) & (0.346) & (0.421) & (0.022) \\
Other Race & 0.119 & 0.122 & 0.132 & 0.010 & 0.121 & 0.115 & -0.006 \\
 & (0.324) & (0.328) & (0.339) & (0.025) & (0.326) & (0.320) & (0.020) \\
White & 0.606 & 0.651 & 0.518 & -0.132*** & 0.639 & 0.507 & -0.133*** \\
 & (0.489) & (0.477) & (0.501) & (0.036) & (0.480) & (0.501) & (0.029) \\
Cohabitating & 0.544 & 0.646 & 0.314 & -0.332*** & 0.634 & 0.279 & -0.354*** \\
 & (0.498) & (0.479) & (0.465) & (0.035) & (0.482) & (0.449) & (0.029) \\
Parent of minor & 0.294 & 0.282 & 0.241 & -0.041 & 0.286 & 0.315 & 0.029 \\
 & (0.456) & (0.450) & (0.429) & (0.033) & (0.452) & (0.465) & (0.028) \\
Caring responsibilities & 0.160 & 0.151 & 0.186 & 0.035 & 0.157 & 0.170 & 0.013 \\
 & (0.367) & (0.358) & (0.390) & (0.027) & (0.364) & (0.376) & (0.022) \\
Not in the labor force & 0.249 & 0.257 & 0.279 & 0.021 & 0.252 & 0.240 & -0.012 \\
 & (0.432) & (0.437) & (0.449) & (0.033) & (0.434) & (0.427) & (0.026) \\
Unemployed in Feb & 0.080 & 0.074 & 0.120 & 0.046* & 0.072 & 0.105 & 0.033* \\
 & (0.272) & (0.262) & (0.326) & (0.024) & (0.259) & (0.307) & (0.019) \\
North-East & 0.152 & 0.148 & 0.164 & 0.015 & 0.149 & 0.162 & 0.013 \\
 & (0.359) & (0.355) & (0.371) & (0.027) & (0.356) & (0.369) & (0.022) \\
Midwest & 0.246 & 0.268 & 0.186 & -0.082** & 0.265 & 0.189 & -0.076*** \\
 & (0.431) & (0.443) & (0.390) & (0.032) & (0.441) & (0.392) & (0.026) \\
South & 0.366 & 0.351 & 0.368 & 0.017 & 0.361 & 0.381 & 0.020 \\
 & (0.482) & (0.478) & (0.483) & (0.036) & (0.480) & (0.486) & (0.029) \\
West & 0.237 & 0.232 & 0.282 & 0.050 & 0.226 & 0.268 & 0.043* \\
 & (0.425) & (0.422) & (0.451) & (0.032) & (0.418) & (0.444) & (0.026) \\
Metropolitan area & 0.865 & 0.861 & 0.877 & 0.016 & 0.862 & 0.877 & 0.015 \\
 & (0.341) & (0.346) & (0.329) & (0.026) & (0.346) & (0.329) & (0.021) \\
No health insurance & 0.082 & 0.067 & 0.100 & 0.033* & 0.072 & 0.113 & 0.041** \\
 & (0.275) & (0.251) & (0.301) & (0.019) & (0.259) & (0.317) & (0.017) \\
Population density in ZCTA & 3,954.007 & 3,889.056 & 4,057.553 & 168.497 & 3,947.825 & 3,972.331 & 24.506 \\
 & (8,920.238) & (9,586.471) & (6,851.486) & (683.266) & (9,525.808) & (6,827.875) & (542.375) \\ 
 \textit{N} & 1,441 & 999 & 220 & 1,441 & 1,076 & 365 & 1,441 
\end{tabular}}
\end{table}

Several demographics are significantly different across groups, indicating that our respondents are not “Missing completely at random (MCAR)”. However, following \cite{fitzgerald1998analysis}, those respondents are “Missing at random (MAR)”, so if such attrition is correlated with observable characteristics but not with our outcomes of interest. The assumption behind this theory is that if attrition occurs randomly within clusters composed by individuals sharing the same observable characteristics, it is possible to correct for potential bias by using post-stratified survey weights. Hence, we compare two models predicting attrition, one including the baseline value of our outcome of interest with one without.

$$Pr(attrition_i | y_{iw1}, X_{iw1}) = \Phi(\alpha + \gamma y_{iw1} + X'_{iw1} \beta ) = \Phi(\Tilde{\alpha} + X'_{iw1} \tilde{\beta} ) = Pr(attrition_i | X_{iw1}) $$

\noindent with $y_{iw1}$ being the baseline outcome, and $X_{iw1}$ a set of demographics. 

If the two models are not statistically different, then it is safe to assume that such attrition occurs at random (MAR) and that it can be corrected. Table \ref{tab:predict_attrition} reports the $\chi^2$ and the p-value associated with a set of likelihood-ratio tests comparing two logistic models predicting attrition, one including the baseline outcome among the independent variables and one without. We repeated this exercise for all outcomes. In all the cases, with the exception of the belief that the government should provide for the elderly, the baseline outcomes cannot predict attrition, suggesting that those observations are missing at random and, thus, will not bias our results. 

\begin{table}[H]
\centering
\caption{Likelihood ratio tests comparing a model predicting attrition including the baseline outcome among the independent variables with one who does not.}
\label{tab:predict_attrition}
\resizebox{\textwidth}{!}{%
\begin{tabular}{l|lcc}
\hline \hline
 & \textbf{Variable} & \begin{tabular}[c]{@{}c@{}}\textbf{Likelihood }\\\textbf{ratio test }\\\textbf{- $\chi^2$}\end{tabular} & \begin{tabular}[c]{@{}c@{}}\textbf{Likelihood }\\\textbf{ratio test~}\\\textbf{- pvalue}\end{tabular} \\ 
\hline
&&&\\
\multirow{21}{*}{\textbf{ Attrition w1-w7~ }} & \multicolumn{3}{l}{\textit{Confidence in people running...}} \\
 & The U.S. Congress and Senate & 0.442 & 0.506 \\
 & The White House & 0.375 & 0.54 \\
 & The Scientific Community & 0.627 & 0.428 \\
 & Financial institutions & 0.383 & 0.536 \\
 & The private sector & 0.685 & 0.408 \\
 & Hospitals & 0.35 & 0.554 \\
 & Health insurance companies & 0.581 & 0.446 \\
 & & & \\
 & Support universal healthcare & 1.229 & 0.268 \\
 & & & \\
 & \multicolumn{3}{l}{\textit{It's a government's responsibility to...}} \\
 & Provide mental healthcare & 0.216 & 0.642 \\
 & Help those affected by natural disasters & 1.007 & 0.316 \\
 & Keep prices under control & 0.5 & 0.48 \\
 & Provide for the elderly & 4.418 & 0.036 \\
 & Provide for the unemployed & 0.003 & 0.96 \\
 & Provide a basic income & 0.045 & 0.832 \\
 & Help industry grow & 0.684 & 0.408 \\
 & Reduce inequality & 0.045 & 0.832 \\
 & Pay university for poor & 0.792 & 0.373 \\ 
\hline
 & & & \\
 & \multicolumn{3}{l}{\textit{The government should...}} \\
\multirow{3}{*}{\textbf{Attrition w4-w7}} & Transfer money to families and businesses & 0.025 & 0.874 \\
 & Do more to protect essential workers & 0.003 & 0.954 \\
 & Spend more on public healthcare & 1.582 & 0.208 \\ \hline
\end{tabular}
}
\end{table}

\textbf{Party affiliation and voting intentions}
\begin{table}[H]
\centering
\caption{Voting intentions by political affiliation collected in June 2020.}
\label{tab:party_vote}
\begin{tabular}{l|cccc|c}\hline
 & Democrats & Republicans & Independents & Others & N votes \\ \hline
Undecided & 5.5\% & 8.0\% & 24.8\% & 34.2\% & 183 \\
Vote Trump & 3.0\% & 82.2\% & 31.8\% & 12.7\% & 335 \\
Vote Biden & 89.9\% & 6.8\% & 37.3\% & 23.6\% & 558 \\
Not voting & 1.8\% & 2.9\% & 6.2\% & 29.6\% & 59 \\ \hline
N parties & 431 & 275 & 311 & 118 & 1135\\ \hline
\end{tabular}
\end{table}

\subsection{Additional robustness checks}
\subsubsection{Different measures of shocks}
\begin{table}[H]
\caption{Preferences for economic and welfare policies - alternative income shock 1}
\label{tab:policy_inc2}
\resizebox{\textwidth}{!}{%
\begin{tabular}{lcccccccccc}
\hline
\hline
 &
 \multicolumn{10}{c}{\begin{tabular}[c]{@{}c@{}}Stronger belief between April and October 2020 that it should be the role of government to:\end{tabular}} \\
  & (1) & (2) & (3) & (4) & (5) & (6) & (7) & (8) & (9) & (10) \\
 & \begin{tabular}[c]{@{}c@{}}Universal \\ health care\end{tabular} &
  \begin{tabular}[c]{@{}c@{}}Help \\ industry \\ grow\end{tabular} &
  \begin{tabular}[c]{@{}c@{}}Keep prices \\ under \\ control\end{tabular} &
  \begin{tabular}[c]{@{}c@{}}Provide for \\the unemployed\end{tabular} &
  \begin{tabular}[c]{@{}c@{}}Provide \\ mental \\ health care\end{tabular} &
  \begin{tabular}[c]{@{}c@{}}Help the \\ elderly\end{tabular} &
  \begin{tabular}[c]{@{}c@{}}Help those \\ affected by \\ natural disasters\end{tabular} &
  \begin{tabular}[c]{@{}c@{}}Guaranteed \\ basic income\end{tabular} &
 \begin{tabular}[c]{@{}c@{}}Reduce \\ income \\ inequality\end{tabular} &
 \begin{tabular}[c]{@{}c@{}}Give financial \\ help to \\ low-income\\ Univ. students\end{tabular} \\ \hline
 & & & & & & & & & & \\
\multirow{2}{*}{\begin{tabular}[c]{@{}l@{}}Income Oct 20\% \\ lower than Apr \end{tabular}} & -0.0471 & -0.128*** & 0.00577 & 0.0425 & -0.0551** & 0.0518* & 0.0285 & 0.0246 & 0.0343 & -0.000268 \\
& (0.0457) & (0.0369) & (0.0428) & (0.0389) & (0.0274) & (0.0311) & (0.0420) & (0.0433) & (0.0386) & (0.0339)
\\
 & & & & & & & & & & \\
 Controls & Yes & Yes & Yes & Yes & Yes & Yes & Yes & Yes & Yes & Yes \\ 
Observations & 1,010 & 1,002 & 1,005 & 1,002 & 1,010 & 1,004 & 1,007 & 999 & 1,006 & 1,003 \\
R-squared & 0.189 & 0.204 & 0.318 & 0.162 & 0.293 & 0.266 & 0.387 & 0.192 & 0.240 & 0.235 \\
Average increase & 0.265 & 0.262 & 0.244 & 0.144 & 0.158 & 0.168 & 0.126 & 0.220 & 0.225 & 0.187 \\
Average decrease & 0.210 & 0.244 & 0.233 & 0.325 & 0.251 & 0.254 & 0.224 & 0.233 & 0.231 & 0.215 \\ \hline
\multicolumn{11}{l}{%
 \begin{minipage}{1.8\textwidth}%
  \textit{Notes}: Standard errors in parentheses. *** p\textless{}0.01, ** p\textless{}0.05, * p\textless{}0.1.\\
  All regressions are OLS regressions that take into account population survey wights and the sampling procedure. The dependent variable is a dummy=1 if the respondent increased their belief that it should be a government's responsibility to provide the following policies. The control variables include: gender, race, age, education, parental status, and caring responsibilities for an elderly or a person with a disability, income in February 2020, housing, labor force participation and employment status in February 2020, health insurance provider, and whether respondents had financial difficulties before the pandemic, the area in which the respondents live, whether it's a metropolitan or rural area, and the population density in the zip code. We also control for whether respondents completed the related surveys in shorter time than the 99$^{th}$ percentile, and ceiling effects. Finally, we include variables considering the media: whether the sources of news consulted are leaning politically or are neutral, the amount of international news consumed and social media usage.
 \end{minipage}%
}\\ 
\end{tabular}%
}
\end{table}

\begin{table}[H]
\caption{Preferences for economic and welfare policies - alternative income shock 2}
\label{tab:policy_inc3}
\resizebox{\textwidth}{!}{%
\begin{tabular}{lcccccccccc}
\hline
\hline
 &
 \multicolumn{10}{c}{\begin{tabular}[c]{@{}c@{}}Stronger belief between April and October 2020 that it should be the role of government to:\end{tabular}} \\
 & (1) & (2) & (3) & (4) & (5) & (6) & (7) & (8) & (9) & (10) \\
 & \begin{tabular}[c]{@{}c@{}}Universal \\ health care\end{tabular} &
  \begin{tabular}[c]{@{}c@{}}Help \\ industry \\ grow\end{tabular} &
  \begin{tabular}[c]{@{}c@{}}Keep prices \\ under \\ control\end{tabular} &
  \begin{tabular}[c]{@{}c@{}}Provide for \\the unemployed\end{tabular} &
  \begin{tabular}[c]{@{}c@{}}Provide \\ mental \\ health care\end{tabular} &
  \begin{tabular}[c]{@{}c@{}}Help the \\ elderly\end{tabular} &
  \begin{tabular}[c]{@{}c@{}}Help those \\ affected by \\ natural disasters\end{tabular} &
  \begin{tabular}[c]{@{}c@{}}Guaranteed \\ basic income\end{tabular} &
 \begin{tabular}[c]{@{}c@{}}Reduce \\ income \\ inequality\end{tabular} &
 \begin{tabular}[c]{@{}c@{}}Give financial \\ help to \\ low-income\\ Univ. students\end{tabular} \\ \hline
 & & & & & & & & & & \\
\% decrease income & -0.0455 & -0.160*** & 0.0283 & 0.0470 & -0.0816* & 0.103* & 0.0254 & 0.0744 & 0.0698 & -0.0292 \\
 & (0.0737) & (0.0596) & (0.0639) & (0.0671) & (0.0460) & (0.0600) & (0.0645) & (0.0749) & (0.0774) & (0.0641) \\
 & & & & & & & & & & \\
 Controls & Yes & Yes & Yes & Yes & Yes & Yes & Yes & Yes & Yes & Yes \\ 
Observations & 1,010 & 1,002 & 1,005 & 1,002 & 1,010 & 1,004 & 1,007 & 999 & 1,006 & 1,003 \\
R-squared & 0.192 & 0.202 & 0.319 & 0.163 & 0.294 & 0.268 & 0.388 & 0.193 & 0.241 & 0.238 \\
Average increase & 0.265 & 0.262 & 0.244 & 0.144 & 0.158 & 0.168 & 0.126 & 0.220 & 0.225 & 0.187 \\
Average decrease & 0.210 & 0.244 & 0.233 & 0.325 & 0.251 & 0.254 & 0.224 & 0.233 & 0.231 & 0.215 \\ \hline
\multicolumn{11}{l}{%
 \begin{minipage}{1.8 \columnwidth}%
  \textit{Notes}: Standard errors in parentheses. *** p\textless{}0.01, ** p\textless{}0.05, * p\textless{}0.1.\\
  All regressions are OLS regressions that take into account population survey wights and the sampling procedure. The dependent variable is a dummy=1 if the respondent increased their belief that it should be a government's responsibility to provide the following policies. The control variables include: gender, race, age, education, parental status, and caring responsibilities for an elderly or a person with a disability, income in February 2020, housing, labor force participation and employment status in February 2020, health insurance provider, and whether respondents had financial difficulties before the pandemic, the area in which the respondents live, whether it's a metropolitan or rural area, and the population density in the zip code. We also control for whether respondents completed the related surveys in shorter time than the 99$^{th}$ percentile, and ceiling effects. Finally, we include variables considering the media: whether the sources of news consulted are leaning politically or are neutral, the amount of international news consumed and social media usage.
 \end{minipage}%
}\\ 
\end{tabular}%
}
\end{table}

\begin{table}[H]
\centering
\caption{Preferences for coronavirus relief policies - alternative income shocks.}
\label{tab:covid_policy_inc23}
\resizebox{\textwidth}{!}{%
\begin{tabular}{lcccccc}\hline \hline
&
 \multicolumn{6}{c}{\begin{tabular}[c]{@{}c@{}}Stronger belief between May and October 2020 that the government should:\end{tabular}} \\
 & (1) & (2) & (3) & (4) & (5) & (6) \\
 &
 \begin{tabular}[c]{@{}c@{}}Spend more on \\ public healthcare \\ to reduce \\ preventable deaths\end{tabular} &
 \begin{tabular}[c]{@{}c@{}}Do more to \\ protect \\ essential workers\end{tabular} &
 \begin{tabular}[c]{@{}c@{}}Transfer money \\ directly to families \\ and businesses \end{tabular} &
 \begin{tabular}[c]{@{}c@{}}Spend more on \\ public healthcare \\ to reduce \\ preventable deaths\end{tabular} &
 \begin{tabular}[c]{@{}c@{}}Do more to \\ protect \\ essential workers\end{tabular} &
 \begin{tabular}[c]{@{}c@{}}Transfer money \\ directly to families \\ and businesses \end{tabular} \\ \hline
 & & & & & & \\
\multirow{2}{*}{\begin{tabular}[c]{@{}l@{}}Income Oct 20\% \\ lower than May \end{tabular}} & 0.132*** & 0.122** & 0.0882** & & & \\
 & (0.0428) & (0.0507) & (0.0444) & & & \\
\% decrease income & & & & 0.259*** & 0.122 & 0.139** \\
 & & & & (0.0831) & (0.0782) & (0.0666) \\
 Controls & Yes & Yes & Yes & Yes & Yes & Yes \\
Observations & 937 & 938 & 942 & 937 & 938 & 942 \\
R-squared & 0.214 & 0.242 & 0.136 & 0.216 & 0.238 & 0.136 \\
Average increase & 0.177 & 0.188 & 0.181 & 0.177 & 0.188 & 0.181 \\
Average decrease & 0.295 & 0.317 & 0.369 & 0.295 & 0.317 & 0.369 \\ \hline
\multicolumn{7}{l}{%
 \begin{minipage}{1.5\columnwidth}%
  \textit{Notes}: Standard errors in parentheses. *** p\textless{}0.01, ** p\textless{}0.05, * p\textless{}0.1.\\
  All regressions are OLS regressions that take into account population survey wights and the sampling procedure. The dependent variable is a dummy=1 if the respondent increased their support for the following policies. The control variables include: gender, race, age, education, parental status, and caring responsibilities for an elderly or a person with a disability, income in February 2020, housing, labor force participation and employment status in February 2020, health insurance provider, and whether respondents had financial difficulties before the pandemic, the area in which the respondents live, whether it's a metropolitan or rural area, and the population density in the zip code. We also control for whether respondents completed the related surveys in shorter time than the 99$^{th}$ percentile, and ceiling effects. Finally, we include variables considering the media: whether the sources of news consulted are leaning politically or are neutral, the amount of international news consumed and social media usage.
 \end{minipage}%
}\\ 
\end{tabular}%
}
\end{table}

\begin{table}[H]
\centering
\caption{Decrease in institutional trust - alternative income shock 1}
\label{tab:trust_inc2}
\resizebox{\textwidth}{!}{%
 \begin{tabular}{lccccccc} \hline \hline
 &
 \multicolumn{7}{c}{\begin{tabular}[c]{@{}c@{}}Lower confidence in people running the following institutions:\end{tabular}} \\
 & (1) & (2) & (3) & (4) & (5) & (6) & (7) \\
& \begin{tabular}[c]{@{}c@{}}Congress \\ \& Senate\end{tabular} & \begin{tabular}[c]{@{}c@{}}White\\ House\end{tabular} & \begin{tabular}[c]{@{}c@{}}Financial \\ institutions \\ \& banks \end{tabular} & \begin{tabular}[c]{@{}c@{}}Private \\ sector\end{tabular} & \begin{tabular}[c]{@{}c@{}}Scientific \\ community\end{tabular} & \begin{tabular}[c]{@{}c@{}}Health \\ insurance \\ companies\end{tabular} & Hospitals \\ \hline
 & & & & & & & \\
\multirow{2}{*}{\begin{tabular}[c]{@{}l@{}}Income Oct 20\% \\ lower than Apr\end{tabular}} & 0.141*** & 0.0388 & 0.0199 & 0.0554 & 0.0453 & -0.00374 & 0.0641 \\
 & (0.0460) & (0.0364) & (0.0463) & (0.0447) & (0.0569) & (0.0365) & (0.0536) \\
 & & & & & & & \\
 Controls & Yes & Yes & Yes & Yes & Yes & Yes & Yes \\
Observations & 1,009 & 1,003 & 1,007 & 1,006 & 1,002 & 1,006 & 1,007 \\
R-squared & 0.185 & 0.376 & 0.122 & 0.129 & 0.096 & 0.127 & 0.109 \\
Average increase & 0.159 & 0.142 & 0.142 & 0.185 & 0.185 & 0.172 & 0.164 \\
Average decrease & 0.351 & 0.312 & 0.299 & 0.247 & 0.286 & 0.284 & 0.306 \\\hline
\multicolumn{8}{l} {
\begin{minipage}{1.12\columnwidth}%
\small \textit{Notes}. Standard errors in parentheses. *** p\textless{}0.01, ** p\textless{}0.05, * p\textless{}0.1. The percentages in the first row report the share of respondents who decreased their Likert-based score of trust in people running the above institutions between the first and last wave of the survey. All regressions are OLS regressions that take into account population survey weights and the sampling procedure. The dependent variable is a dummy=1 if the respondent reduced their confidence in the people running the following institutions. The control variables include: gender, race, age, education, parental status, caring responsibilities for an elderly or a person with a disability, baseline income in February 2020, cohabitation with a partner, labor force participation and employment status in February 2020, health insurance provider, if the respondent had financial difficulties before the pandemic, macro-region, metro vs. rural, the population density at the zip code, and two dummy variables indicating if they consume at least 30min a week of international news and if they consume news from social media. We also control for whether respondents completed the survey in a shorter time than the 99$^{th}$ percentile as well as ceiling effects. Finally, we include variables considering the media: whether the sources of news consulted are leaning politically or are neutral, the amount of international news consumed and social media usage. \end{minipage}
}
\end{tabular}%
}
\end{table}

\begin{table}[H]
\centering
\caption{Decrease in institutional trust - alternative income shock 2}
\label{tab:trust_inc3}
\resizebox{\textwidth}{!}{%
 \begin{tabular}{lccccccc} \hline \hline
 &
 \multicolumn{7}{c}{\begin{tabular}[c]{@{}c@{}}Lower confidence in people running the following institutions:\end{tabular}} \\
 & (1) & (2) & (3) & (4) & (5) & (6) & (7) \\
& \begin{tabular}[c]{@{}c@{}}Congress \\ \& Senate\end{tabular} & \begin{tabular}[c]{@{}c@{}}White\\ House\end{tabular} & \begin{tabular}[c]{@{}c@{}}Financial \\ institutions \\ \& banks \end{tabular} & \begin{tabular}[c]{@{}c@{}}Private \\ sector\end{tabular} & \begin{tabular}[c]{@{}c@{}}Scientific \\ community\end{tabular} & \begin{tabular}[c]{@{}c@{}}Health \\ insurance \\ companies\end{tabular} & Hospitals \\ \hline
 & & & & & & &  \\
\% decrease income & 0.229** & -0.0133 & -0.0239 & 0.115 & 0.100 & 0.0179 & 0.175* \\
 & (0.0949) & (0.0692) & (0.0829) & (0.0694) & (0.115) & (0.0834) & (0.102) \\
 & & & & & & & \\
 Controls & Yes & Yes & Yes & Yes & Yes & Yes & Yes \\
Observations & 1,009 & 1,003 & 1,007 & 1,006 & 1,002 & 1,006 & 1,007 \\
R-squared & 0.187 & 0.377 & 0.123 & 0.131 & 0.098 & 0.128 & 0.113 \\
Average increase & 0.159 & 0.142 & 0.142 & 0.185 & 0.185 & 0.172 & 0.164 \\
Average decrease & 0.351 & 0.312 & 0.299 & 0.247 & 0.286 & 0.284 & 0.306 \\ \hline
\multicolumn{8}{l} {
\begin{minipage}{1.12\columnwidth}%
\small \textit{Notes}. Standard errors in parentheses. *** p\textless{}0.01, ** p\textless{}0.05, * p\textless{}0.1. The percentages in the first row report the share of respondents who increased their Likert-based score of trust in people running the above institutions between the first and last wave of the survey. All regressions are OLS regressions that take into account population survey weights and the sampling procedure. The dependent variable is a dummy=1 if the respondent reduced their confidence in the people running the following institutions. The control variables include: gender, race, age, education, parental status, caring responsibilities for an elderly or a person with a disability, baseline income in February 2020, cohabitation with a partner, labor force participation and employment status in February 2020, health insurance provider, if the respondent had financial difficulties before the pandemic, macro-region, metro vs. rural, the population density at the zip code, and two dummy variables indicating if they consume at least 30min a week of international news and if they consume news from social media. We also control for whether respondents completed the survey in a shorter time than the 99$^{th}$ percentile as well as ceiling effects. Finally, we include variables considering the media: whether the sources of news consulted are leaning politically or are neutral, the amount of international news consumed and social media usage. \end{minipage}
}
\end{tabular}%
}
\end{table}

\newpage 

\subsection{Sample balance excluding problematic observations}

Some of the reported incomes in wave 7 were in stark contrast with the ones previously recorded. Therefore, we engaged in a thorough data cleaning process. By comparing the respondent’s and their partner’s reported income with what previously stated, we identified respondents whose earnings were substantially misaligned. We then proceeded with checking whether they had changed marital status, or had gone through some major life changes. We then flagged those for whom we were not able to justify such a large gap. Most of them simply had not reported their incomes or had done so in an apparently random way. Hence, we proceeded to remove them from the sample, assuming that they had not read carefully the questionnaire, potentially compromising our analysis. As shown in table \ref{tab:balancetable_prob}, some demographic characteristics are significantly correlated with reporting a potentially incorrect income, hence this is not a random phenomenon and it might bias our results.

\begin{table}[H]
\centering
\caption{Differences in demographics across individuals with ``problematic'' income and those without irregularities.}
\label{tab:balancetable_prob}
\resizebox*{!}{0.95\textheight}{%
\begin{tabular}{lccc} \hline \hline
 & (1) & (2) & (3) \\
 & Mean non problematic & Mean problematic & Diff \\ \hline
Republican & 0.268 & 0.186 & -0.082 \\
 & (0.443) & (0.393) & (0.059) \\
Democrat & 0.398 & 0.373 & -0.025 \\
 & (0.490) & (0.488) & (0.065) \\
Independent/ non-voter & 0.334 & 0.441 & 0.107* \\
 & (0.472) & (0.501) & (0.063) \\
Woman & 0.478 & 0.424 & -0.055 \\
 & (0.500) & (0.498) & (0.066) \\
Age: 18-29 & 0.224 & 0.220 & -0.003 \\
 & (0.417) & (0.418) & (0.055) \\
Age: 30-44 & 0.270 & 0.458 & 0.188*** \\
 & (0.444) & (0.502) & (0.059) \\
Age: 45-59 & 0.226 & 0.153 & -0.073 \\
 & (0.418) & (0.363) & (0.055) \\
Age: 60+ & 0.281 & 0.169 & -0.111* \\
 & (0.450) & (0.378) & (0.059) \\
Less than HS & 0.031 & 0.085 & 0.054** \\
 & (0.174) & (0.281) & (0.024) \\
High school & 0.161 & 0.186 & 0.026 \\
 & (0.367) & (0.393) & (0.049) \\
Some college & 0.413 & 0.390 & -0.023 \\
 & (0.493) & (0.492) & (0.065) \\
Bachelor + & 0.395 & 0.339 & -0.056 \\
 & (0.489) & (0.477) & (0.065) \\
I income q & 0.175 & 0.847 & 0.672*** \\
 & (0.380) & (0.363) & (0.050) \\
II income q & 0.204 & 0.068 & -0.136** \\
 & (0.403) & (0.254) & (0.053) \\
III income q & 0.207 & 0.017 & -0.190*** \\
 & (0.405) & (0.130) & (0.053) \\
IV income q & 0.208 & 0.017 & -0.191*** \\
 & (0.406) & (0.130) & (0.053) \\
V income q & 0.205 & 0.051 & -0.155*** \\
 & (0.404) & (0.222) & (0.053) \\
\multirow{2}{*}{\begin{tabular}[c]{@{}l@{}}Financial hardship\\ pre-COVID-19\end{tabular}} & 0.281 & 0.305 & 0.024 \\
 & (0.450) & (0.464) & (0.060) \\
African American & 0.111 & 0.153 & 0.041 \\
 & (0.315) & (0.363) & (0.042) \\
Hispanic & 0.157 & 0.271 & 0.114** \\
 & (0.364) & (0.448) & (0.049) \\
Other Race & 0.117 & 0.169 & 0.052 \\
 & (0.322) & (0.378) & (0.043) \\
White & 0.614 & 0.407 & -0.208*** \\
 & (0.487) & (0.495) & (0.065) \\
Cohabitating & 0.543 & 0.559 & 0.016 \\
 & (0.498) & (0.501) & (0.066) \\
Parent of minor & 0.292 & 0.339 & 0.047 \\
 & (0.455) & (0.477) & (0.061) \\
Caring responsibilities & 0.159 & 0.186 & 0.027 \\
 & (0.366) & (0.393) & (0.049) \\
Not in the labor force & 0.259 & 0.000 & -0.259*** \\
 & (0.438) & (0.000) & (0.057) \\
Unemployed in Feb & 0.051 & 0.288 & 0.237*** \\
 & (0.219) & (0.457) & (0.031) \\
North-East & 0.151 & 0.169 & 0.018 \\
 & (0.358) & (0.378) & (0.048) \\
Midwest & 0.247 & 0.220 & -0.026 \\
 & (0.431) & (0.418) & (0.057) \\
South & 0.365 & 0.373 & 0.007 \\
 & (0.482) & (0.488) & (0.064) \\
West & 0.237 & 0.237 & 0.001 \\
 & (0.425) & (0.429) & (0.057) \\
Metropolitan area & 0.865 & 0.881 & 0.017 \\
 & (0.342) & (0.326) & (0.045) \\
No health insurance & 0.076 & 0.220 & 0.144*** \\
 & (0.265) & (0.418) & (0.036) \\
Population density ZCTA & 3,812.121 & 6,875.387 & 3,063.266*** \\
 & (8,453.292) & (16,072.607) & (1,181.512) \\ \hline
Observations & 1,382 & 59 & 1,441 \\ \hline
\end{tabular}%
}
\end{table}

Thus, borrowing from the attrition-related literature, we study whether excluding these observations might bias our results. The probability of providing imprecise information is significantly correlated with income and education, with low-income and less educated respondents being more likely to misreport, and the same holds true for individuals who declared in previous waves to have been unemployed before the pandemic crisis and those with no health insurance. Hence, these observations are not “Missing completely at random (MCAR)”. However, following \cite{fitzgerald1998analysis}, as above, we test whether the attrition generated by excluding such “problematic respondents” is “Missing at random (MAR)”, so if such attrition is correlated with observable characteristics, but not with our outcomes of interest.

\par As above, we perform a set of likelihood-ratio tests comparing models predicting being a ``problematic observation'' including our baseline outcomes or not. The $\chi^2$ and the p-values of such tests are reported in table \ref{tab:predict_problem} and confirm that removing such observations from our sample should not bias results.

\begin{table}[H]
\centering
\caption{Likelihood ratio tests comparing a model predicting being ``problematic'' including the baseline outcome among the independent variables with one who does not}
\label{tab:predict_problem}
\begin{tabular}{lcc} \hline \hline
 \textbf{Baseline outcome} & \begin{tabular}[c]{@{}c@{}}\textbf{Likelihood }\\\textbf{ratio test }\\\textbf{- $\chi^2$}\end{tabular} & \begin{tabular}[c]{@{}c@{}}\textbf{Likelihood }\\\textbf{ratio test~}\\\textbf{- pvalue}\end{tabular} \\ 
\hline
&&\\
\textit{Confidence in people running...}\\
The U.S. Congress and Senate & 0.889 & 0.346 \\
The White House & 1.400 & 0.237 \\
The scientific community & 1.574 & 0.210 \\
Financial institutions & 1.189 & 0.275 \\
The private sector & 0.750 & 0.386 \\
Hospitals & 1.920 & 0.166 \\
Health insurance companies & 0.729 & 0.393 \\
 & & \\
Support for universal healthcare & 1.801 & 0.180 \\
 & & \\
\multicolumn{3}{l}{\textit{It's a government's responsibility to...}} \\
Provide mental healthcare & 0.009 & 0.926 \\
Help those affected by natural disasters & 1.145 & 0.285 \\
Keep prices under control & 0.628 & 0.428 \\
Provide for the elderly & 0.858 & 0.354 \\
Provide for the unemployed & 0.204 & 0.651 \\
Provide a basic income & 0.116 & 0.734 \\
Help industry grow & 0.122 & 0.727 \\
Reduce inequality & 0.169 & 0.681 \\
Pay university for poor & 2.389 & 0.122 \\
 & & \\
\multicolumn{3}{l}{\textit{The government should...}} \\
Transfer money to families and businesses & 0.940 & 0.332 \\
Do more to protect essential workers & 0.069 & 0.793 \\
Spend more on public healthcare & 0.978 & 0.323 \\ \hline
\end{tabular}
\end{table}

\subsubsection{Alternative outcomes}  
\begin{table}[H]
\centering
\caption{The effect of shocks and media on welfare policy preferences}
\label{tab:decr_policy}
\resizebox{\textwidth}{!}{%
\begin{tabular}{lcccccccccc} \hline \hline
 & (1) & (2) & (3) & (4) & (5) & (6) & (7) & (8) & (9) & (10) \\
& \multicolumn{10}{c}{\begin{tabular}[c]{@{}c@{}} Lower support between April and October 2020 that it's a government responsibility to :\end{tabular}} \\
 & \begin{tabular}[c]{@{}c@{}}Universal \\ health care\end{tabular} & \begin{tabular}[c]{@{}c@{}}Industrial \\ growth\end{tabular} & \begin{tabular}[c]{@{}c@{}}Control \\ prices\end{tabular} & Unemployed & \begin{tabular}[c]{@{}c@{}}Mental\\ health care\end{tabular} & Elderly & \begin{tabular}[c]{@{}c@{}}Natural \\ disasters\end{tabular} & \begin{tabular}[c]{@{}c@{}}Basic \\ income\end{tabular} & Inequality & University \\
 & & & & & & & & & & \\ \hline
Republican & 0.0926*** & 0.0448 & 0.0592 & 0.0670* & -0.00209 & 0.00286 & 0.135*** & 0.0559* & 0.00831 & 0.0205 \\
 & (0.0328) & (0.0399) & (0.0404) & (0.0404) & (0.0358) & (0.0343) & (0.0429) & (0.0313) & (0.0492) & (0.0453) \\
Democrat & -0.0700* & -0.0442 & -0.0752** & -0.0928*** & -0.0704* & -0.0328 & 0.00370 & -0.0920*** & -0.0872** & -0.0454 \\
 & (0.0404) & (0.0404) & (0.0335) & (0.0354) & (0.0377) & (0.0394) & (0.0359) & (0.0333) & (0.0434) & (0.0421) \\
Lost 20\% income & -0.0377 & 0.0218 & -0.0402 & 0.00826 & 0.000558 & -0.0358 & -0.0514 & 0.0200 & -0.00114 & -0.0175 \\
 & (0.0367) & (0.0288) & (0.0384) & (0.0324) & (0.0403) & (0.0448) & (0.0348) & (0.0264) & (0.0364) & (0.0343) \\
Knows hospitalized & -0.0489* & 0.0197 & -0.0231 & -0.0850** & 0.0269 & 0.0110 & -0.0434 & -0.0721** & -0.0766** & -0.0677** \\
 & (0.0266) & (0.0349) & (0.0267) & (0.0349) & (0.0393) & (0.0304) & (0.0275) & (0.0298) & (0.0318) & (0.0316) \\
Var consumer exp & 0.0279 & 0.0133 & -0.0440 & 0.00594 & 0.0398 & 0.00936 & -0.00619 & -0.0214 & -0.0516** & 0.00134 \\
 & (0.0215) & (0.0216) & (0.0275) & (0.0227) & (0.0309) & (0.0184) & (0.0428) & (0.0257) & (0.0234) & (0.0333) \\
Incr COVID-19 cases & 0.0226 & 0.0204 & -0.0201 & -0.00851 & 0.00633 & -0.00965 & 0.00768 & -0.0125 & 0.0159 & 0.00860 \\
 & (0.0175) & (0.0219) & (0.0151) & (0.0252) & (0.0158) & (0.0179) & (0.0168) & (0.0169) & (0.0203) & (0.0210) \\
Rep leaning news & 0.0575 & 0.00567 & 0.0510 & 0.0951** & 0.0411 & 0.0437 & 0.0358 & -0.00361 & 0.106* & -0.0212 \\
 & (0.0387) & (0.0413) & (0.0428) & (0.0454) & (0.0492) & (0.0478) & (0.0464) & (0.0434) & (0.0547) & (0.0387) \\
Dem leaning news & -0.0287 & -0.0439 & 0.0533 & -0.0160 & 0.00200 & -0.0521 & 0.0563 & -0.00989 & -0.0215 & -0.0411 \\
 & (0.0378) & (0.0409) & (0.0409) & (0.0371) & (0.0479) & (0.0412) & (0.0405) & (0.0383) & (0.0390) & (0.0400) \\
Constant & 0.527*** & 0.155 & 0.386** & 0.240 & 0.383* & 0.305** & 0.300** & 0.547*** & 0.283** & 0.491*** \\
 & (0.107) & (0.123) & (0.160) & (0.206) & (0.208) & (0.136) & (0.134) & (0.156) & (0.110) & (0.135) \\
 & & & & & & & & & & \\
Observations & 1,010 & 1,002 & 1,005 & 1,002 & 1,010 & 1,004 & 1,007 & 999 & 1,006 & 1,003 \\
R-squared & 0.173 & 0.095 & 0.106 & 0.159 & 0.070 & 0.111 & 0.113 & 0.173 & 0.166 & 0.125 \\
Controls & Yes & Yes & Yes & Yes & Yes & Yes & Yes & Yes & Yes & Yes \\
Average increase & 0.265 & 0.262 & 0.244 & 0.144 & 0.158 & 0.168 & 0.126 & 0.220 & 0.225 & 0.187 \\
Average decrease & 0.210 & 0.244 & 0.233 & 0.325 & 0.251 & 0.254 & 0.224 & 0.233 & 0.231 & 0.215\\ \hline
\multicolumn{11}{l}{%
 \begin{minipage}{1.55\columnwidth}%
  \small \textit{Notes}: Standard errors in parentheses. *** p\textless{}0.01, ** p\textless{}0.05, * p\textless{}0.1.\\
  All regressions are OLS regressions that take into account population survey wights and the sampling procedure. The dependent variable is a dummy=1 if the respondent decreased their belief that it should be the government's responsibility to provide the following policies. The control variables include: gender, race, age, education, parental status, and caring responsibilities for an elderly or a person with a disability, income in February 2020, housing, labor force participation and employment status in February 2020, health insurance provider, and whether respondents had financial difficulties before the pandemic, the area in which the respondents live, whether it's a metropolitan or rural area, and the population density in the zip code. We also control for whether respondents completed the related surveys in a shorter time than the 99$^{th}$ percentile, and ceiling effects.
 \end{minipage}%
}\\ 
\end{tabular}%
}
\end{table}

\begin{table}[H]
\centering
\caption{The effect of shocks and media on temporary relief policies}
\label{tab:decr_covid_policy}
\resizebox{0.8\textwidth}{!}{%
\begin{tabular}{lccc} \hline \hline
& \multicolumn{3}{c}{\begin{tabular}[c]{@{}c@{}} Lower support between May and October 2020 to :\end{tabular}} \\
 & (1) & (2) & (3) \\
 &
 \begin{tabular}[c]{@{}c@{}}Spend more on \\ public healthcare \\ to reduce \\ preventable deaths\end{tabular} &
 \begin{tabular}[c]{@{}c@{}}Do more to \\ protect \\ essential workers\end{tabular} &
 \begin{tabular}[c]{@{}c@{}}Transfer money \\ directly to families \\ and businesses \end{tabular} \\ \hline
 & & & \\
Republican & 0.0126 & -0.0334 & 0.00861 \\
 & (0.0614) & (0.0605) & (0.0465) \\
Democrat & -0.0991** & -0.110** & -0.0248 \\
 & (0.0423) & (0.0430) & (0.0356) \\
Lost 20\% income & -0.0614** & -0.0179 & -0.0345 \\
 & (0.0305) & (0.0434) & (0.0434) \\
Knows hospitalized & -0.00322 & -0.0167 & -0.0484 \\
 & (0.0430) & (0.0420) & (0.0420) \\
Var consumer expenditures & 0.0156 & -0.0180* & 0.00570 \\
 & (0.00948) & (0.0106) & (0.0127) \\
Incr COVID-19 cases & -0.0362* & -0.00335 & 0.0161 \\
 & (0.0193) & (0.0209) & (0.0198) \\
Republican leaning news & 0.0890 & -0.0217 & 0.0195 \\
 & (0.0597) & (0.0482) & (0.0541) \\
Democratic leaning news & 0.0613 & 0.0185 & 0.00416 \\
 & (0.0473) & (0.0592) & (0.0500) \\
Constant & 0.0822 & 0.328* & 0.233 \\
 & (0.121) & (0.183) & (0.152) \\
 & & & \\
Observations & 937 & 938 & 942 \\
R-squared & 0.108 & 0.089 & 0.123 \\
Controls & Yes & Yes & Yes \\
Average increase & 0.177 & 0.188 & 0.181 \\
Average decrease & 0.295 & 0.317 & 0.369 \\ \hline
\multicolumn{4}{l}{%
 \begin{minipage}{\columnwidth}%
  \small \textit{Notes}: Standard errors in parentheses. *** p\textless{}0.01, ** p\textless{}0.05, * p\textless{}0.1.\\
  All regressions are OLS regressions that take into account population survey wights and the sampling procedure. The dependent variable is a dummy=1 if the respondent decreased their support for the following policies. The control variables include: gender, race, age, education, parental status, and caring responsibilities for an elderly or a person with a disability, income in February 2020, housing, labor force participation and employment status in February 2020, health insurance provider, and whether respondents had financial difficulties before the pandemic, the area in which the respondents live, whether it's a metropolitan or rural area, and the population density in the zip code. We also control for whether respondents completed the related surveys in shorter time than the 99$^{th}$ percentile, and ceiling effects.
 \end{minipage}%
}\\ 
\end{tabular}%
}
\end{table}

\begin{table}[H]
\centering
\caption{The effect of shocks and media on trust in institutions}
\label{tab:incr_trust}
\resizebox{\textwidth}{!}{%
\begin{tabular}{lccccccc} \hline \hline
 & \multicolumn{7}{c}{Increased confidence in people running the following institutions:} \\
 & (1) & (2) & (3) & (4) & (5) & (6) & (7) \\
 & \begin{tabular}[c]{@{}c@{}}Congress \\ \& Senate\end{tabular} & \begin{tabular}[c]{@{}c@{}}White\\ House\end{tabular} & \begin{tabular}[c]{@{}c@{}}Financial\\ institutions\\ \& banks\end{tabular} & \multicolumn{1}{l}{\begin{tabular}[c]{@{}l@{}}Private\\ sector\end{tabular}} & \begin{tabular}[c]{@{}c@{}}Scientific\\ community\end{tabular} & \begin{tabular}[c]{@{}c@{}}Health \\ insurance\\ companies\end{tabular} & Hospitals \\ \hline
 & & & & & & & \\
Republican & 0.0993** & 0.0835** & -0.0282 & 0.0583 & -0.0164 & 0.00667 & 0.0120 \\
 & (0.0400) & (0.0410) & (0.0299) & (0.0361) & (0.0335) & (0.0383) & (0.0274) \\
Democrat & -0.0360 & -0.0904*** & -0.0340 & -0.00803 & 0.128*** & 0.0698** & 0.00902 \\
 & (0.0279) & (0.0241) & (0.0415) & (0.0425) & (0.0400) & (0.0335) & (0.0344) \\
Lost 20\% income & -0.0805*** & -0.00858 & -0.0227 & -0.0686** & 0.0272 & -0.0312 & 0.00872 \\
 & (0.0308) & (0.0245) & (0.0256) & (0.0292) & (0.0259) & (0.0294) & (0.0246) \\
Knows hospitalized & -0.0455** & 0.00549 & 0.0175 & 0.00614 & -0.0113 & -0.0563** & -0.00141 \\
 & (0.0229) & (0.0240) & (0.0335) & (0.0339) & (0.0306) & (0.0237) & (0.0301) \\
Var consumer expenditures & -0.0311 & -0.0294 & -0.0158 & 0.00985 & 0.00741 & 0.0120 & 0.0498** \\
 & (0.0321) & (0.0181) & (0.0149) & (0.0176) & (0.0180) & (0.0197) & (0.0212) \\
Incr COVID-19 cases & -0.00608 & 0.0202 & 0.0172 & -0.0156 & -0.0181 & -0.00496 & 0.0113 \\
 & (0.0123) & (0.0145) & (0.0146) & (0.0182) & (0.0163) & (0.0189) & (0.0177) \\
Republican leaning news & 0.0595* & 0.00690 & -0.0207 & -0.0189 & -0.0985*** & 0.0394 & -0.0732** \\
 & (0.0356) & (0.0428) & (0.0314) & (0.0397) & (0.0342) & (0.0378) & (0.0315) \\
Democratic leaning news & 0.0130 & -0.0865*** & 0.0226 & -0.000460 & -0.0478 & -0.0809** & -0.0466 \\
 & (0.0306) & (0.0303) & (0.0399) & (0.0383) & (0.0323) & (0.0386) & (0.0378) \\
Constant & 0.298* & 0.195* & 0.0111 & 0.136 & 0.0624 & 0.0511 & 0.144 \\
 & (0.161) & (0.103) & (0.105) & (0.129) & (0.111) & (0.136) & (0.148) \\
 & & & & & & & \\
Observations & 1,009 & 1,003 & 1,007 & 1,006 & 1,002 & 1,006 & 1,007 \\
R-squared & 0.116 & 0.177 & 0.079 & 0.072 & 0.185 & 0.107 & 0.116 \\
Controls & Yes & Yes & Yes & Yes & Yes & Yes & Yes \\
Avg increase dep. var. & 0.159 & 0.142 & 0.142 & 0.185 & 0.185 & 0.172 & 0.164 \\
Avg decrease dep. var. & 0.351 & 0.312 & 0.299 & 0.247 & 0.286 & 0.284 & 0.306 \\ \hline
\multicolumn{8}{l} {
\begin{minipage}{1.3\columnwidth}%
\small \textit{Notes}. Standard errors in parentheses. *** p\textless{}0.01, ** p\textless{}0.05, * p\textless{}0.1. The percentages in the first row report the share of respondents who increased their Likert-based score of trust in people running the above institutions between the first and last wave of the survey. All regressions are OLS regressions that take into account population survey weights and the sampling procedure. The dependent variable is a dummy=1 if the respondent increased their confidence in the people running the following institutions. The control variables include: gender, race, age, education, parental status, caring responsibilities for an elderly or a person with a disability, baseline income in February 2020, cohabitation with a partner, labor force participation and employment status in February 2020, health insurance provider, if the respondent had financial difficulties before the pandemic, macro-region, metro vs. rural, the population density at the zip code, and two dummy variables indicating if they consume at least 30min a week of international news and if they consume news from social media. We also control for whether respondents completed the survey in a shorter time than the 99$^{th}$ percentile as well as ceiling effects.\end{minipage}
}
\end{tabular}%
}
\end{table}

\subsubsection{Entropy weights}
\begin{table}[H]
\centering
\caption{Difference in the average demographic characteristics between people having lost at least 20\% of their household income between any two months from March 2020 to October 2020 and those who have not.}
\label{tab:balancetable_shock}
\resizebox*{!}{0.95\textheight}{%
\begin{tabular}{lccc} \hline \hline
 & (1) & (2) & (3) \\ 
& Mean no shock & Mean shocks & Difference \\ \hline
Republican & 0.266 & 0.246 & -0.020 \\
 & (0.442) & (0.431) & (0.028) \\
Democrat & 0.364 & 0.406 & 0.042 \\
 & (0.482) & (0.492) & (0.031) \\
Independent/ non-voter & 0.370 & 0.348 & -0.022 \\
 & (0.483) & (0.477) & (0.031) \\
Woman & 0.444 & 0.490 & 0.046 \\
 & (0.497) & (0.501) & (0.032) \\
Age: 18-29 & 0.167 & 0.262 & 0.095*** \\
 & (0.373) & (0.440) & (0.025) \\
Age: 30-44 & 0.290 & 0.285 & -0.004 \\
 & (0.454) & (0.452) & (0.029) \\
Age: 45-59 & 0.239 & 0.194 & -0.045* \\
 & (0.427) & (0.396) & (0.027) \\
Age: 60+ & 0.304 & 0.259 & -0.045 \\
 & (0.460) & (0.439) & (0.029) \\
Less than HS & 0.030 & 0.024 & -0.007 \\
 & (0.171) & (0.152) & (0.010) \\
High school & 0.156 & 0.178 & 0.022 \\
 & (0.363) & (0.383) & (0.024) \\
Some college & 0.383 & 0.385 & 0.002 \\
 & (0.487) & (0.487) & (0.031) \\
Bachelor + & 0.431 & 0.414 & -0.017 \\
 & (0.496) & (0.493) & (0.032) \\
I income q & 0.222 & 0.236 & 0.014 \\
 & (0.416) & (0.425) & (0.027) \\
II income q & 0.140 & 0.230 & 0.091*** \\
 & (0.347) & (0.422) & (0.024) \\
III income q & 0.184 & 0.196 & 0.012 \\
 & (0.388) & (0.398) & (0.025) \\
IV income q & 0.205 & 0.202 & -0.003 \\
 & (0.404) & (0.402) & (0.026) \\
V income q & 0.249 & 0.136 & -0.113*** \\
 & (0.433) & (0.343) & (0.026) \\
\begin{tabular}[c]{@{}l@{}}Financial hardship\\ pre-COVID-19\end{tabular} & 0.252 & 0.338 & 0.086*** \\
 & (0.434) & (0.474) & (0.029) \\
African American & 0.097 & 0.110 & 0.013 \\
 & (0.296) & (0.313) & (0.019) \\
Hispanic & 0.121 & 0.170 & 0.049** \\
 & (0.326) & (0.376) & (0.022) \\
Other Race & 0.127 & 0.110 & -0.017 \\
 & (0.333) & (0.313) & (0.021) \\
White & 0.656 & 0.610 & -0.046 \\
 & (0.476) & (0.488) & (0.031) \\
Cohabitating & 0.670 & 0.568 & -0.102*** \\
 & (0.471) & (0.496) & (0.031) \\
Parent of minor & 0.284 & 0.291 & 0.007 \\
 & (0.451) & (0.455) & (0.029) \\
Caring responsibilities & 0.141 & 0.186 & 0.045* \\
 & (0.348) & (0.390) & (0.023) \\
Not in the labor force & 0.291 & 0.181 & -0.110*** \\
 & (0.455) & (0.385) & (0.027) \\
Unemployed in Feb & 0.049 & 0.063 & 0.014 \\
 & (0.216) & (0.243) & (0.014) \\
North-East & 0.147 & 0.152 & 0.005 \\
 & (0.354) & (0.359) & (0.023) \\
Midwest & 0.275 & 0.246 & -0.029 \\
 & (0.447) & (0.431) & (0.028) \\
South & 0.353 & 0.374 & 0.021 \\
 & (0.478) & (0.485) & (0.031) \\
West & 0.225 & 0.228 & 0.003 \\
 & (0.418) & (0.420) & (0.027) \\
Metropolitan area & 0.847 & 0.887 & 0.040* \\
 & (0.360) & (0.316) & (0.022) \\
No health insurance & 0.058 & 0.098 & 0.040** \\
 & (0.234) & (0.297) & (0.017) \\
Population density ZCTA & 3,951.735 & 3,940.752 & -10.983 \\
 & (10,430.189) & (7,635.053) & (607.622) \\ \hline
Observations & 694 & 382 & 1,076 \\ \hline
\end{tabular}%
}
\end{table}

\begin{table}[H]
\centering
\caption{The effect of shocks and media on welfare policy preferences, with entropy weights}
\label{tab:policy_entropy}
\resizebox{\textwidth}{!}{%
\begin{tabular}{lcccccccccc} \hline \hline
 & (1) & (2) & (3) & (4) & (5) & (6) & (7) & (8) & (9) & (10) \\
 & \begin{tabular}[c]{@{}c@{}}Universal \\ healthcare\end{tabular} & \begin{tabular}[c]{@{}c@{}}Provide\\ basic \\ income\end{tabular} & \begin{tabular}[c]{@{}c@{}}Provide\\ for\\ unemployed\end{tabular} & \begin{tabular}[c]{@{}c@{}}Provide\\ for the\\ elderly\end{tabular} & \begin{tabular}[c]{@{}c@{}}Reduce\\ inequality\end{tabular} & \begin{tabular}[c]{@{}c@{}}Help\\ low-income\\ students\end{tabular} & \begin{tabular}[c]{@{}c@{}}Help \\ disasters'\\ victims\end{tabular} & \begin{tabular}[c]{@{}c@{}}Provide\\ Mental \\ healthcare\end{tabular} & \begin{tabular}[c]{@{}c@{}}Control \\ prices\end{tabular} & \begin{tabular}[c]{@{}c@{}}Help\\ industry\\ grow\end{tabular} \\ \hline
 & & & & & & & & & & \\
Republican & -0.00101 & 0.0675 & -0.00611 & -0.0444 & 0.00921 & 0.0233 & 0.00465 & 0.0126 & 0.0120 & -0.00327 \\
 & (0.0452) & (0.0447) & (0.0364) & (0.0366) & (0.0380) & (0.0389) & (0.0356) & (0.0474) & (0.0438) & (0.0304) \\
Democrat & 0.0277 & 0.106*** & 0.0371 & 0.0245 & 0.0211 & -0.00586 & -0.0316 & 0.0418 & 0.0907** & 0.0531 \\
 & (0.0355) & (0.0329) & (0.0350) & (0.0330) & (0.0324) & (0.0294) & (0.0207) & (0.0398) & (0.0432) & (0.0343) \\
Lost 20\% income & 0.0394 & -0.0710** & -0.0115 & 0.0230 & -0.0292 & 0.0473* & 0.0189 & 0.0423 & 0.0273 & -0.00824 \\
 & (0.0370) & (0.0282) & (0.0293) & (0.0313) & (0.0215) & (0.0271) & (0.0307) & (0.0271) & (0.0287) & (0.0270) \\
Knows hospitalized & 0.0108 & -0.00149 & 0.0421 & -0.00672 & 0.0312 & 0.0582** & 0.00142 & -0.0137 & -0.00910 & 0.0360 \\
 & (0.0360) & (0.0334) & (0.0327) & (0.0253) & (0.0291) & (0.0242) & (0.0224) & (0.0361) & (0.0352) & (0.0268) \\
Consumer exp - Apr & -0.104 & -0.0471 & -0.488** & -0.0688 & -0.0939 & 0.112 & 0.0860 & -0.362 & -0.0231 & -0.302 \\
 & (0.275) & (0.260) & (0.231) & (0.204) & (0.222) & (0.252) & (0.192) & (0.267) & (0.244) & (0.202) \\
Var consumer expenditures & -0.0492** & -0.00574 & 0.0349* & -0.00225 & -0.0116 & 0.00536 & 0.0409** & 0.00502 & 0.0495*** & 0.0147 \\
 & (0.0198) & (0.0187) & (0.0194) & (0.0163) & (0.0215) & (0.0218) & (0.0166) & (0.0222) & (0.0175) & (0.0166) \\
Incr COVID-19 cases & -0.0123 & -0.0237 & -0.0263 & 0.00922 & 0.00630 & -0.0127 & -0.0159 & 0.0279 & -0.0135 & -0.000533 \\
 & (0.0215) & (0.0182) & (0.0165) & (0.0165) & (0.0174) & (0.0162) & (0.0125) & (0.0192) & (0.0177) & (0.0154) \\
Rep leaning news & -0.155*** & -0.0298 & 0.00209 & -0.00629 & -0.0821** & -0.102*** & -0.000805 & -0.0871** & -0.0429 & -0.0457 \\
 & (0.0481) & (0.0429) & (0.0505) & (0.0422) & (0.0407) & (0.0345) & (0.0282) & (0.0336) & (0.0446) & (0.0464) \\
Dem leaning news & 0.0123 & 0.0450 & 0.0236 & -0.00231 & -0.0654** & -0.00787 & 0.0151 & 0.0319 & 0.00811 & -0.0433 \\
 & (0.0409) & (0.0422) & (0.0348) & (0.0284) & (0.0259) & (0.0312) & (0.0277) & (0.0342) & (0.0426) & (0.0340) \\
Constant & 0.136 & 0.0564 & 0.0913 & 0.0945 & 0.471*** & 0.350*** & 0.471*** & -0.0126 & 0.237 & 0.307* \\
 & (0.143) & (0.161) & (0.128) & (0.107) & (0.103) & (0.124) & (0.147) & (0.120) & (0.165) & (0.181) \\
 & & & & & & & & & & \\
Observations & 1,010 & 1,002 & 1,005 & 1,002 & 1,010 & 1,004 & 1,007 & 999 & 1,006 & 1,003 \\
R-squared & 0.202 & 0.217 & 0.325 & 0.175 & 0.293 & 0.286 & 0.403 & 0.223 & 0.237 & 0.256 \\
Controls & Yes & Yes & Yes & Yes & Yes & Yes & Yes & Yes & Yes & Yes \\
Average increase & 0.282 & 0.247 & 0.239 & 0.144 & 0.162 & 0.173 & 0.128 & 0.226 & 0.235 & 0.196 \\
Average decrease & 0.202 & 0.250 & 0.226 & 0.334 & 0.245 & 0.250 & 0.215 & 0.238 & 0.226 & 0.198 \\ \hline
\multicolumn{11}{l}{%
 \begin{minipage}{1.65\columnwidth}%
  \small \textit{Notes}: Standard errors in parentheses. *** p\textless{}0.01, ** p\textless{}0.05, * p\textless{}0.1.\\
   All regressions are OLS regressions. Observations have been re-weighted with entropy weights, so that the group of individuals who incurred an income shock and the group that did not are balanced in terms of a set of demographics. The dependent variable is a dummy=1 if the respondent increased their belief that it's a government's responsibility to provide the following policies. The control variables include: gender, race, age, education, parental status, and caring responsibilities for an elderly or a person with a disability, income in February 2020, housing, labor force participation and employment status in February 2020, health insurance provider, and whether respondents had financial difficulties before the pandemic, the area in which the respondents live, whether it's a metropolitan or rural area, and the population density in the zip code. We also control for whether respondents completed the related surveys in shorter time than the 99$^{th}$ percentile, and ceiling effects.
 \end{minipage}%
}\\ 
\end{tabular}%
}
\end{table}

\begin{table}[H]
\centering
\caption{The effect of shocks and media on support for coronavirus relief policies, with entropy weights}
\label{tab:covid_policy_entropy}
\resizebox{0.82\textwidth}{!}{%
\begin{tabular}{lccc} \hline \hline
 &
 \multicolumn{3}{c}{\begin{tabular}[c]{@{}c@{}}Stronger belief between May and October 2020 \\ that the government should:\end{tabular}} \\
 & (1) & (2) & (3) \\
 &
 \begin{tabular}[c]{@{}c@{}}Spend more on \\ public healthcare \\ to reduce \\ preventable deaths\end{tabular} &
 \begin{tabular}[c]{@{}c@{}}Do more to \\ protect \\ essential workers\end{tabular} &
 \begin{tabular}[c]{@{}c@{}}Transfer money \\ directly to families \\ and businesses \end{tabular} \\ \hline
 & & & \\
Republican & -0.0746 & 0.0388 & -0.0396 \\
 & (0.0546) & (0.0448) & (0.0437) \\
Democrat & 0.126*** & 0.0553* & 0.0114 \\
 & (0.0418) & (0.0333) & (0.0391) \\
Lost 20\% income & 0.0526 & 0.0723** & 0.0531 \\
 & (0.0330) & (0.0330) & (0.0330) \\
Knows hospitalized & 0.0253 & -0.00846 & -0.00535 \\
 & (0.0350) & (0.0245) & (0.0287) \\
Var consumer expenditures & 0.0109* & 0.0106 & 0.0186** \\
 & (0.00569) & (0.0101) & (0.00788) \\
Consumer exp - May & -0.0352 & -0.0655 & -0.0292 \\
 & (0.202) & (0.160) & (0.166) \\
Incr COVID-19 cases & 0.0206 & 0.00730 & -0.00836 \\
 & (0.0185) & (0.0186) & (0.0165) \\
Rep leaning news & -0.000672 & -0.122*** & -0.0160 \\
 & (0.0481) & (0.0402) & (0.0388) \\
Dem leaning news & -0.0409 & 0.0198 & 0.00916 \\
 & (0.0302) & (0.0354) & (0.0388) \\
Constant & 0.288* & 0.112 & 0.133 \\
 & (0.162) & (0.137) & (0.0982) \\
 & & & \\
Observations & 937 & 938 & 942 \\
R-squared & 0.246 & 0.272 & 0.162 \\
Controls & Yes & Yes & Yes \\
Average increase & 0.192 & 0.193 & 0.182 \\
Average decrease & 0.265 & 0.304 & 0.349 \\ \hline
\multicolumn{4}{l}{%
 \begin{minipage}{\columnwidth}%
  \small \textit{Notes}: Standard errors in parentheses. *** p\textless{}0.01, ** p\textless{}0.05, * p\textless{}0.1.\\
  All regressions are OLS regressions. Observations have been re-weighted with entropy weights, so that the group of individuals who incurred an income shock and the group that did not are balanced in terms of a set of demographics. The dependent variable is a dummy=1 if the respondent increased their support for the following policies. The control variables include: gender, race, age, education, parental status, and caring responsibilities for an elderly or a person with a disability, income in February 2020, housing, labor force participation and employment status in February 2020, health insurance provider, and whether respondents had financial difficulties before the pandemic, the area in which the respondents live, whether it's a metropolitan or rural area, and the population density in the zip code. We also control for whether respondents completed the related surveys in shorter time than the 99$^{th}$ percentile, and ceiling effects.
 \end{minipage}%
}\\ 
\end{tabular}%
}
\end{table}

\begin{table}[H]
\centering
\caption{The effect of shocks and media on institutional trust, with entropy weights}
\label{tab:trust_entropy}
\resizebox{\textwidth}{!}{%
\begin{tabular}{lccccccc} \hline \hline
 & (1) & (2) & (3) & (4) & (5) & (6) & (7) \\
 & \begin{tabular}[c]{@{}c@{}}Congress \\ \& Senate\end{tabular} & \begin{tabular}[c]{@{}c@{}}White\\ House\end{tabular} & \begin{tabular}[c]{@{}c@{}}Financial\\ institutions\end{tabular} & \begin{tabular}[c]{@{}c@{}}Private\\ sector\end{tabular} & \begin{tabular}[c]{@{}c@{}}Scientific\\ community\end{tabular} & \begin{tabular}[c]{@{}c@{}}Health \\ insurance \\ companies\end{tabular} & Hospitals \\ \hline
 & & & & & & & \\
Republican & -0.101** & -0.149*** & 0.0801 & 0.0350 & 0.00470 & -0.0680 & 0.0492 \\
 & (0.0453) & (0.0384) & (0.0500) & (0.0486) & (0.0460) & (0.0502) & (0.0447) \\
Democrat & 0.0198 & 0.0813** & 0.0456 & 0.0235 & -0.108** & -0.0555 & -0.0456 \\
 & (0.0419) & (0.0367) & (0.0483) & (0.0458) & (0.0431) & (0.0440) & (0.0401) \\
Lost 20\% income & 0.0613 & 0.0230 & 0.0642* & 0.0604* & -0.0226 & 0.0360 & 0.0453 \\
 & (0.0387) & (0.0298) & (0.0342) & (0.0319) & (0.0289) & (0.0374) & (0.0345) \\
Knows hospitalized & 0.0423 & 0.0496 & 0.00730 & 0.0305 & 0.0582 & 0.0281 & 0.0274 \\
 & (0.0392) & (0.0370) & (0.0399) & (0.0418) & (0.0483) & (0.0400) & (0.0391) \\
Var consumer expenditures & 0.00381 & 0.0399* & -0.00335 & -0.0319 & 0.0210 & -0.0755*** & -0.0694** \\
 & (0.0364) & (0.0203) & (0.0316) & (0.0223) & (0.0358) & (0.0226) & (0.0287) \\
Consumer exp - Apr & 0.0773 & -0.110 & 0.120 & 0.0417 & 0.0353 & -0.0248 & 0.408* \\
 & (0.227) & (0.224) & (0.281) & (0.243) & (0.294) & (0.286) & (0.228) \\
Incr COVID-19 cases & -0.00307 & -0.0140 & 0.00569 & 0.0209 & 0.0431* & -0.00472 & -0.0186 \\
 & (0.0238) & (0.0181) & (0.0215) & (0.0205) & (0.0224) & (0.0223) & (0.0253) \\
Rep leaning news & -0.137*** & -0.142*** & -0.0800 & -0.0577 & 0.0570 & 0.0592 & 0.134** \\
 & (0.0514) & (0.0540) & (0.0532) & (0.0401) & (0.0601) & (0.0663) & (0.0543) \\
Dem leaning news & -0.0352 & 0.0236 & -0.0105 & -0.0437 & -0.0195 & 0.0630 & 0.0267 \\
 & (0.0389) & (0.0325) & (0.0465) & (0.0428) & (0.0462) & (0.0497) & (0.0454) \\
Constant & 0.497*** & 0.717*** & 0.435** & -0.0354 & 0.356* & 0.190 & 0.422*** \\
 & (0.186) & (0.140) & (0.171) & (0.123) & (0.194) & (0.175) & (0.156) \\
 & & & & & & & \\
Observations & 1,009 & 1,003 & 1,007 & 1,006 & 1,002 & 1,006 & 1,007 \\
R-squared & 0.202 & 0.398 & 0.116 & 0.152 & 0.103 & 0.141 & 0.132 \\
Controls & Yes & Yes & Yes & Yes & Yes & Yes & Yes \\
Average increase & 0.156 & 0.139 & 0.147 & 0.172 & 0.194 & 0.176 & 0.167 \\
Average decrease & 0.360 & 0.323 & 0.308 & 0.264 & 0.278 & 0.298 & 0.326 \\ \hline
\multicolumn{8}{l}{%
 \begin{minipage}{1.3\columnwidth}%
  \small \textit{Notes}: Standard errors in parentheses. *** p\textless{}0.01, ** p\textless{}0.05, * p\textless{}0.1.\\
  All regressions are OLS regressions. Observations have been re-weighted with entropy weights, so that the group of individuals who incurred an income shock and the group that did not are balanced in terms of a set of demographics. The dependent variable is a dummy=1 if the respondent decreased trust in the above institutions. The control variables include: gender, race, age, education, parental status, and caring responsibilities for an elderly or a person with a disability, income in February 2020, housing, labor force participation and employment status in February 2020, health insurance provider, and whether respondents had financial difficulties before the pandemic, the area in which the respondents live, whether it's a metropolitan or rural area, and the population density in the zip code. We also control for whether respondents completed the related surveys in shorter time than the 99$^{th}$ percentile, and ceiling effects.
 \end{minipage}%
 }
\end{tabular}%
}
\end{table}

\subsubsection{Voting intentions}
\begin{table}[H]
\centering
\caption{The effect of shocks and media on welfare policy preferences, considering voting intentions}
\label{tab:policy_vote}
\resizebox{\textwidth}{!}{%
\begin{tabular}{lcccccccccc} \hline \hline
 & (1) & (2) & (3) & (4) & (5) & (6) & (7) & (8) & (9) & (10) \\
 & \begin{tabular}[c]{@{}c@{}}Universal \\ healthcare\end{tabular} & \begin{tabular}[c]{@{}c@{}}Provide\\ basic \\ income\end{tabular} & \begin{tabular}[c]{@{}c@{}}Provide\\ for\\ unemployed\end{tabular} & \begin{tabular}[c]{@{}c@{}}Provide\\ for the\\ elderly\end{tabular} & \begin{tabular}[c]{@{}c@{}}Reduce\\ inequality\end{tabular} & \begin{tabular}[c]{@{}c@{}}Help\\ low-income\\ students\end{tabular} & \begin{tabular}[c]{@{}c@{}}Help \\ disasters'\\ victims\end{tabular} & \begin{tabular}[c]{@{}c@{}}Provide\\ Mental \\ healthcare\end{tabular} & \begin{tabular}[c]{@{}c@{}}Control \\ prices\end{tabular} & \begin{tabular}[c]{@{}c@{}}Help\\ industry\\ grow\end{tabular} \\ \hline
 & & & & & & & & & & \\
Vote Trump & -0.0192 & -0.0104 & 0.00930 & -0.0844 & -0.0206 & -0.0349 & -0.0527 & -0.103* & -0.0967* & -0.00126 \\
 & (0.0501) & (0.0495) & (0.0488) & (0.0519) & (0.0439) & (0.0393) & (0.0389) & (0.0526) & (0.0516) & (0.0547) \\
Vote Biden & 0.0715 & 0.0146 & 0.0885** & 0.0438 & 0.0219 & 0.0434 & 0.0141 & -0.00890 & 0.0573 & 0.0512 \\
 & (0.0453) & (0.0435) & (0.0394) & (0.0417) & (0.0372) & (0.0377) & (0.0293) & (0.0470) & (0.0491) & (0.0367) \\
Lost 20\% income & 0.0296 & -0.0735** & -0.00396 & 0.0213 & -0.0221 & 0.0404 & 0.0157 & 0.0425 & 0.0395 & 0.00831 \\
 & (0.0355) & (0.0283) & (0.0313) & (0.0310) & (0.0218) & (0.0290) & (0.0315) & (0.0263) & (0.0259) & (0.0273) \\
Knows hospitalized & 0.0172 & -0.0263 & 0.0437 & -0.0216 & 0.0178 & 0.0510* & 0.00711 & -0.0137 & -0.0253 & 0.0333 \\
 & (0.0305) & (0.0331) & (0.0297) & (0.0235) & (0.0272) & (0.0259) & (0.0257) & (0.0382) & (0.0329) & (0.0255) \\
Consumer exp - Apr & -0.177 & -0.0712 & -0.444** & -0.0673 & -0.113 & 0.227 & -0.110 & -0.297 & -0.0852 & -0.357** \\
 & (0.263) & (0.250) & (0.211) & (0.181) & (0.195) & (0.277) & (0.185) & (0.270) & (0.222) & (0.173) \\
Var consumer expenditures & -0.0489** & -0.0216 & 0.0365** & -0.00438 & -0.0127 & 0.0100 & 0.0366** & -0.00814 & 0.0249 & 0.00375 \\
 & (0.0211) & (0.0232) & (0.0184) & (0.0173) & (0.0169) & (0.0221) & (0.0156) & (0.0198) & (0.0164) & (0.0166) \\
Incr COVID-19 cases & -0.00722 & -0.0111 & -0.0231 & 0.00894 & 0.00212 & -0.00413 & -0.00695 & 0.0200 & -0.0185 & 0.00273 \\
 & (0.0196) & (0.0157) & (0.0141) & (0.0159) & (0.0160) & (0.0152) & (0.0103) & (0.0177) & (0.0148) & (0.0138) \\
Rep leaning news & -0.106** & -0.0271 & 0.00715 & 0.0144 & -0.0818** & -0.0745*** & 0.0490* & -0.0506 & -0.0445 & -0.0744 \\
 & (0.0487) & (0.0460) & (0.0467) & (0.0405) & (0.0370) & (0.0244) & (0.0289) & (0.0352) & (0.0400) & (0.0489) \\
Dem leaning news & 0.0117 & 0.0506 & 0.00769 & -0.0201 & -0.0605** & -0.0445 & -0.00203 & 0.0129 & 0.00440 & -0.0335 \\
 & (0.0394) & (0.0417) & (0.0347) & (0.0254) & (0.0265) & (0.0283) & (0.0299) & (0.0306) & (0.0420) & (0.0364) \\
Constant & 0.146 & 0.189 & 0.137 & 0.106 & 0.410*** & 0.368*** & 0.410*** & 0.121 & 0.291** & 0.249 \\
 & (0.131) & (0.178) & (0.126) & (0.0949) & (0.109) & (0.122) & (0.142) & (0.127) & (0.143) & (0.156) \\
 & & & & & & & & & & \\
Observations & 1,012 & 1,004 & 1,007 & 1,004 & 1,012 & 1,006 & 1,009 & 1,001 & 1,008 & 1,005 \\
R-squared & 0.188 & 0.195 & 0.302 & 0.167 & 0.263 & 0.260 & 0.371 & 0.208 & 0.226 & 0.226 \\
Controls & Yes & Yes & Yes & Yes & Yes & Yes & Yes & Yes & Yes & Yes \\
Average increase & 0.265 & 0.262 & 0.244 & 0.144 & 0.158 & 0.168 & 0.126 & 0.220 & 0.225 & 0.187 \\
Average decrease & 0.210 & 0.244 & 0.233 & 0.325 & 0.251 & 0.254 & 0.224 & 0.233 & 0.231 & 0.215\\ \hline
\multicolumn{11}{l}{%
 \begin{minipage}{1.6\columnwidth}%
  \small \textit{Notes}: Standard errors in parentheses. *** p\textless{}0.01, ** p\textless{}0.05, * p\textless{}0.1.\\
  All regressions are OLS regressions that take into account population survey wights and the sampling procedure. The dependent variable is a dummy=1 if the respondent increased their belief that it's a government's responsibility to provide the following policies. The control variables include: gender, race, age, education, parental status, and caring responsibilities for an elderly or a person with a disability, income in February 2020, housing, labor force participation and employment status in February 2020, health insurance provider, and whether respondents had financial difficulties before the pandemic, the area in which the respondents live, whether it's a metropolitan or rural area, and the population density in the zip code. We also control for whether respondents completed the related surveys in shorter time than the 99$^{th}$ percentile, and ceiling effects.
 \end{minipage}%
}\\ 
\end{tabular}%
}
\end{table}

\begin{table}[H]
\centering
\caption{The effect of shocks and media on support for coronavirus relief policies, considering voting intentions}
\label{tab:covid_policy_vote}
\resizebox{0.83\textwidth}{!}{%
\begin{tabular}{lccc} \hline \hline
 &
 \multicolumn{3}{c}{\begin{tabular}[c]{@{}c@{}}Stronger belief between May and October 2020 \\ that the government should:\end{tabular}} \\
 & (1) & (2) & (3) \\
 &
 \begin{tabular}[c]{@{}c@{}}Spend more on \\ public healthcare \\ to reduce \\ preventable deaths\end{tabular} &
 \begin{tabular}[c]{@{}c@{}}Do more to \\ protect \\ essential workers\end{tabular} &
 \begin{tabular}[c]{@{}c@{}}Transfer money \\ directly to families \\ and businesses \end{tabular} \\ \hline
 & & & \\
Vote Biden & 0.0584 & 0.0442 & 0.0439 \\
 & (0.0408) & (0.0448) & (0.0470) \\
Vote Trump & -0.0505 & -0.0160 & -0.0824* \\
 & (0.0554) & (0.0586) & (0.0418) \\
Lost 20\% income & 0.0685** & 0.0745** & 0.0515* \\
 & (0.0311) & (0.0339) & (0.0297) \\
Knows hospitalized & -0.00551 & -0.0274 & -0.0239 \\
 & (0.0303) & (0.0227) & (0.0271) \\
Var consumer expenditures & 0.0132** & 0.00992 & 0.0105 \\
 & (0.00517) & (0.00988) & (0.00651) \\
Consumer exp - May & -0.0749 & 0.0332 & -0.00547 \\
 & (0.160) & (0.150) & (0.134) \\
Incr COVID-19 cases & 0.0152 & 0.00912 & -0.00963 \\
 & (0.0154) & (0.0158) & (0.0140) \\
Rep leaning news & -0.0164 & -0.0822** & -0.00142 \\
 & (0.0513) & (0.0398) & (0.0406) \\
Dem leaning news & -0.00259 & 0.00661 & -0.0272 \\
 & (0.0365) & (0.0365) & (0.0375) \\
Constant & 0.239 & 0.174 & 0.212** \\
 & (0.164) & (0.154) & (0.101) \\
 & & & \\
Observations & 939 & 940 & 944 \\
R-squared & 0.191 & 0.234 & 0.144 \\
Controls & Yes & Yes & Yes \\
Average increase & 0.177 & 0.188 & 0.181 \\
Average decrease & 0.295 & 0.317 & 0.369 \\\hline
\multicolumn{4}{l}{%
 \begin{minipage}{\columnwidth}%
  \small \textit{Notes}: Standard errors in parentheses. *** p\textless{}0.01, ** p\textless{}0.05, * p\textless{}0.1.\\
  All regressions are OLS regressions that take into account population survey wights and the sampling procedure. The dependent variable is a dummy=1 if the respondent increased their support for the following policies. The control variables include: gender, race, age, education, parental status, and caring responsibilities for an elderly or a person with a disability, income in February 2020, housing, labor force participation and employment status in February 2020, health insurance provider, and whether respondents had financial difficulties before the pandemic, the area in which the respondents live, whether it's a metropolitan or rural area, and the population density in the zip code. We also control for whether respondents completed the related surveys in shorter time than the 99$^{th}$ percentile, and ceiling effects.
 \end{minipage}%
}\\ 
\end{tabular}
}
\end{table}

\begin{table}[H]
\centering
\caption{The effect of shocks and media on institutional trust, considering voting intentions}
\label{tab:trust_vote}
\resizebox{\textwidth}{!}{%
\begin{tabular}{lccccccc} \hline \hline
 & (1) & (2) & (3) & (4) & (5) & (6) & (7) \\
 & \begin{tabular}[c]{@{}c@{}}Congress \\ \& Senate\end{tabular} & \begin{tabular}[c]{@{}c@{}}White\\ House\end{tabular} & \begin{tabular}[c]{@{}c@{}}Financial\\ institutions\end{tabular} & \begin{tabular}[c]{@{}c@{}}Private\\ sector\end{tabular} & \begin{tabular}[c]{@{}c@{}}Scientific\\ community\end{tabular} & \begin{tabular}[c]{@{}c@{}}Health \\ insurance \\ companies\end{tabular} & Hospitals \\ \hline
 & & & & & & & \\
Vote Trump & -0.110 & -0.362*** & -0.0460 & -0.0495 & 0.0921* & 0.0221 & -0.0781 \\
 & (0.0668) & (0.0439) & (0.0496) & (0.0407) & (0.0470) & (0.0553) & (0.0517) \\
Vote Biden & 0.0188 & 0.0582 & -0.00593 & -0.0353 & -0.105** & 0.0173 & -0.143*** \\
 & (0.0548) & (0.0425) & (0.0434) & (0.0353) & (0.0466) & (0.0525) & (0.0520) \\
Lost 20\% income & 0.0612 & 0.0199 & 0.0717** & 0.0579* & -0.0268 & 0.0337 & 0.0612* \\
 & (0.0391) & (0.0275) & (0.0332) & (0.0306) & (0.0297) & (0.0369) & (0.0316) \\
Knows hospitalized & 0.0261 & 0.0238 & 0.0256 & 0.0163 & 0.0592 & 0.0239 & 0.0467 \\
 & (0.0364) & (0.0320) & (0.0376) & (0.0404) & (0.0441) & (0.0353) & (0.0374) \\
Var consumer expenditures & -0.000110 & 0.0227 & -0.00346 & -0.0155 & 0.0140 & -0.0623*** & -0.0626** \\
 & (0.0366) & (0.0236) & (0.0311) & (0.0212) & (0.0342) & (0.0212) & (0.0246) \\
Consumer exp - Apr & 0.0558 & -0.210 & -0.00700 & -0.173 & 0.000239 & -0.00286 & 0.359* \\
 & (0.220) & (0.203) & (0.279) & (0.204) & (0.260) & (0.248) & (0.213) \\
Incr COVID-19 cases & -0.00900 & -0.0133 & 0.00495 & 0.0291 & 0.0346 & -0.00753 & -0.0157 \\
 & (0.0216) & (0.0147) & (0.0197) & (0.0202) & (0.0210) & (0.0189) & (0.0234) \\
Constant & 0.507** & 0.848*** & 0.515*** & -0.00572 & 0.263 & 0.145 & 0.495*** \\
 & (0.207) & (0.140) & (0.161) & (0.102) & (0.176) & (0.162) & (0.164) \\
 & & & & & & & \\
Observations & 1,011 & 1,005 & 1,009 & 1,008 & 1,004 & 1,008 & 1,009 \\
R-squared & 0.165 & 0.419 & 0.115 & 0.126 & 0.100 & 0.123 & 0.100 \\
Controls & Yes & Yes & Yes & Yes & Yes & Yes & Yes \\
Average increase & 0.159 & 0.142 & 0.142 & 0.185 & 0.185 & 0.172 & 0.164 \\
Average decrease & 0.351 & 0.312 & 0.299 & 0.247 & 0.286 & 0.284 & 0.306 \\ \hline
\multicolumn{8}{l}{%
 \begin{minipage}{1.25\columnwidth}%
  \small \textit{Notes}: Standard errors in parentheses. *** p\textless{}0.01, ** p\textless{}0.05, * p\textless{}0.1.\\
  All regressions are OLS regressions that take into account population survey wights and the sampling procedure. The dependent variable is a dummy=1 if the respondent decreased trust in the above institutions. The control variables include: gender, race, age, education, parental status, and caring responsibilities for an elderly or a person with a disability, income in February 2020, housing, labor force participation and employment status in February 2020, health insurance provider, and whether respondents had financial difficulties before the pandemic, the area in which the respondents live, whether it's a metropolitan or rural area, and the population density in the zip code. We also control for whether respondents completed the related surveys in shorter time than the 99$^{th}$ percentile, and ceiling effects.
 \end{minipage}%
 }
\end{tabular}%
}
\end{table}

\subsubsection{Fixed effects}
\begin{landscape}

\begin{table}[H]
\centering
\caption{The effect of shocks on welfare policy preferences, using panel data and individual fixed effects.}
\label{tab:fe_policy_1}
\resizebox{1.2\textwidth}{!}{%
\begin{tabular}{lccccccccc} \hline \hline
 & (1) & (2) & (3) & (4) & (5) & (6) & (7) & (8) & (9) \\
 & All & Dem & Rep & All & Dem & Rep & All & Dem & Rep \\
& \begin{tabular}[c]{@{}c@{}}Support for \\ universal \\ healthcare\end{tabular} & \begin{tabular}[c]{@{}c@{}}Support for \\ universal \\ healthcare\end{tabular} & \begin{tabular}[c]{@{}c@{}}Support for \\ universal \\ healthcare\end{tabular} & \begin{tabular}[c]{@{}c@{}}Gov should \\ help industry \\ grow\end{tabular} & \begin{tabular}[c]{@{}c@{}}Gov should \\ help industry \\ grow\end{tabular} & \begin{tabular}[c]{@{}c@{}}Gov should \\ help industry \\ grow\end{tabular} & \begin{tabular}[c]{@{}c@{}}Gov should \\ control\\ prices\end{tabular} & \begin{tabular}[c]{@{}c@{}}Gov should \\ control\\ prices\end{tabular} & \begin{tabular}[c]{@{}c@{}}Gov should \\ control\\ prices\end{tabular} \\ \hline
 & & & & & & & & & \\
Lost 20\% income & -0.00209 & 0.0248 & -0.0437 & -0.0658* & -0.112* & -0.0368 & -0.0150 & -0.0241 & 0.0180 \\
 & (0.0345) & (0.0565) & (0.0523) & (0.0378) & (0.0619) & (0.0677) & (0.0333) & (0.0416) & (0.0823) \\
Knows hospitalized & 0.0353 & 0.0413 & 0.0439 & -0.0207 & 0.0102 & -0.109 & 0.00647 & -0.0192 & -0.0105 \\
 & (0.0301) & (0.0513) & (0.0564) & (0.0381) & (0.0604) & (0.0793) & (0.0336) & (0.0451) & (0.0746) \\
Variation consumer exp & -0.0502 & 0.0940 & -0.0264 & 0.00155 & -0.105 & 0.0920 & 0.209 & 0.126 & 0.303* \\
 & (0.0948) & (0.199) & (0.104) & (0.0998) & (0.260) & (0.119) & (0.138) & (0.212) & (0.160) \\
log COVID-19 cases & -0.00601 & 0.0269 & 0.00331 & -0.0105 & -0.00989 & 0.00140 & 0.0180 & 0.0102 & -0.0188 \\
 & (0.0116) & (0.0217) & (0.0181) & (0.0131) & (0.0250) & (0.0238) & (0.0142) & (0.0227) & (0.0252) \\
End April & 0.0149 & -0.0346 & -0.0245 & 0.00657 & 0.0964* & -0.0261 & -0.0485* & 0.0239 & -0.106** \\
 & (0.0264) & (0.0470) & (0.0330) & (0.0289) & (0.0563) & (0.0553) & (0.0271) & (0.0489) & (0.0492) \\
October & 0.0407 & -0.109 & -0.00731 & 0.0771 & 0.130 & 0.0395 & -0.103 & -0.0212 & -0.0158 \\
 & (0.0608) & (0.116) & (0.0999) & (0.0690) & (0.140) & (0.123) & (0.0704) & (0.127) & (0.123) \\
Constant & 0.485*** & 0.657*** & 0.128** & 0.725*** & 0.696*** & 0.665*** & 0.782*** & 0.853*** & 0.851*** \\
 & (0.0521) & (0.102) & (0.0647) & (0.0480) & (0.101) & (0.0862) & (0.0737) & (0.0984) & (0.111) \\
 & & & & & & & & & \\
Observations & 2,530 & 967 & 668 & 2,512 & 958 & 662 & 2,522 & 963 & 668 \\
R-squared & 0.002 & 0.009 & 0.007 & 0.004 & 0.017 & 0.013 & 0.006 & 0.008 & 0.030 \\
Number of respondents & 864 & 330 & 228 & 864 & 330 & 228 & 863 & 330 & 228\\ \hline
\multicolumn{10}{l}{%
 \begin{minipage}{1.2\columnwidth}%
  \small \textit{Notes}: Standard errors in parentheses. *** p\textless{}0.01, ** p\textless{}0.05, * p\textless{}0.1.\\
  All regressions are OLS regressions with the data organized in a panel structure and fixed effects at the individual level. The dependent variable is a dummy=1 if the respondent believes that it's a government responsibility to provide the following welfare. Col (1), (4) and (7) consider the whole sample, col. (2), (5) and (8) the subsample of Democrats, and col. (3), (6) and (9) the subsample of Republicans.
 \end{minipage}%
}\\ 
\end{tabular}%
}
\end{table}

\begin{table}[H]
\centering
\caption{The effect of shocks on welfare policy preferences, using panel data and individual fixed effects.}
\label{tab:fe_policy_2}
\resizebox{1.2\textwidth}{!}{%
\begin{tabular}{lccccccccc} \hline \hline
 & (1) & (2) & (3) & (4) & (5) & (6) & (7) & (8) & (9) \\
 & All & Dem & Rep & All & Dem & Rep & All & Dem & Rep \\
& \begin{tabular}[c]{@{}c@{}}Gov should\\ provide for\\ the unemployed\end{tabular} & \begin{tabular}[c]{@{}c@{}}Gov should\\ provide for\\ the unemployed\end{tabular} & \begin{tabular}[c]{@{}c@{}}Gov should\\ provide for\\ the unemployed\end{tabular} & \begin{tabular}[c]{@{}c@{}}Gov should\\ provide mental\\ healthcare\end{tabular} & \begin{tabular}[c]{@{}c@{}}Gov should\\ provide mental\\ healthcare\end{tabular} & \begin{tabular}[c]{@{}c@{}}Gov should\\ provide mental\\ healthcare\end{tabular} & \begin{tabular}[c]{@{}c@{}}Gov should\\ provide for\\ the elderly\end{tabular} & \begin{tabular}[c]{@{}c@{}}Gov should\\ provide for\\ the elderly\end{tabular} & \begin{tabular}[c]{@{}c@{}}Gov should\\ provide for\\ the elderly\end{tabular} \\ \hline
 & & & & & & & & & \\
Lost 20\% income & -0.0622* & -0.0635 & -0.0704 & -0.0189 & -0.0193 & -0.0312 & 0.0107 & -0.0521 & 0.109* \\
 & (0.0351) & (0.0565) & (0.0622) & (0.0297) & (0.0421) & (0.0682) & (0.0316) & (0.0462) & (0.0662) \\
Knows hospitalized & 0.0636** & 0.0366 & 0.0801 & -0.0351 & -0.0852** & -0.0172 & 0.0408 & 0.0911** & -0.0695 \\
 & (0.0322) & (0.0426) & (0.0777) & (0.0289) & (0.0394) & (0.0904) & (0.0308) & (0.0451) & (0.0749) \\
Variation consumer exp & -0.163* & -0.0938 & -0.303** & -0.145 & 0.0150 & -0.0988 & -0.271** & -0.0803 & -0.217 \\
 & (0.0968) & (0.229) & (0.142) & (0.131) & (0.167) & (0.221) & (0.130) & (0.209) & (0.178) \\
log COVID-19 cases & -0.00812 & -0.00549 & 0.0116 & 0.00614 & 0.0267 & 0.0145 & 0.0207 & 0.00333 & 0.0370 \\
 & (0.0143) & (0.0188) & (0.0352) & (0.0109) & (0.0163) & (0.0243) & (0.0136) & (0.0169) & (0.0312) \\
 End April & -0.0394* & 0.00778 & -0.0831* & -0.0327 & -0.0438 & -0.0672 & -0.0582** & -0.0463 & -0.152*** \\
 & (0.0229) & (0.0469) & (0.0463) & (0.0251) & (0.0427) & (0.0645) & (0.0254) & (0.0384) & (0.0510) \\
October & -0.0384 & -0.0166 & -0.148 & 0.0179 & -0.0686 & -0.0336 & -0.0801 & -0.0529 & -0.193 \\
 & (0.0699) & (0.120) & (0.159) & (0.0599) & (0.0864) & (0.147) & (0.0694) & (0.104) & (0.155) \\
Constant & 0.683*** & 0.857*** & 0.337*** & 0.806*** & 0.842*** & 0.674*** & 0.687*** & 0.920*** & 0.528*** \\
 & (0.0512) & (0.0715) & (0.120) & (0.0580) & (0.0849) & (0.0975) & (0.0589) & (0.0729) & (0.103) \\
 & & & & & & & & & \\
Observations & 2,515 & 961 & 666 & 2,527 & 964 & 669 & 2,516 & 963 & 664 \\
R-squared & 0.043 & 0.026 & 0.091 & 0.010 & 0.023 & 0.012 & 0.028 & 0.034 & 0.058 \\
Number of respondents & 864 & 330 & 228 & 864 & 330 & 228 & 864 & 330 & 228\\ \hline
\multicolumn{10}{l}{%
 \begin{minipage}{1.38\columnwidth}%
  \small \textit{Notes}: Standard errors in parentheses. *** p\textless{}0.01, ** p\textless{}0.05, * p\textless{}0.1.\\
  All regressions are OLS regressions with the data organized in a panel structure and fixed effects at the individual level. The dependent variable is a dummy=1 if the respondent believes that it's a government responsibility to provide the following welfare. Col (1), (4) and (7) consider the whole sample, col. (2), (5) and (8) the subsample of Democrats, and col. (3), (6) and (9) the subsample of Republicans.
 \end{minipage}%
}\\ \end{tabular}%
}
\end{table}

\begin{table}[H]
\centering
\caption{The effect of shocks on welfare policy preferences, using panel data and individual fixed effects.}
\label{tab:fe_policy_3}
\resizebox{1.25\textwidth}{!}{%
\begin{tabular}{lcccccccccccc} \hline \hline
 & (1) & (2) & (3) & (4) & (5) & (6) & (7) & (8) & (9) & (10) & (11) & (12) \\
 & All & Dem & Rep & All & Dem & Rep & All & Dem & Rep & All & Dem & Rep \\
& \begin{tabular}[c]{@{}c@{}}Gov should \\ help affected \\ by disasters\end{tabular} & \begin{tabular}[c]{@{}c@{}}Gov should \\ help affected \\ by disasters\end{tabular} & \begin{tabular}[c]{@{}c@{}}Gov should \\ help affected \\ by disasters\end{tabular} & \begin{tabular}[c]{@{}c@{}}Gov should \\ provide a\\ basic income\end{tabular} & \begin{tabular}[c]{@{}c@{}}Gov should \\ provide a\\ basic income\end{tabular} & \begin{tabular}[c]{@{}c@{}}Gov should \\ provide a\\ basic income\end{tabular} & \begin{tabular}[c]{@{}c@{}}Gov should \\ reduce\\ inequality\end{tabular} & \begin{tabular}[c]{@{}c@{}}Gov should \\ reduce\\ inequality\end{tabular} & \begin{tabular}[c]{@{}c@{}}Gov should \\ reduce\\ inequality\end{tabular} & \begin{tabular}[c]{@{}c@{}}Gov should\\ pay university\\ for the poor\end{tabular} & \begin{tabular}[c]{@{}c@{}}Gov should\\ pay university\\ for the poor\end{tabular} & \begin{tabular}[c]{@{}c@{}}Gov should\\ pay university\\ for the poor\end{tabular} \\ \hline
 & & & & & & & & & & & & \\
Lost 20\% income & 0.00640 & 0.00282 & 0.0188 & -0.0146 & -0.0549 & 0.000214 & -0.0119 & -0.0939 & 0.124* & 0.00174 & -0.0166 & 0.0370 \\
 & (0.0259) & (0.0377) & (0.0571) & (0.0338) & (0.0656) & (0.0456) & (0.0369) & (0.0578) & (0.0705) & (0.0296) & (0.0287) & (0.0731) \\
Knows hospitalized & 0.0263 & 0.0189 & -0.00458 & 0.0545 & 0.0212 & 0.0302 & 0.0589* & 0.136*** & -0.103 & 0.0226 & 0.0479 & -0.0122 \\
 & (0.0177) & (0.0286) & (0.0431) & (0.0332) & (0.0613) & (0.0528) & (0.0347) & (0.0506) & (0.0786) & (0.0270) & (0.0315) & (0.0727) \\
Variation consumer exp & -0.112 & 0.209 & -0.138 & 0.0104 & 0.195 & 0.0806 & -0.0663 & 0.0479 & 0.0460 & -0.0151 & 0.202 & 0.114 \\
 & (0.106) & (0.154) & (0.161) & (0.122) & (0.236) & (0.120) & (0.123) & (0.271) & (0.124) & (0.119) & (0.125) & (0.183) \\
log COVID-19 cases & 0.00725 & -0.00462 & 0.00386 & 0.00692 & 0.0118 & 0.0232 & -0.0159 & -0.0248 & 0.00131 & -0.0205 & -0.0128 & -0.00513 \\
 & (0.00758) & (0.0126) & (0.0145) & (0.0126) & (0.0260) & (0.0183) & (0.0144) & (0.0207) & (0.0333) & (0.0134) & (0.0112) & (0.0242) \\
 End April & -0.0217 & -0.0283 & -0.0299 & 0.0257 & 0.0292 & 0.0135 & 0.00309 & 0.0350 & -0.0239 & 0.0127 & -0.0193 & -0.0342 \\
 & (0.0171) & (0.0340) & (0.0343) & (0.0254) & (0.0518) & (0.0481) & (0.0245) & (0.0437) & (0.0434) & (0.0292) & (0.0217) & (0.0672) \\
October & -0.0336 & -0.0810 & -0.0204 & -0.0331 & -0.0732 & -0.0845 & 0.0631 & 0.0975 & -0.00570 & 0.0510 & -0.0398 & -0.0394 \\
 & (0.0365) & (0.0664) & (0.0837) & (0.0617) & (0.131) & (0.0968) & (0.0739) & (0.122) & (0.155) & (0.0682) & (0.0546) & (0.135) \\
Constant & 0.899*** & 1.073*** & 0.892*** & 0.450*** & 0.678*** & 0.122* & 0.639*** & 0.938*** & 0.273** & 0.851*** & 1.051*** & 0.645*** \\
 & (0.0493) & (0.0794) & (0.0681) & (0.0634) & (0.121) & (0.0668) & (0.0561) & (0.0987) & (0.112) & (0.0499) & (0.0554) & (0.0948) \\
 & & & & & & & & & & & & \\
Observations & 2,522 & 961 & 668 & 2,512 & 961 & 664 & 2,516 & 961 & 664 & 2,519 & 963 & 662 \\
R-squared & 0.010 & 0.017 & 0.011 & 0.006 & 0.010 & 0.012 & 0.005 & 0.026 & 0.028 & 0.005 & 0.013 & 0.003 \\
Number of respondents & 864 & 330 & 228 & 864 & 330 & 228 & 864 & 330 & 228 & 864 & 330 & 228 \\ \hline
\multicolumn{13}{l}{%
 \begin{minipage}{1.6\columnwidth}%
  \small \textit{Notes}: Standard errors in parentheses. *** p\textless{}0.01, ** p\textless{}0.05, * p\textless{}0.1.\\
  All regressions are OLS regressions with the data organized in a panel structure and fixed effects at the individual level. The dependent variable is a dummy=1 if the respondent believes that it's a government responsibility to provide the following welfare. Col (1), (4), (7) and (10) consider the whole sample, col. (2), (5), (8) and (11) the subsample of Democrats, and col. (3), (6), (9) and (12) the subsample of Republicans.
 \end{minipage}%
}\\ 
\end{tabular}%
}
\end{table}

\begin{table}[H]
\centering
\caption{The effect of shocks on support for coronavirus relief policies, using panel data and individual fixed effects.}
\label{tab:fe_covid_policy}
\resizebox{1.2\textwidth}{!}{%
\begin{tabular}{lccccccccc} \hline \hline
 & (1) & (2) & (3) & (4) & (5) & (6) & (7) & (8) & (9) \\
 & All & Dem & Rep & All & Dem & Rep & All & Dem & Rep \\
& \begin{tabular}[c]{@{}c@{}}Gov should \\ increase health\\ expenditures\end{tabular} & \begin{tabular}[c]{@{}c@{}}Gov should \\ increase health\\ expenditures\end{tabular} & \begin{tabular}[c]{@{}c@{}}Gov should \\ increase health\\ expenditures\end{tabular} & \begin{tabular}[c]{@{}c@{}}Gov should\\ protect essential \\ workers\end{tabular} & \begin{tabular}[c]{@{}c@{}}Gov should\\ protect essential \\ workers\end{tabular} & \begin{tabular}[c]{@{}c@{}}Gov should\\ protect essential \\ workers\end{tabular} & \begin{tabular}[c]{@{}c@{}}Gov should\\ transfer money\\ to privates\end{tabular} & \begin{tabular}[c]{@{}c@{}}Gov should\\ transfer money\\ to privates\end{tabular} & \begin{tabular}[c]{@{}c@{}}Gov should\\ transfer money\\ to privates\end{tabular} \\ \hline
 & & & & & & & & & \\
Lost 20\% income & 0.141*** & 0.0645 & 0.160** & 0.0401 & 0.0472 & 0.0129 & 0.0829* & 0.114* & -0.0400 \\
 & (0.0425) & (0.0700) & (0.0767) & (0.0490) & (0.0579) & (0.104) & (0.0478) & (0.0677) & (0.0927) \\
Knows hospitalized & 0.0501 & 0.0445 & 0.0650 & 0.0137 & 0.0361 & 0.0378 & 0.0165 & -0.0715 & 0.0875 \\
 & (0.0535) & (0.0499) & (0.113) & (0.0526) & (0.0398) & (0.0880) & (0.0484) & (0.0674) & (0.0630) \\
Variation consumer exp & -0.0869 & -0.171 & 0.132 & 0.206 & -0.144 & 0.412 & 0.0345 & -0.471 & 0.0912 \\
 & (0.141) & (0.289) & (0.163) & (0.187) & (0.289) & (0.255) & (0.152) & (0.335) & (0.211) \\
log COVID-19 cases & 0.0320 & 0.0201 & 0.0241 & -0.00512 & -0.0185 & 0.0272 & 0.0405** & 0.0235 & 0.0259 \\
 & (0.0198) & (0.0285) & (0.0351) & (0.0198) & (0.0196) & (0.0360) & (0.0194) & (0.0276) & (0.0326) \\
 October & -0.201*** & -0.0293 & -0.246* & -0.129* & 0.0174 & -0.237 & -0.303*** & -0.139 & -0.261** \\
 & (0.0752) & (0.0992) & (0.143) & (0.0775) & (0.0889) & (0.145) & (0.0759) & (0.108) & (0.123) \\
Constant & 0.445*** & 0.658*** & 0.247 & 0.789*** & 0.931*** & 0.418** & 0.357*** & 0.484*** & 0.281** \\
 & (0.0964) & (0.159) & (0.165) & (0.0994) & (0.117) & (0.167) & (0.0939) & (0.159) & (0.141) \\
 & & & & & & & & & \\
Observations & 1,658 & 634 & 439 & 1,660 & 635 & 440 & 1,663 & 635 & 441 \\
R-squared & 0.059 & 0.016 & 0.074 & 0.041 & 0.017 & 0.037 & 0.107 & 0.103 & 0.124 \\
Number of respondents & 862 & 330 & 227 & 863 & 330 & 228 & 863 & 330 & 228\\ \hline
\multicolumn{10}{l}{%
 \begin{minipage}{1.4\columnwidth}%
  \small \textit{Notes}: Standard errors in parentheses. *** p\textless{}0.01, ** p\textless{}0.05, * p\textless{}0.1.\\
  All regressions are OLS regressions with the data organized in a panel structure and fixed effects at the individual level. The dependent variable is a dummy=1 if the respondent support the provision of the following policies. Col (1), (4) and (7) consider the whole sample, col. (2), (5) and (8) the subsample of Democrats, and col. (3), (6) and (9) the subsample of Republicans.
 \end{minipage}%
}\\ 
\end{tabular}%
}
\end{table}

\begin{table}[H]
\centering
\caption{The effect of shocks on institutional trust, using panel data and individual fixed effects.}
\label{tab:fe_trust_1}
\resizebox{1.2\textwidth}{!}{%
\begin{tabular}{lcccccccccccc} \hline \hline
 & (1) & (2) & (3) & (4) & (5) & (6) & (7) & (8) & (9) & (10) & (11) & (12) \\
 & All & Dem & Rep & All & Dem & Rep & All & Dem & Rep & All & Dem & Rep \\
& \begin{tabular}[c]{@{}c@{}}Confidence in \\ federal \\ government\end{tabular} & \begin{tabular}[c]{@{}c@{}}Confidence in \\ federal \\ government\end{tabular} & \begin{tabular}[c]{@{}c@{}}Confidence in \\ federal \\ government\end{tabular} & \begin{tabular}[c]{@{}c@{}}Confidence in \\ President Trump\end{tabular} & \begin{tabular}[c]{@{}c@{}}Confidence in \\ President Trump\end{tabular} & \begin{tabular}[c]{@{}c@{}}Confidence in \\ President Trump\end{tabular} & \begin{tabular}[c]{@{}c@{}}Confidence in \\ banks and financial\\ institutions\end{tabular} & \begin{tabular}[c]{@{}c@{}}Confidence in \\ banks and financial\\ institutions\end{tabular} & \begin{tabular}[c]{@{}c@{}}Confidence in \\ banks and financial\\ institutions\end{tabular} & \begin{tabular}[c]{@{}c@{}}Confidence in \\ private sector\end{tabular} & \begin{tabular}[c]{@{}c@{}}Confidence in \\ private sector\end{tabular} & \begin{tabular}[c]{@{}c@{}}Confidence in \\ private sector\end{tabular} \\ \hline
 & & & & & & & & & & & & \\
Lost 20\% income & -0.00300 & 0.000264 & 0.0481 & -0.00473 & 0.0121 & 0.0249 & -0.0363 & -0.0566 & -0.0148 & -0.0464** & -0.0264 & -0.0901 \\
 & (0.0162) & (0.0203) & (0.0458) & (0.0209) & (0.0233) & (0.0530) & (0.0255) & (0.0356) & (0.0656) & (0.0231) & (0.0266) & (0.0640) \\
Knows hospitalized & -0.0116 & -0.0107 & 0.0116 & -0.00649 & 0.00575 & -0.0696 & 0.0350 & 0.000119 & 0.0521 & 0.00690 & 0.0454 & 0.0471 \\
 & (0.0205) & (0.0373) & (0.0490) & (0.0176) & (0.0101) & (0.0497) & (0.0284) & (0.0443) & (0.0847) & (0.0243) & (0.0353) & (0.0562) \\
Variation consumer exp & -0.0267 & -0.0797 & -0.0347 & 0.0476 & 0.113 & -0.0269 & -0.0926 & -0.232 & -0.187 & 0.139 & -0.107 & 0.0992 \\
 & (0.0376) & (0.0954) & (0.0726) & (0.0565) & (0.0893) & (0.101) & (0.0925) & (0.243) & (0.181) & (0.109) & (0.107) & (0.197) \\
log COVID-19 cases & -0.0145** & -0.0171 & -0.0267 & 0.00198 & 0.00247 & 0.00752 & -0.00468 & -0.0151 & -0.0146 & -0.0203* & -0.0100 & -0.0228 \\
 & (0.00708) & (0.0108) & (0.0179) & (0.00892) & (0.00534) & (0.0208) & (0.0110) & (0.0144) & (0.0275) & (0.0105) & (0.0159) & (0.0247) \\
 End April & 0.00202 & 0.00265 & 0.0125 & -0.0201 & -0.0304* & -0.00996 & 0.000852 & 0.0447 & -0.0115 & 0.0224 & 0.0107 & 0.00136 \\
 & (0.0145) & (0.0255) & (0.0386) & (0.0167) & (0.0165) & (0.0387) & (0.0237) & (0.0415) & (0.0534) & (0.0233) & (0.0380) & (0.0504) \\
June & 0.0159 & 0.0216 & 0.0595 & -0.0458 & -0.0404 & -0.0663 & -0.0466 & -0.00473 & -0.0305 & -0.00649 & 0.0199 & -0.00930 \\
 & (0.0224) & (0.0380) & (0.0523) & (0.0281) & (0.0311) & (0.0638) & (0.0363) & (0.0592) & (0.0901) & (0.0368) & (0.0553) & (0.0870) \\
October & 0.0366 & 0.0357 & 0.0850 & -0.0341 & -0.0648* & 0.000568 & -0.0330 & 0.0699 & -0.0475 & 0.0245 & 0.00336 & 0.0548 \\
 & (0.0339) & (0.0546) & (0.0881) & (0.0404) & (0.0344) & (0.0976) & (0.0542) & (0.0870) & (0.138) & (0.0542) & (0.0804) & (0.141) \\
Constant & 0.127*** & 0.121** & 0.199*** & 0.245*** & 0.0525 & 0.548*** & 0.196*** & 0.175* & 0.345*** & 0.332*** & 0.123** & 0.515*** \\
 & (0.0280) & (0.0546) & (0.0659) & (0.0353) & (0.0318) & (0.0775) & (0.0458) & (0.0996) & (0.101) & (0.0523) & (0.0569) & (0.0902) \\
 & & & & & & & & & & & & \\
Observations & 3,316 & 1,266 & 865 & 3,309 & 1,262 & 864 & 3,314 & 1,265 & 865 & 3,315 & 1,260 & 868 \\
R-squared & 0.007 & 0.017 & 0.010 & 0.005 & 0.010 & 0.015 & 0.024 & 0.041 & 0.040 & 0.012 & 0.018 & 0.013 \\
Number of respondents & 863 & 330 & 228 & 864 & 330 & 228 & 863 & 330 & 228 & 864 & 330 & 228 \\ \hline
\multicolumn{13}{l}{%
 \begin{minipage}{1.9\columnwidth}%
  \small \textit{Notes}: Standard errors in parentheses. *** p\textless{}0.01, ** p\textless{}0.05, * p\textless{}0.1.\\
  All regressions are OLS regressions with the data organized in a panel structure and fixed effects at the individual level. The dependent variable is a dummy=1 if the respondent trust the people running the following institutions. Col (1), (4) and (7) consider the whole sample, col. (2), (5) and (8) the subsample of Democrats, and col. (3), (6) and (9) the subsample of Republicans.
 \end{minipage}%
}\\ 
\end{tabular}%
}
\end{table}

\begin{table}[H]
\centering
\caption{The effect of shocks on institutional trust, using panel data and individual fixed effects.}
\label{tab:fe_trust_2}
\resizebox{\textwidth}{!}{%
\begin{tabular}{lccccccccc} \hline \hline
 & (1) & (2) & (3) & (4) & (5) & (6) & (7) & (8) & (9) \\
 & All & Dem & Rep & All & Dem & Rep & All & Dem & Rep \\
& \begin{tabular}[c]{@{}c@{}}Confidence in \\ scientific \\ community\end{tabular} & \begin{tabular}[c]{@{}c@{}}Confidence in \\ scientific \\ community\end{tabular} & \begin{tabular}[c]{@{}c@{}}Confidence in \\ scientific \\ community\end{tabular} & \begin{tabular}[c]{@{}c@{}}Confidence in \\ health insurance \\ companies\end{tabular} & \begin{tabular}[c]{@{}c@{}}Confidence in \\ health insurance \\ companies\end{tabular} & \begin{tabular}[c]{@{}c@{}}Confidence in \\ health insurance \\ companies\end{tabular} & \begin{tabular}[c]{@{}c@{}}Confidence in \\ hospitals\end{tabular} & \begin{tabular}[c]{@{}c@{}}Confidence in \\ hospitals\end{tabular} & \begin{tabular}[c]{@{}c@{}}Confidence in \\ hospitals\end{tabular} \\ \hline
 & & & & & & & & & \\
Lost 20\% income & -0.0225 & -0.0725* & 0.115 & -0.0160 & -0.0128 & -0.00546 & -0.0186 & 0.00162 & 0.00111 \\
 & (0.0341) & (0.0433) & (0.0846) & (0.0207) & (0.0254) & (0.0477) & (0.0329) & (0.0569) & (0.0656) \\
Knows hospitalized & -0.0358 & -0.0278 & -0.0464 & 0.000791 & 0.00247 & -0.00756 & -0.0195 & 0.00954 & -0.0494 \\
 & (0.0317) & (0.0369) & (0.0733) & (0.0221) & (0.0397) & (0.0445) & (0.0335) & (0.0509) & (0.0700) \\
Variation consumer exp & 0.00124 & 0.406* & -0.0301 & 0.0262 & -0.138 & 0.322*** & 0.0758 & 0.221 & 0.0307 \\
 & (0.0909) & (0.208) & (0.118) & (0.102) & (0.144) & (0.121) & (0.118) & (0.227) & (0.184) \\
log COVID-19 cases & -0.00104 & -0.00466 & 0.0229 & -0.00520 & -0.0193 & -0.00513 & 0.0142 & 0.00930 & 0.00791 \\
 & (0.0121) & (0.0181) & (0.0255) & (0.0106) & (0.0175) & (0.0245) & (0.0131) & (0.0215) & (0.0296) \\
 End April & -0.0462* & -0.0113 & -0.167*** & -0.0417* & 0.0197 & -0.124** & -0.109*** & -0.0803 & -0.137*** \\
 & (0.0242) & (0.0405) & (0.0496) & (0.0242) & (0.0445) & (0.0521) & (0.0288) & (0.0496) & (0.0493) \\
June & -0.113*** & -0.133** & -0.303*** & -0.0429 & 0.0271 & -0.164** & -0.241*** & -0.220** & -0.252*** \\
 & (0.0411) & (0.0662) & (0.0797) & (0.0398) & (0.0710) & (0.0692) & (0.0493) & (0.0899) & (0.0888) \\
October & -0.0565 & -0.0388 & -0.301** & -0.0339 & 0.0473 & -0.120 & -0.158** & -0.143 & -0.165 \\
 & (0.0588) & (0.0962) & (0.127) & (0.0556) & (0.0980) & (0.117) & (0.0679) & (0.113) & (0.149) \\
Constant & 0.606*** & 0.845*** & 0.457*** & 0.195*** & 0.185*** & 0.393*** & 0.662*** & 0.775*** & 0.683*** \\
 & (0.0538) & (0.102) & (0.0885) & (0.0551) & (0.0626) & (0.103) & (0.0602) & (0.101) & (0.114) \\
 & & & & & & & & & \\
Observations & 3,306 & 1,260 & 865 & 3,314 & 1,260 & 868 & 3,317 & 1,261 & 869 \\
R-squared & 0.026 & 0.036 & 0.082 & 0.010 & 0.016 & 0.040 & 0.050 & 0.033 & 0.062 \\
Number of respondents & 863 & 330 & 228 & 864 & 330 & 228 & 864 & 330 & 228 \\ \hline
\multicolumn{10}{l}{%
 \begin{minipage}{1.4\columnwidth}%
  \small \textit{Notes}: Standard errors in parentheses. *** p\textless{}0.01, ** p\textless{}0.05, * p\textless{}0.1.\\
  All regressions are OLS regressions with the data organized in a panel structure and fixed effects at the individual level. The dependent variable is a dummy=1 if the respondent trust the people running the following institutions. Col (1), (4) and (7) consider the whole sample, col. (2), (5) and (8) the subsample of Democrats, and col. (3), (6) and (9) the subsample of Republicans.
 \end{minipage}%
}
\end{tabular}%
}
\end{table}

\end{landscape}

\subsubsection{Logit}

\begin{table}[H]
\centering
\caption{The effect of shocks and media on welfare policy preferences - logistic regression model}
\label{tab:policy_logit}
\resizebox{\textwidth}{!}{%
\begin{tabular}{lcccccccccc} \hline \hline
 & (1) & (2) & (3) & (4) & (5) & (6) & (7) & (8) & (9) & (10) \\
 & \begin{tabular}[c]{@{}c@{}}Universal \\ healthcare\end{tabular} & \begin{tabular}[c]{@{}c@{}}Provide\\ basic \\ income\end{tabular} & \begin{tabular}[c]{@{}c@{}}Provide\\ for\\ unemployed\end{tabular} & \begin{tabular}[c]{@{}c@{}}Provide\\ for the\\ elderly\end{tabular} & \begin{tabular}[c]{@{}c@{}}Reduce\\ inequality\end{tabular} & \begin{tabular}[c]{@{}c@{}}Help\\ low-income\\ students\end{tabular} & \begin{tabular}[c]{@{}c@{}}Help \\ disasters'\\ victims\end{tabular} & \begin{tabular}[c]{@{}c@{}}Provide\\ Mental \\ healthcare\end{tabular} & \begin{tabular}[c]{@{}c@{}}Control \\ prices\end{tabular} & \begin{tabular}[c]{@{}c@{}}Help\\ industry\\ grow\end{tabular} \\ \hline
 & & & & & & & & & & \\
Republican & 0.0115 & 0.234 & -0.106 & -0.423 & 0.0411 & 0.365 & 0.250 & -0.0379 & -0.0923 & 0.0393 \\
 & (0.269) & (0.239) & (0.262) & (0.322) & (0.301) & (0.349) & (0.338) & (0.271) & (0.291) & (0.254) \\
Democrat & 0.0935 & 0.567** & 0.397 & 0.200 & 0.456 & 0.337 & -0.000504 & 0.244 & 0.832*** & 0.191 \\
 & (0.222) & (0.226) & (0.260) & (0.268) & (0.327) & (0.343) & (0.395) & (0.270) & (0.278) & (0.303) \\
Lost 20\% income & 0.137 & -0.507** & -0.0786 & 0.232 & -0.242 & 0.416* & 0.0806 & 0.297 & 0.280 & 0.136 \\
 & (0.230) & (0.197) & (0.216) & (0.309) & (0.248) & (0.244) & (0.375) & (0.193) & (0.190) & (0.206) \\
Knows hospitalized & 0.117 & -0.0703 & 0.413* & -0.316 & 0.212 & 0.537** & 0.273 & -0.206 & -0.0965 & 0.387* \\
 & (0.199) & (0.204) & (0.236) & (0.255) & (0.264) & (0.246) & (0.359) & (0.316) & (0.242) & (0.225) \\
Consumer exp - Apr & -1.025 & -0.197 & -3.028 & -1.074 & -0.522 & 3.196 & -2.723 & -1.578 & -0.518 & -2.736* \\
 & (1.719) & (1.438) & (1.861) & (1.749) & (2.194) & (2.450) & (1.930) & (1.772) & (1.713) & (1.545) \\
Var consumer expenditures & -0.285** & -0.226* & 0.215 & -0.0314 & -0.253 & -0.00731 & 0.626** & -0.0232 & 0.269** & 0.0654 \\
 & (0.115) & (0.128) & (0.181) & (0.172) & (0.172) & (0.190) & (0.292) & (0.144) & (0.127) & (0.200) \\
Incr COVID-19 cases & -0.0571 & -0.117 & -0.155 & 0.150 & -0.00545 & 0.0152 & -0.196 & 0.0901 & -0.105 & 0.0586 \\
 & (0.119) & (0.103) & (0.113) & (0.146) & (0.145) & (0.168) & (0.153) & (0.125) & (0.107) & (0.124) \\
Republican leaning news & -0.798*** & -0.257 & -0.0948 & -0.0899 & -0.915*** & -0.963*** & 0.189 & -0.511** & -0.412 & -0.682* \\
 & (0.266) & (0.269) & (0.320) & (0.306) & (0.339) & (0.231) & (0.363) & (0.210) & (0.258) & (0.372) \\
Democratic leaning news & 0.0883 & 0.164 & 0.210 & -0.236 & -0.527* & -0.279 & -0.292 & 0.0885 & 0.114 & -0.174 \\
 & (0.235) & (0.217) & (0.331) & (0.271) & (0.295) & (0.267) & (0.538) & (0.229) & (0.264) & (0.279) \\
Constant & -2.076** & -1.800 & -3.218*** & -3.648*** & 0.138 & 0.259 & 0.535 & -3.711** & -1.449 & -1.600 \\
 & (0.819) & (1.214) & (1.217) & (1.182) & (1.268) & (1.210) & (1.281) & (1.433) & (1.055) & (1.126) \\
 & & & & & & & & & & \\
Observations & 751 & 789 & 585 & 721 & 532 & 554 & 332 & 777 & 689 & 691 \\
Controls & Yes & Yes & Yes & Yes & Yes & Yes & Yes & Yes & Yes & Yes \\
Average increase & 0.265 & 0.262 & 0.244 & 0.144 & 0.158 & 0.168 & 0.126 & 0.220 & 0.225 & 0.187 \\
Average decrease & 0.210 & 0.244 & 0.233 & 0.325 & 0.251 & 0.254 & 0.224 & 0.233 & 0.231 & 0.215 \\ \hline
\multicolumn{11}{l}{%
 \begin{minipage}{1.6\columnwidth}%
  \small \textit{Notes}: Standard errors in parentheses. *** p\textless{}0.01, ** p\textless{}0.05, * p\textless{}0.1.\\
  All regressions are logistic regressions that take into account population survey wights and the sampling procedure. The dependent variable is a dummy=1 if the respondent increased their belief that it's a government's responsibility to provide the following policies. The control variables include: gender, race, age, education, parental status, and caring responsibilities for an elderly or a person with a disability, income in February 2020, housing, labor force participation and employment status in February 2020, health insurance provider, and whether respondents had financial difficulties before the pandemic, the area in which the respondents live, whether it's a metropolitan or rural area, and the population density in the zip code. We also control for whether respondents completed the related surveys in shorter time than the 99$^{th}$ percentile, and ceiling effects.
 \end{minipage}%
}\\
\end{tabular}%
}
\end{table}

\begin{table}[H]
\centering
\caption{The effect of shocks and media on support for coronavirus relief policies - logistic regression model}
\label{tab:covid_policy_logit}
\resizebox{0.8\textwidth}{!}{%
\begin{tabular}{lccc} \hline \hline
 &
 \multicolumn{3}{c}{\begin{tabular}[c]{@{}c@{}}Stronger belief between May and October 2020 \\ that the government should:\end{tabular}} \\
 & (1) & (2) & (3) \\
 &
 \begin{tabular}[c]{@{}c@{}}Spend more on \\ public healthcare \\ to reduce \\ preventable deaths\end{tabular} &
 \begin{tabular}[c]{@{}c@{}}Do more to \\ protect \\ essential workers\end{tabular} &
 \begin{tabular}[c]{@{}c@{}}Transfer money \\ directly to families \\ and businesses \end{tabular} \\ \hline
 & & & \\
Republican & -0.296 & 0.492 & -0.333 \\
 & (0.337) & (0.299) & (0.280) \\
Democrat & 1.201*** & 0.704** & -0.0875 \\
 & (0.320) & (0.316) & (0.261) \\
Lost 20\% income & 0.492* & 0.555* & 0.490** \\
 & (0.260) & (0.282) & (0.227) \\
Knows hospitalized & 0.265 & 0.0465 & -0.122 \\
 & (0.253) & (0.194) & (0.199) \\
Var consumer expenditures & 0.110*** & 0.0846 & 0.113** \\
 & (0.0394) & (0.0918) & (0.0499) \\
Consumer exp - May & -0.583 & -0.374 & -0.177 \\
 & (1.222) & (1.440) & (1.065) \\
Incr COVID-19 cases & 0.170 & -0.0222 & -0.0619 \\
 & (0.104) & (0.139) & (0.107) \\
Republican leaning news & 0.177 & -0.674* & -0.263 \\
 & (0.317) & (0.362) & (0.311) \\
Democratic leaning news & -0.105 & 0.0723 & -0.107 \\
 & (0.311) & (0.380) & (0.268) \\
Constant & -2.162* & -3.002** & -1.891** \\
 & (1.146) & (1.155) & (0.773) \\
 & & & \\
Observations & 610 & 553 & 742 \\
Controls & Yes & Yes & Yes \\
Average increase & 0.177 & 0.188 & 0.181 \\
Average decrease & 0.295 & 0.317 & 0.369 \\ \hline
\multicolumn{4}{l}{%
 \begin{minipage}{\columnwidth}%
  \small \textit{Notes}: Standard errors in parentheses. *** p\textless{}0.01, ** p\textless{}0.05, * p\textless{}0.1.\\
  All regressions are logistic regressions. Observations have been re-weighted with entropy weights, so that the group of individuals who incurred an income shock and the group that did not are balanced in terms of a set of demographics. The dependent variable is a dummy=1 if the respondent increased their support for the following policies. The control variables include: gender, race, age, education, parental status, and caring responsibilities for an elderly or a person with a disability, income in February 2020, housing, labor force participation and employment status in February 2020, health insurance provider, and whether respondents had financial difficulties before the pandemic, the area in which the respondents live, whether it's a metropolitan or rural area, and the population density in the zip code. We also control for whether respondents completed the related surveys in shorter time than the 99$^{th}$ percentile, and ceiling effects.
 \end{minipage}%
}\\ 
\end{tabular}%
}
\end{table}

\begin{table}[H]
\centering
\caption{The effect of shocks and media on institutional trust - logistic regression model}
\label{tab:trust_logit}
\resizebox{\textwidth}{!}{%
\begin{tabular}{lccccccc} \hline \hline
 & (1) & (2) & (3) & (4) & (5) & (6) & (7) \\
 & \begin{tabular}[c]{@{}c@{}}Congress \\ \& Senate\end{tabular} & \begin{tabular}[c]{@{}c@{}}White\\ House\end{tabular} & \begin{tabular}[c]{@{}c@{}}Financial\\ institutions\end{tabular} & \begin{tabular}[c]{@{}c@{}}Private\\ sector\end{tabular} & \begin{tabular}[c]{@{}c@{}}Scientific\\ community\end{tabular} & \begin{tabular}[c]{@{}c@{}}Health \\ insurance \\ companies\end{tabular} & Hospitals \\ \hline
 & & & & & & & \\
Republican & -0.445* & -0.480** & 0.246 & 0.183 & 0.105 & -0.202 & 0.195 \\
 & (0.234) & (0.229) & (0.226) & (0.295) & (0.232) & (0.237) & (0.192) \\
Democrat & 0.151 & 0.930*** & 0.228 & 0.0589 & -0.581*** & -0.204 & -0.245 \\
 & (0.188) & (0.289) & (0.263) & (0.251) & (0.201) & (0.218) & (0.206) \\
Lost 20\% income & 0.327* & 0.0742 & 0.406** & 0.348* & -0.131 & 0.218 & 0.285* \\
 & (0.196) & (0.227) & (0.183) & (0.186) & (0.158) & (0.190) & (0.165) \\
Knows hospitalized & 0.241 & 0.404 & 0.168 & 0.179 & 0.265 & 0.127 & 0.200 \\
 & (0.188) & (0.256) & (0.192) & (0.247) & (0.228) & (0.185) & (0.187) \\
Var consumer expenditures & -0.0357 & 0.256 & 0.0117 & -0.0929 & 0.0576 & -0.295*** & -0.255** \\
 & (0.185) & (0.165) & (0.183) & (0.126) & (0.172) & (0.101) & (0.116) \\
Consumer exp - Apr & 1.078 & -1.526 & 0.118 & -1.375 & -0.0892 & -0.159 & 1.620 \\
 & (1.181) & (1.686) & (1.675) & (1.243) & (1.517) & (1.418) & (1.298) \\
Incr COVID-19 cases & -0.0356 & -0.0897 & 0.00596 & 0.194* & 0.184 & -0.0353 & -0.104 \\
 & (0.107) & (0.131) & (0.112) & (0.117) & (0.112) & (0.109) & (0.124) \\
Republican leaning news & -0.638** & -0.563* & -0.469 & -0.385 & 0.271 & 0.258 & 0.485** \\
 & (0.254) & (0.289) & (0.285) & (0.261) & (0.301) & (0.315) & (0.240) \\
Democratic leaning news & -0.0559 & 0.611* & -0.119 & -0.233 & -0.0985 & 0.315 & 0.137 \\
 & (0.215) & (0.325) & (0.236) & (0.234) & (0.277) & (0.247) & (0.243) \\
Constant & 0.399 & 1.041 & -0.235 & -3.007*** & -1.050 & -1.866** & -0.604 \\
 & (0.952) & (1.042) & (0.908) & (0.689) & (0.865) & (0.923) & (0.734) \\
 & & & & & & & \\
Observations & 862 & 631 & 917 & 927 & 976 & 881 & 981 \\
Controls & Yes & Yes & Yes & Yes & Yes & Yes & Yes \\
Average increase & 0.159 & 0.142 & 0.142 & 0.185 & 0.185 & 0.172 & 0.164 \\
Average decrease & 0.351 & 0.312 & 0.299 & 0.247 & 0.286 & 0.284 & 0.306 \\ \hline
\multicolumn{8}{l}{%
 \begin{minipage}{1.25\columnwidth}%
  \small \textit{Notes}: Standard errors in parentheses. *** p\textless{}0.01, ** p\textless{}0.05, * p\textless{}0.1.\\
  All regressions are OLS regressions that take into account population survey wights and the sampling procedure. The dependent variable is a dummy=1 if the respondent decreased trust in the above institutions. The control variables include: gender, race, age, education, parental status, and caring responsibilities for an elderly or a person with a disability, income in February 2020, housing, labor force participation and employment status in February 2020, health insurance provider, and whether respondents had financial difficulties before the pandemic, the area in which the respondents live, whether it's a metropolitan or rural area, and the population density in the zip code. We also control for whether respondents completed the related surveys in shorter time than the 99$^{th}$ percentile, and ceiling effects.
 \end{minipage}%
 }
\end{tabular}%
}
\end{table}

\newpage

\end{document}